\newcommand{\pa}{\partial}
\newcommand{\be}{\begin{equation}}
\newcommand{\ee}{\end{equation}}
\newcommand{\bea}{\begin{eqnarray}}
\newcommand{\eea}{\end{eqnarray}}
\def \ci{\cite}
\def \YY {{\rm Y}}
\def \lra {\leftrightarrow}
\newcommand{\nn}{\nonumber}
\newcommand{\p}[1]{(\ref{#1})}
\newcommand{\bt}[1]{{\bar t}}

\newcommand{\ellK}{{\rm K}}

\newcommand{\ellE}{{\rm E}}

\def \sm {{sigma-model}}

\newcommand{\sn}{\mathop{\mathrm{sn}}\nolimits}
\newcommand{\cn}{\mathop{\mathrm{cn}}\nolimits}
\newcommand{\dn}{\mathop{\mathrm{dn}}\nolimits}

\documentclass[12pt]{article}
\input epsf
\usepackage{epsfig,color,latexsym}
\def \sql {{\sqrt{\l}}\ }
\def \del{\partial}
\def \a {\alpha}
\def \aa {{\a'}}
\def\g{\gamma}
\def\s{\sigma}
\def\z{\zeta}
\def\zi{\zeta_1}
\def\zii{\zeta_2}
\def\ov{\over}
\def\la{\label}
\def\I{{\cal I}}
\def\J{{\cal J}}

\def \om {\omega}

\def\E{{\cal E}}
\def\w{\omega}
\def\b{\beta}
\def\l{\lambda}
\def\eps{\epsilon}
\def\vep{\varepsilon}
\def \adss{$AdS_5 \times S^5$\ } 

\def \r { \rho}
\def \sql {\sqrt{\lambda} }
\def \t {\theta}
\def \p {\phi}
\def \vp {\varphi}
\def \Om {\Omega}
\def \ads {{$AdS_5$}}
\def \ov {\over}
\def \s{\sigma}
\def \pa{\partial}
\def \ta{\tau}
\def \sh {\sinh}
\def \ha {{1 \over 2}}

\def \la{\label}

\def \k {\kappa}
\def\foot{\footnote}

\def \const {{\rm const}}
 
\def \J {{\cal J}}
\def \L {\Lambda}
\def\rr {{\rm r}}

\def \sa {\sum_{a=1}^2} 
\newcommand{\rf}[1]{(\ref{#1})}
\renewcommand{\theequation}{\thesection.\arabic{equation}}
\renewcommand{\thefootnote}{\fnsymbol{footnote}}

\def\appendix#1{
  \addtocounter{section}{1}
  \setcounter{equation}{0}
  \renewcommand{\thesection}{\Alph{section}}
  \section*{Appendix \thesection\protect\indent \parbox[t]{11.15cm}
  {#1} }
  \addcontentsline{toc}{section}{Appendix \thesection\ \ \ #1}
  }

\newcommand{\eq}[1]{(\ref{#1})}

\def \four{{\textstyle {1\ov 4}}}
 \def \third { \textstyle {1\ov 3}}
\def\det{\hbox{det}}
\def\be{\begin{equation}}
\def\ee{\end{equation}}

\def \ci {\cite}

\def \foot {\footnote}
\def \bi{\bibitem}
\def \tr {{\rm tr}}
\def \ha {{1 \over 2}}
\def \td {\tilde}

\def \ci{\cite}
\def \N {{\cal N}}
\def \ww {\Omega} 
\def \const {{\rm const}}

\def \ss {\sum_{i=1}^3 }

\def \t {\tau} 
\def\S{{\cal S} }

\def \nn {\nu}
\def \XX {{\rm X}}
\def \Om {\Omega}
\def \vom {{\bar \omega}}
\def \Y{{\rm Y}} 
\def \zz {{\rm z}}
\def \rL {{L}}
\def \ab {{\rm a}}

\def \n {\nu}

\begin{document}

%\vspace*{1cm}
\null\vskip-24pt 
\vskip-1pt
\hfill
%OHSTPY-HEP-T-03-???
%Imperial/TP/2-03/2
\vskip-1pt
%\hfill {\tt hep-th/0311004}
\vskip0.2truecm
\begin{center}
\vskip 0.2truecm {\Large\bf
Spinning strings and AdS/CFT duality
}
%\\
\vskip 1.5truecm
%\vfill
{\bf A.A. Tseytlin\footnote{Also at Imperial College London
and  Lebedev
 Institute, Moscow.}\\
\vskip 0.4truecm
\vskip .2truecm
 {\it Department of Physics,
The Ohio State University,\\
Columbus, OH 43210-1106, USA}\\
 }
\end{center}
\vskip 0.5truecm
\vskip 0.2truecm \noindent\centerline{\bf Abstract}
\vskip .2truecm
We review  a special  class of   semiclassical string states
in $AdS_5 \times S^5$ that have a regular 
expansion of their  energy in integer powers of the ratio of the 
square of string tension (`t Hooft coupling) 
and the square of large angular momentum in $S^5$. 
They  allow one to  quantitatively check 
the AdS/CFT duality in non-supersymmetric sector of states and 
also help to uncover the role of   integrable  structures 
on the two sides of the string theory -- gauge theory  duality. 

\vskip 0.5truecm
\vskip 0.5truecm
\vskip 0.5truecm
{\it  To appear in  Ian Kogan Memorial Volume,  
 ``From Fields to Strings: Circumnavigating Theoretical Physics'', 
  M. Shifman, A. Vainshtein, and J. Wheater, eds.
  (World Scientific, 2004).}

\def \Y{{\rm Y}}

\def\[{\begin{equation}}
\def\]{\end{equation}}
\def\<{\begin{myeqnarray}}
\def\>{\end{myeqnarray}}

\def \lc {light-cone\ }
\def \sm {$\s$-model }
\def \la {\langle}
\def \ra {\rangle}
\def \e {{\rm e}} \def \four{{\textstyle{1\ov 4}}}
\def \ov {\over}
\def \we { \wedge}
\def \F {{\mathcal F}}
\def \ep {\epsilon}
\def \k {\kappa}
\def \N {{\mathcal N}}
\def \L {{\mathcal L}}
\def \K {{\mathcal K}}
\def \I {{\mathcal I}}
\def \J {{\mathcal J}}

\def \a {\alpha}
\def \E {{\mathcal E}}
\def \b {\beta}
\def \g {\gamma}
\def \G {\Gamma}
\def \d {\delta}
\def \l {\lambda}
\def \La {\Lambda}
\def \m {\mu}
\def \n {\nu}
\def \s {\sigma}
\def \S {\Sigma}
\def \r {\rho}
\def \t {\theta}
\def \ta {\tau}
\def \p {\phi}
\def \P { \Phi}
\def \vp {\varphi}
\def \ev {\varepsilon}
\def \ps {\psi}
\def \rL   {{\rm L}}
\def \frac#1#2{{ #1 \over #2}}
\def \td {\tilde}
\def \M {{\mathcal M}}
\def \aa {{\a'}}

\def \adss {$AdS_5 \times S^5\ $}
\def \ads {$AdS_5$\ }
\def \pw {plane wave\ }
\def \N {{\mathcal N}}
\def \lc {light-cone\ }
\def \ta { \tau}
\def \s { \sigma }
\def  \sqf {\l^{1/4}}
\def \sg {\sqrt {g }}
\def \vp {\varphi}
\def \fourth {{1 \ov 4}}
\def \fo  {{{\textstyle {1 \ov 4}}}}
\def \inv {^{-1}}
 \def \diag {{\rm diag}} \def \td { \tilde }

\def \ab {{\rm a}} 

\def \la {\label}
\def \alpr {\alpha'}
\def \om {\omega}
\def \del{\partial}
\def \Tr {{\rm Tr}}
\def \R {R^{(2)} }
\def \la {\label}
\def \tr {{\rm tr}}
\def \ha {{1 \over 2}}
\def \ov {\over}
\def \JJ {{\mathcal J}} 
\def \Om {\tilde \Omega}
\def \ome {\Omega}

\def \z {\zeta}
\def \w  {\omega}
\def \oo  {\omega}
\def \ww { {\rm w} }
\def \www { {\rm w}_{21} }
\def \Ev {E_{\rm vac}}
\def \sql {{\sqrt{\l}}\ }
\def \tri {{\textstyle {1\ov 3}}}
\def \ta {\tau}
\def \tdr {{\td \r}}
\def \ap {\approx}
\def \isql {{\textstyle { 1 \ov \sqrt{\l}}}}

\def \D {{\Delta}}
\def \ta {\tau}

\def \sp {$S^5$ }

\def \kk {{\rm k}}
\def \po {{\psi_0}}
\def \sql {{\sqrt{\l}}\ }
\def \del{\partial}

\def\E{{\mathcal E}}
\def\w{\omega}
\def\b{\beta}
\def\l{\lambda}
\def\eps{\epsilon}
\def\vep{\varepsilon}
\def \rL {{\rm L}} 
%%%%%%%%%%%%%%%%%%%%%%%

\def \ads {{$AdS_5$}}
\def \ov {\over}
\def \s{\sigma}
\def \pa{\partial}
\def \ta{\tau}
\def \sh {\sinh}
\def \ha {{1 \over 2}}

\def \k {\kappa}
\def\foot{\footnote}
\def \const {{\rm const}}

\def \four{{\textstyle {1\ov 4}}}
 \def \third { \textstyle {1\ov 3}}
\def\det{\hbox{det}}
\def\be{\begin{equation}}
\def\ee{\end{equation}}

\def \foot {\footnote}
\def \bi{\bibitem}
\def \tr {{\rm tr}}
\def \ha {{1 \over 2}}
\def \td {\tilde}
\def \ci{\cite}
\def \N {{\mathcal N}}
\def \ww {\Omega} 
\def \const {{\rm const}}
\def \ss {\sum_{i=1}^3 }
\def \t {\tau} 
\def\S{{\mathcal S} }
\def \nn {\nu}
\def \XX {{\rm X}}

\newcommand{\nln}{\nonumber\\}
\newcommand{\nl}{\nonumber\\&&\mathord{}}
\newcommand{\nlnum}{\\&&\mathord{}}
\newcommand{\nle}{\nonumber\\&=&\mathrel{}}
\newcommand{\earel}[1]{\mathrel{}&#1&\mathrel{}}
\newenvironment{myeqnarray}{\arraycolsep0pt\begin{eqnarray}}{\end{eqnarray}\ignorespacesafterend}
\newenvironment{myeqnarray*}{\arraycolsep0pt\begin{eqnarray*}}{\end{eqnarray*}\ignorespacesafterend}

\newcommand{\lrbrk}[1]{\left(#1\right)}
\newcommand{\bigbrk}[1]{\bigl(#1\bigr)}
\newcommand{\brk}[1]{(#1)}
\newcommand{\vev}[1]{\langle#1\rangle}
\newcommand{\normord}[1]{\mathopen{:}#1\mathclose{:}}
\newcommand{\lrvev}[1]{\left\langle#1\right\rangle}
\newcommand{\bigvev}[1]{\bigl\langle#1\bigr\rangle}
\newcommand{\bigcomm}[2]{\big[#1,#2\big]}
\newcommand{\comm}[2]{[#1,#2]}
\newcommand{\lrcomm}[2]{\left[#1,#2\right]}
\newcommand{\acomm}[2]{\{#1,#2\}}
\newcommand{\bigacomm}[2]{\big\{#1,#2\big\}}
\newcommand{\gcomm}[2]{[#1,#2\}}
\newcommand{\lrabs}[1]{\left|#1\right|}
\newcommand{\abs}[1]{|#1|}
\newcommand{\bigabs}[1]{\bigl|#1\bigr|}
\newcommand{\bigeval}[1]{#1\big|}
\newcommand{\eval}[1]{#1|}
\newcommand{\lreval}[1]{\left.#1\right|}

%%%%%%%%%%%%%%%%%%%%%%%%%
\def \lra {\leftrightarrow}

\def \vom {{\bar \omega}}
\def \E {{\mathcal  E}} \def \J {{\mathcal  J}}
\def \YY {{\rm Y}}

\def \d {\del} 

\def \sms {sigma models\ }
\def \sm {sigma model\ }
\def \L {\Lambda}
\def \gl {\ell} 
\def \Q {{\mathcal Q}}
\def \tr {{\rm tr\ }}
\def\z{\zeta}
\def\zi{\zeta_1}
\def\zii{\zeta_2}
\def\K{\mbox{K}}    
\def\eE{\mbox{E}}   \def \vt {\vartheta} 
\def \vr {\varrho}
%%%%%%%%%%%%%%%%%%%%%%%%%%%%%%%%%%%%%%%%%%%%
\def \wup {w}

\renewcommand{\theequation}{\thesection.\arabic{equation}}
\renewcommand{\thefootnote}{\fnsymbol{footnote}}

\def \sn {{\rm sn}}
\def \dn {{\rm dn}}\def \cn {{\rm cn}}

\def \vo{\omega}
\def \wup {w}
\def \half {\ha}
\def \cN {{\mathcal N}} \def \rN {{\rm N}}
\def \rr {\vr}

\def \rr{{\rm r}}
\def \zz{{\rm z}}
\def \q{{\rm q}} 
\def \sa {\sum_{a=1}^2}

\newpage

\tableofcontents
\newpage

%%%%%%%%%%%%%%%%%%%%%%%%%%%%%%%%%%%%%%%%%%%%
\setcounter{equation}{0}
%%%%%%%%%%%%%%
%%%%%%%%%%%%%%%%%%%%%%%%%%%%%%%%%%%
\section{Introduction }
%%%%%%%%%%%%%%%%%%%%%%%%%%%

Better understanding of the duality between type IIB superstring theory 
in \adss  and planar limit of $\cN=4$ supersymmetric Yang-Mills 
theory \ci{dua} and  extending it  to less supersymmetric cases 
may allow us to find   simple  string-theoretic 
 descriptions  of  various 
dynamical aspects of gauge theories, from high-energy scattering 
to confinement. 
This AdS/CFT duality is usually  viewed   as an example 
of strong coupling -- weak  coupling duality:   while the large $N$ 
perturbative expansion in SYM  theory  assumes 
that  the 't Hooft coupling $\l= g^2_{\rm YM} N $ is small, 
the string perturbative (inverse tension)
expansion applies for $ \sql= {R^2\ov \a'} \gg 1$. 
In general, observables (like scaling dimensions, 
correlation functions, finite temperature free energy, etc.) 
depend on $\l$ through 
 non-trivial functions $f(\l)$ of the couplings, with the perturbative SYM  
 and string theories describing  opposite asymptotic regions. 
 For special ``protected''   BPS  observables the 
 dependence on $\l$ may become trivial 
  due to  supersymmetry 
 and then  can be directly reproduced on the two sides of the duality
 \ci{aha}.  
 
Checking duality  beyond the BPS cases  remains  a
  challenge. The Green-Schwarz superstring action in \adss 
  appears to have  a complicated structure  \ci{Met},
   so finding, e.g.,  its full  spectrum exactly in $\l$ seems  hard
   at present. 
  One may hope to by-pass the 
computation of non-trivial 
 functions $f(\l)$ by considering special limits
involving other parameters or quantum numbers besides $\l$.
A specific progress can be made by concentrating  
on a particular  (and basic) 
class of observables which should be related according to AdS/CFT: 
a spectrum of energies of single-string states (in global 
\adss coordinates)  and scaling dimensions of the corresponding
single-trace  gauge-invariant local operators in  SYM theory.
These may carry quantum numbers like 
$SO(2,4) \times SO(6)$ spins of string states
or 
powers  of scalar fields 
and  covariant derivatives 
in the  SYM operators. 
A remarkable recent  development   initiated in  \ci{bmn} 
(which in turn was inspired by \ci{papa,meet})
is based on the idea that for a  special subset of string/SYM states 
parametrized by large quantum numbers  \ci{sas,gkp}
%$Q$ 
there may be new interesting  limits 
%(symbolically, $Q \to \infty$, ${\l \ov Q^2}={\rm fixed} < 1$) 
in which certain quantum corrections 
%(e.g., inverse tension corrections on the string side)
 may be  suppressed. 
One  may then be able to 
 check the AdS/CFT for such special non-supersymmetric states 
  by comparing  the corresponding   string  energies 
to the  perturbative gauge-theory  scaling dimensions. 

In the BMN case \ci{bmn} (see  \ci{bmnr} for reviews)
one concentrates on a particular 
``semiclassical'' \ci{gkp}  sector of  near-BPS 
states represented by small closed strings with center of mass 
moving along a large circle of $S^5$ with angular momentum $J \gg1 $. 
 The  SYM operators 
 are of the type tr$(Z^J...)$,\ $Z=\Phi_5 + i \Phi_6$
where $\Phi_M$ are $SO(6)$ scalars  and dots stand for a small number of
other SYM  fields or covariant derivatives. 
By considering the limit $J \to \infty$, ${\l \ov J^2}=$fixed\ 
one is able to  establish a precise correspondence between the 
energies of the string states  and   scaling dimensions of the 
corresponding SYM operators \ci{bmn,zan} (for a complete 
list of references see \ci{bmnr}). 
The reason why  this is possible  can be understood by
 interpreting 
this sector of states as ``semiclassical''  states \ci{gkp}
corresponding to quadratic fluctuations near point-like string 
running along a geodesic in  $S^5$ with  
angular velocity $w= {J\ov \sql}$. 
One is then able to argue \ci{ft1,tsec} that 
higher than 1-loop string \sm corrections to the leading 
(``quadratic''   or ``plane-wave'') 
 string energies are suppressed  in the 
 limit $J \to \infty, \ w$=fixed.
 
One may hope to apply a similar reasoning 
to other, far from BPS,  semiclassical sectors of string states.
For example, considering  a string rotating with large spin $S$ in $AdS_5$ 
one discovers \ci{gkp}   a  new {\it qualitative}  test of AdS/CFT: 
the agreement between the dependence of
the string  energy $E$ on  large  spin $S$  and 
the spin dependence 
of the  anomalous dimension of  twist 2 
 gauge-invariant SYM operators:
% (${S\ov\sql}\gg 1$):  
$ E= S +  f(\l) \ln S + ... $.
Here $f(\l) = b_0 \sql + b_1 + { b_2\ov  \sql} + ...$
on perturbative string side 
 and  $f(\l) = a_1 \l + a_2\l^2  + ...$
on perturbative gauge theory  side  \ci{gkp,ft1,kot}. 
According to the AdS/CFT duality the   two expansions  
 must  represent
 different asymptotics of the same function. Checking  this in a 
 precise manner 
  is obviously hard since that  would require first finding 
 all terms in
 the respective perturbative series and then  resumming them. 

For other    semiclassical string states one might   expect 
to find similar ``interpolation in $\l$'' pattern, 
precluding the  direct {\it quantitative} comparison 
to perturbative  SYM theory. 
Remarkably, as was noticed  in \ci{ft2,ft3},  
there are exceptions: for certain multispin string states 
(with at least one large $S^5$ spin component $J$) 
the classical energy has a {\it regular} expansion in $\l\ov J^2$ 
while   quantum  superstring  \sm corrections are  suppressed 
in the limit $J \to \infty, \ {\l\ov J^2}=$fixed. 
It was proposed \ci{ft2} that for such states one can  carry out 
the  precise test of the  AdS/CFT duality in a  non-BPS sector
by comparing the  ${\l\ov J^2 }\ll 1$ 
expansion of the {\it classical} string energy 
with  the corresponding {\it quantum}  anomalous dimensions
in  perturbative SYM theory.

This was indeed successfully accomplished 
in a series of  recent papers  \ci{mz2,ft4,afrt,bfst,AS,mz3}. 
The main technical problem -- how to find eigenvalues of  
anomalous dimension matrix for ``long'' (large $J$) 
scalar operators --  was solved (at the one-loop level) 
using the interpretation 
of the  anomalous dimension matrix as 
an integrable spin-chain Hamiltonian 
\ci{mz1,BM}. 
%%%%%%%%%%%%%%%%%%%%%%%%%%%%%%%%%%%%%%%%%%%%%%%
\foot{Integrable spin chain
connection  was uncovered  and  extensively studied 
previously in  QCD context.
%(or asymptotically free gauge theories in general). 
 In particular, the Regge asymptotics  
of scattering amplitudes was described by evolution equations
that were related to the $SL(2,C)$ Heisenberg spin chain
\ci{lip}. More importantly for the present discussion, the 
one-loop anomalous dimensions of certain 
(quasipartonic) composite operators were  related 
to the energies of the $SL(2,R)$ XXX Heisenberg 
spin chain \ci{bra,beli}.
Similar  relations hold in other asymptotically free 
gauge theories, in particular,  
 supersymmetric theories
\ci{belm,lipk}. 
The role of  conformal symmetry 
in QCD and these   integrability relations were reviewed 
in \ci{bkm}.
More recent work relating 
integrability of light-cone 
QCD operators to gauge/string duality 
appeared in \ci{gor,gorr,gorrr}. 
In  $\N$=4 SYM theory  viewed as a particular gauge theory
with adjoint matter the above QCD-inspired work 
 implies that 
the (1-loop, large $N$) anomalous dimension  matrix 
for the minimal-twist operators (like
 tr$(\bar \Phi D^S \Phi) + ...$, \ $D= D_0 + D_3$)  should be the same as 
 the Hamiltonian of the $SL(2,R)$ XXX spin chain. 
Independently, 
 it was observed in \ci{mz1} that 
 the one-loop planar anomalous dimension matrix   in the 
 pure-scalar sector of operators tr$(\Phi_{M_1}...
 \Phi_{M_J})$ can be interpreted as a  Hamiltonian of 
an integrable $SO(6)$ spin chain. 
Ref.\ci{BM} generalized these facts to  all superconformal
operators to claim that 
the complete  one-loop planar dilatation operator 
of $\N$=4 SYM is
equivalent to a Hamiltonian of an integrable $SU(2,2|4)$ 
(super) spin chain. More recent work \ci{bel}
addressed the same problem using the original (operators on 
light-cone)  QCD approach, i.e. considering   the  subsector of 
supermultiplets of quasipartonic operators ( 
tr$(D^{s_1}\Phi_{M_1} ... D^{s_n}\Phi_{M_n})+ ...$, etc.)
with the conclusion that in this case the one-loop 
dilatation operator  coincides with the Hamiltonian of 
$SL(2|4)$ spin chain. The relation between the approaches 
of \ci{BM} and \ci{bel} and also whether the $SL(2)$
integrability in the twist 2  sector may be somehow 
 related by supersymmetry 
 %in the context of $\N$=4 SYM theory 
   to $SO(6)$ integrability in the pure-scalar sector 
seems worth  clarifying further.}
 This allowed one  to find  the one-loop anomalous dimensions
 by applying  the Bethe ansatz \ci{FF} techniques.
 The leading order $\l \ov J$ terms in the energies
  of particular string solutions were then reproduced 
  as one-loop anomalous dimensions on the SYM side 
  by choosing  particular Bethe root distributions in the
  ``thermodynamic'' limit  of 
  ``long'' ($J \to \infty$) 
  operators. There is some evidence \ci{bfst} that
  the  correspondence
  extends, as one of course expects,  to the next $\l^2\ov J^3 $  order, 
  but checking this explicitly  and going beyond the 2-loop level 
  remains an important open   problem.

\bigskip 

Our aim here will be to review  a class of such   classical
string solutions in \adss 
\ci{ft1,ft2,ft3,ft4,afrt,art}
whose energy $E$ has a regular 
expansion in integer powers of 
 $\l$ (i.e.  the square of the effective 
 string tenson) divided over the square of the  total  $S^5$
 spin $J$, and for which quantum \sm corrections to  $E$ 
 should be suppressed in the $J \to \infty $ limit. 
 
\bigskip 

%%%%%%%%%%%%%%%%%%%%%%%%%%%%%%%%%%%%%%%%%%%%%%%%%%%%%%%%%%%%%%%%
Let us first  make some general comments on the structure of this 
semiclassical expansion for the string energy. 
The form of a 
 classical  solution cannot  depend on the value of the string tension, 
 i.e. on $\sql$, which appears as  a factor in front 
the string  action 
$I= {\sql \ov 4 \pi} \int d^2 \xi \ G_{\m\n} (x) \del_a
X^\m \del^a  X^\n+... $. Thus the classical energy 
can be written as  $E= \sql \E(w)$, where $w$ stands for all 
constant parameters 
that enter the classical solution. These parameters should be 
 fixed in the standard sigma model loop ($\a' \sim {1 \ov \sql} $) expansion.
The classical values of the  integrals of motion  like  the 
$S^5$ and $AdS_5$ angular momentum   components  are also 
proportional to the string tension, e.g.,  
$J= \sql \J(w)$  (they    take  integer values 
 in the full quantum theory).
Expressed in terms of these integrals   the  classical 
energy is 
$E=\sql \E ( \J) =\sql  \E ( {J\ov \sql}).$

  In the limit of 
 large values of semiclassical parameters and the corresponding quantum
 charges  the classical energy  of a string  solution in any
$AdS_p \times S^m $  space 
goes as {\it  linear} function of $J$, i.e.  
$E = J + ... $.  This  linear behaviour \ci{vesa}
(seen explicitly on  examples of particular solutions 
 \ci{vega,gkp,ft1,rus,mina})  is different from the flat-space
 Regge  one 
 $E \sim \sqrt J $ and 
is  a consequence of  the constant curvature  of 
$AdS$ space. This  is   consistent with the  AdS/CFT duality:  
one expects that the large $J $  expression for 
 the full dimension  of  the corresponding gauge-theory  operator
 should start with its  canonical  dimension.
 % ( total spin).  

We would like  to identify a class of  special
 classical string solutions in \adss   
 whose energy  has a particular
 dependence on conserved  charges 
 that allows for a direct comparison with anomalous dimensions on the 
 perturbative SYM side. 
 While such extended string solutions turn out to   have 
  {\it several} conserved global  charges, here for 
   notational simplicity  we shall keep track 
  of just one of them  -- total $S^5$ 
angular momentum  $J=\sql \J$. 
% (explicit examples will be discussed in detail  below). 
For the solutions we will be interested in 
 the classical
energy $E =\sql \E$  should  have  the following expansion  
in large classical parameter   $ \J \gg 1$ 
\be\la{kokl}
  \E = 
  \J \big(1  + { c_1 \ov \J^2} + { c_2 \ov \J^4} + ...\big) , \ \ee 
i.e. $\E\ov \J$  should  have  an  
expansion  in  {\it even}  inverse powers of $\J$. 
The coefficients 
$c_i$ may be functions of ratios of conserved charges that
are finite in the large-charge limit. 
 Equivalently, for $
{1 \ov \J} = {\sql \ov J} \ll 1$ 
\be\la{jojk}
E = \ J\ \big(1 + { c_1 \l \ov J^2} + { c_2 \l^2 \ov J^4} + ...\big)
=   J  + { c_1 \l \ov J} + { c_2 \l^2 \ov J^3} + ... \ ,  \ee
which   formally 
looks like an 
expansion in positive  integer powers of $\l$. 
Rotating string solutions with this  property  were indeed found in 
\ci{ft1,ft2,ft4,afrt,art} and will be reviewed below. 

Furthermore, let us assume that in such  cases  the  string \sm 
loop corrections to the energy 
which  in general can be computed in the standard inverse tension
expansion 
\be \la{yte}
E_{\rm tot}
= \sql \E(\J) + \E_1(\J) + { 1 \ov \sql} \E_2(\J) + 
 {1 \ov (\sql)^2} \E_3(\J) + .... 
 =E+ \sum_{n=1}^\infty E_n 
 \    \ee 
 should  have  the following  specific form of their  expansion 
  in  $\J \gg 1$
\be \la{tqp}
\E_n(\J) = { d_{n1} \ov \J^{n+1}}  + { d_{n2} \ov \J^{n+3}}  + ...  \ ,
\ \ \ \ \ \ \   n=1,2,...\ \  .  \ee
This behaviour was verified in \ci{ft3}  for $n=1$  on a particular  example 
of a solution \ci{ft2} satisfying \rf{jojk}.
Eq.\rf{tqp} implies that the $n$-loop term in the quantum-corrected energy 
\rf{yte} will be given,  for  ${\l\ov J^2 } \ll 1$, by  
\be \la{oppu}
 E_n= {1 \ov (\sql)^{n-1}} \E_n(\J) = 
 {  d_{n1}\l \ov J^{n+1}}  +  {  d_{n2} \l^2 \ov J^{n+3}}  + ...  \ . \ee
In general, the   energy $ E_{\rm tot}= E_{\rm tot}( J, \l) $
should be  some function of $J$ and the 
string tension
but if  the above assumptions \rf{jojk} and \rf{oppu} are  true,  
 it 
 will  be given by the following 
{\it double expansion} in $ {\l\ov J^2}$ {\it and}  $1 \ov J $:
\be \la{iipy}
E_{\rm tot}= J\bigg[ 1  + \sum_{k=1}^\infty  ({ \l\ov J^2})^k 
\bigg( c_k + \sum_{n=1}^\infty  { d_{nk}\ov J^n} \bigg) \bigg] 
\ . \ee
Then   if we first take the limit of  $J \gg 1 $
 for fixed ${\l\ov J^2}$,  
 all quantum \sm corrections  will be suppressed 
and the full energy 
$E_{\rm tot}$ will be given   just by  its {\it classical}
 part $E$ \rf{jojk}. 
%Note that it is not  necessary to assume 
%that ${\l\ov J^2} < 1 $ to suppress all quantum corrections, 
%provided $J \gg 1$ and  \rf{tqp} is satisfied. 
%  can then further assume that ${\l\ov J^2} < 1 $ 
%and develop a perturbative expansion in this latter parameter. 

In the BMN case \ci{bmn},  where one expands \ci{gkp,ft1}
near a point-like BPS 
string state,  the ground-state energy is not renormalized, i.e. $
 E_{\rm tot}=E= J$, but the double 
 expansion similar  to the one for $E_{\rm tot}/J$ 
 in  \rf{iipy}, namely, 
 $ E_{\rm fluct}= J +   \sum_{k=0}^\infty 
  ({ \l\ov J^2})^k 
\big( h_k + \sum_{n=1}^\infty  { f_{nk}\ov J^n} \big) $, 
  applies  to energies of string fluctuations near the
   geodesic, 
  i.e. to energies of excited string modes \ci{ft1,tsec,par,cal}.
 In the limit  $J \to \infty$ 
 their energies are then determined by 
 the quadratic (``one-loop'' or ``plane-wave'') approximation. 
 
 The general conditions for the validity of the expansions \rf{jojk} and
\rf{oppu}  remain
 to be clarified (some observations made 
in  recent papers \ci{matt,mik} relating large $\J$ 
limit to an ultra-relativistic limit 
 may turn out to be useful for that). 
 In particular, the ``regularity'' of the expansion 
 of the  energy  in $\l$ \rf{jojk} may apply not
only to multi-spin rotating but also to $S^5$ pulsating \ci{mina,mz3}
solutions.

%%%%%%%%%%%%%%%%%%%%%%%%%%%%%%%%%%%%%%%%%%%%%%%%%%%%%%%%

In order to  test the AdS/CFT duality one should reproduce the same 
 expression for the (quantum-corrected) $AdS_5$ string 
 energy 
$E_{\rm tot} (\l,J)$  \rf{iipy}  as the exact 
 scaling dimension $\Delta (\l,J)$
of the corresponding SYM  operator, i.e.  as 
a particular eigenvalue  of the dilatation operator  having the 
 same global charges (i.e. belonging to the same 
  $SO(2,4) \times SO(6)$
 representation as the string state).
  Given that \rf{iipy}
looks like an expansion in the `t Hooft coupling $\l$
it is  natural to expect   that the perturbative ($\l \ll 1$) 
expansion for 
$\Delta (\l,J)$ can be organized in the following way
\be \la{ity}
\Delta (\l,J) = J  +  \sum_{k=1}^\infty  q_k (J)\ \l^k \ , 
\ee
where the functions $q_k (J)$ should have the 
following form   for  $J\gg 1$ 
\be \la{jz}
q_k (J) = {1 \ov J^{2k-1} }\big(a_{k} + {a_{k1} \ov J } + ...\big) \ . 
\ee
Assuming that  this is indeed 
the case and taking the $J\to \infty$ limit,
 one could then directly 
 compare the classical part 
 of the  energy in \rf{iipy} 
expanded in $\l \ov J^2$ with the sum of the leading  ($J \gg 1$) 
terms  at each order of expansion of $\Delta$ in powers of $\l$.
The AdS/CFT implies then that 
 the two expressions should  coincide, i.e. that $c_k=a_k$. 
The {\it classical} string energy  should thus represent 
the leading $J\to \infty$ term 
in the  {\it quantum}  SYM scaling  dimension.

In particular, the coefficient $c_1$ 
of the first subleading (order $\l$) term in the classical string energy 
\rf{jojk} should  match the coefficient $a_{1}$ in 
 the one-loop SYM term in \rf{ity},\rf{jz}.
 This was indeed  verified  on specific examples in 
 \ci{mz2,ft4,afrt,bfst,mz3}. There is also a numerical evidence 
 \ci{bfst}
 that this matching extends  to the $\l^2$ term, i.e.
 $c_2= a_{2}$. 
 
 The $J \to \infty$  behavior \rf{jz} of the one-loop correction
 to the anomalous dimensions was checked  using the spin chain
 relation and the  Bethe
 ansatz   for particular large R-charge or  large spin operators 
 \ci{mz2,bfst,mz3}. The general proof of \rf{jz}  which should 
 follow from  a higher-loop structure of the 
 dilatation operator \ci{beik,bei3}  and should be heavily 
 based  on  the superconformal symmetry of the $\cN=4$ SYM theory
 remains to be given.

%%%%%%%%%%%%%%%%%%%%%%%%%%%%%%%%%%%%%%%%%%%%%%%%

\bigskip

Let us now summarize the contents of the following sections. 
In section 2 we shall write down the bosonic part of the superstring action 
in \adss and the corresponding integrals of motion 
as a preparation for a discussion of classical 
finite energy closed string solutions which carry several 
$SO(2,4) \times SO(6)$ spin components. 
In section 3 we shall consider 
the special case of $SO(6)$ invariant \sm (embedded into string 
theory by adding time direction from $AdS_5$) 
and  briefly review its integrability  (local and non-local
conserved charges, etc.). 

Then in section 4 we shall concentrate on   a particular 
class of semiclassical string states rotating in $S^5$ with three 
angular momenta $J_i$ and show that for  the rotating string
ansatz the $R_t \times S^5$ \sm reduces to a 
well-known one-dimensional integrable system -- Neumann-Rosochatius (NR)
system. Its special case is the $n=3$ Neumann system describing an oscillator 
on a 2-sphere. This relation allows one to classify the corresponding 
rotating string solutions, which, as in flat space, can be of folded 
or circular type. 

In section 5 we shall study a
 simple special class of circular rotating string  solutions on $S^5$ 
 whose  energy  has a regular large-spin 
expansion as in \rf{jojk}.  We shall  also  determine (in section 
 5.3) 
the 
spectrum of  quadratic fluctuations near these circular solutions 
pointing out some analogy with the point-like (BMN) case. 
In section 5.4 we shall consider the one-loop string \sm  correction 
to the energy of a particular solution (with two equal spins); 
this one-loop correction   indeed  turns out to be suppressed 
in the large spin limit,  in agreement with \rf{iipy}. 

The discussion of sections 4 and 5 will be generalized 
in section 6 to the case of  states 
represented by semiclassical strings 
 rotating in both $S^5$ and $AdS_5$ 
and thus carrying 3+2 spin quantum numbers. They are again described 
by a generalised NR integrable system. 
While the energy of strings rotating only in $AdS_5$ 
is non-analytic in $\l$ (section 6.1), the expansion \rf{jojk} is
true for  circular strings having  large $S^5$ spin components. 

Similar conclusions apply to other multi-spin  solutions 
of the NR system representing folded and circular strings 
with more complicated (``inhomogeneous'') dependence on the 
string coordinate $\s$.  In particular, we consider 
 a class of two-spin 
solutions for which the Neumann  system 
degenerates to a sine-Gordon one 
and, as a result, the solutions 
are expressed in terms of the  elliptic functions (section 7). 
 The classical energy  can then  be  found as a solution of the  two 
  parametric equations  involving elliptic integrals
  and has again a regular expansion as in \rf{jojk}.
 
 Section 8 will contain a summary of some open problems  and possible
 generalizations, including a brief discussion of pulsating string 
 solutions. 

%%%%%%%%%%%%%%%%%%%%%%%%%%%%%%%%%%%%%%%%%%%%
\setcounter{equation}{0}
%%%%%%%%%%%%%%%%%%%%%%%%%%%%%%%%%%%%%%%%%%%%
\setcounter{footnote}{0}
%%%%%%%%%%%%%%%%%%%%%%%%%%%%%%%%%%%
\section{Closed superstrings in $AdS_5 \times S^5$:\\
\  classical solutions }
%%%%%%%%%%%%%%%%%%%%%%%%%%%

Superstrings in \adss can be  described by a Green-Schwarz
 action \ci{Met}  which   defines a consistent perturbation theory near each 
semiclassical string configuration, e.g., point-like 
massless geodesic in a light-cone type  gauge as in \ci{thorn,ft1}
or extended string configurations  as in \ci{dgt,ft1,ft2,ft3}.

The bosonic part of the action in the conformal gauge is the  sum of the two 
  coset-space  sigma models ($AdS_5= SO(2,4)/SO(1,4)$ and 
  $S^5=SO(6)/SO(5)$ ones)
 \be \la{ac}
I= - { \sql  \ov 4\pi }
\int d\tau d\sigma  \ \big[ G^{(AdS_5)}_{mn}(x)
\del_a x^m \del^a  x^n\ + \    G^{(S^5)}_{pq}(y)  \del_a y^p
\del^a y^q \big] \ . \ee
%\be\la{ten}\ \ \ \sql \equiv  { R^2 \ov
%\aa}  \ . \ee 
The effective string tension $T_{\rm eff} = \frac{\sql}{2\pi}=
\frac{R^2}{2\pi\a' }$
is related to the `t Hooft coupling $\l= g^2_{\rm YM} N$ 
on the SYM side of the string/gauge theory duality \ci{dua}. 
  The $AdS_5$ and $S^5$  parts of the action 
  are  ``coupled'' at the classical 
  level through the conformal
  gauge constraints. 
  
  The classical conformal 
  invariance of this \sm is preserved at 
   the quantum level  after addition of fermions with coupling 
   to the metric and R-R 5-form background \ci{Met}.  
    There are  quadratic and quartic  fermionic terms 
  in the action  
 (in a particular gauge). 
 The quadratic   part of the
fermionic Lagrangian   can be written as 
(see, e.g.,  \ci{Met,dgt,ft1})
\be \la{gff}
L_F=
i (\eta^{ab }\delta^{IJ} -
\ep^{ab } s^{IJ} ) \bar \vt^I \vr_a D_b \vt^J   \ ,\ \ \ \ \ \
\vr_a \equiv \G_{A}  e^A_a \ ,
\ee
where
 $I,J=1,2$,  $s^{IJ}=$diag(1,-1), and
  $\vr_a\equiv \G_{A}  E^{A}_\m \del_a X^\m
  $ are the projections of the 10-d Dirac
matrices.
Here $X^\a$ are the  string coordinates
(given functions of $\tau$ and $\s$ for a particular classical
solution)
corresponding to the  \ads\ ($\m=0,1,2,3,4$) and $S^5$
($\m=5,6,7,8,9$)
factors.
The covariant derivative $D_a$
can be put into  the  following form
\be  \la{form}
D_a\vt^I   = (\delta^{IJ} {\rm D}_a
- { i \ov 2 } \epsilon^{IJ}  \G_* \vr_a ) \vt^J\ ,
\ \ \ \ \ \  \G_* \equiv i \G_{01234} \ , \ \
\G_*^2 =1 \ ,\ee  where
 $  {\rm D}_a = \del_a
+\fourth    \omega^{AB}_a\Gamma_{AB}  ,\ $ 
 $ \omega^{AB}_a \equiv   \del_a  X^\a \omega^{AB}_\a$
 and
 the ``mass term''  originates from the R-R  5-form
 coupling.
   
Here we  will be interested mostly 
in the classical bosonic finite-energy solutions 
for closed strings in \adss space and ignore 
 the fermions.     To study these bosonic solutions 
      it is useful 
to rewrite the  action \rf{ac} in the form 
\be
I={ \sql }\int d\tau \int^{2\pi}_0 {d\sigma\ov 2 \pi} 
 \ (  L_{AdS} + L_S )\ 
, \la{Lsp} \ee
where
\bea
\la{SL}
L_{AdS}&=&-\frac{1}{2}\eta^{PQ}\pa_a Y_P\pa^a Y_Q + 
\frac{1}{2}
\td \Lambda (\eta^{PQ}Y_P Y_Q+1)\, , \\ 
L_S&=&-\frac{1}{2}\pa_a X_M\pa^a X_M+\frac{1}{2}\Lambda 
(X_M X_M-1)\,  .
\la{SSL}
\eea
We use $(-+)$ signature on the world sheet and 
  $X_M$, $M=1,\ldots , 6$  and $Y_P$, 
 $P=0,\ldots , 5$ are  
the embedding coordinates of ${R}^6$ 
with the Euclidean metric $\delta_{MN}$ in $L_S$ and 
 with  $\eta_{PQ}=(-1,+1,+1,+1,+1,-1)$ 
 in $L_{AdS}$,  respectively.
 $\Lambda$  and $\td \Lambda$ are the Lagrange multiplier
 functions of $\tau$ and $\s$.
The action \eq{Lsp} is  to be supplemented 
 with the usual conformal gauge  constraints
 expressing the vanishing of the  total 2-d  energy-momentum 
 tensor 
 \be \la{cv}
 \eta^{PQ} ( \dot{Y}_P\dot{Y}_Q+  Y'_P  Y'_Q ) 
 +   \dot{X}_M\dot{X}_M+ X_M ' X_M ' =0 \ , \ee
 \be \la{cvv}
 \eta^{PQ} \dot{Y}_P {Y'}_Q+ \dot X_M  X_M '  =0 \ , \ee
 where 
 \be\la{coo}
  \eta^{PQ}Y_P Y_Q=- 1\ , \ \ \ \ \ \ \ \  X_M X_M =1 \ . \ee 
 We shall assume that the world sheet is a cylinder, i.e. 
 impose the
 closed string periodicity conditions 
 \be \la{peri}
 Y_P (\s + 2\pi) = Y_P (\s) \ , \ \ \ \ \ \ \ \ 
  X_M (\s + 2\pi) = X_M (\s) \ . \ee 
 The classical  equations that  follow from \rf{Lsp} 
 can be written as 
 \be \la{ade}
 \del^a \del_a Y_P - \td \L Y_ P =0 \ , \ \ \ \ \ \ \ 
 \td \L =  \eta^{PQ} \del^a  Y_P \del_a Y_Q  \ ,\ \ \ \ 
  \eta^{PQ}Y_P Y_Q=- 1\ ,\ee 
 \be \la{sse}
 \del^a \del_a X_M   +  \L  X_M  =0 \ , \ \ \ \ \ \ \ 
 \L = \del^a X_M  \del_a  X_M  \ , \ \ \ \  X_M X_M =1 \ . \ee  
 The action is invariant under the $O(2,4)$ and $O(6)$  global 
 symmetries with the corresponding conserved (on-shell) 
 charges being 
 \be \la{cha}
 S_{PQ}= \sqrt{\lambda } \int_0^{2\pi } {d\s\over 2\pi }
\ (Y_P \dot Y_Q  -  Y_Q\dot Y_P)\ , \ee
\be \la{ss}
J_{MN}= \sqrt{\lambda } \int_0^{2\pi } {d\s\over 2\pi }
\ (X_M\dot  X_N  -  X_N\dot X_M)\ .
\ee
We are interested  in finding ``spinning'' solutions 
that have non-zero values of these charges. 
The physical  target-space  interpretation of the
solutions depends on a particular choice of coordinates
(that solve \rf{coo})  in  $AdS_5$ and $S^5$. 
One natural (``global coordinate'')  choice is 
$$
\YY_1\equiv Y_1+ iY_2 = \sinh \r \ \sin \theta \  e^{i \phi_1}\
,
 \ \ \ \ \
\YY_2 \equiv 
Y_3 + i Y_4  = \sinh \r \ \cos \theta \  e^{i  \phi_2}\  ,  
$$
\be 
\YY_0 \equiv Y_5 + i Y_0 = \cosh \r \  e^{i t }\ , \ \ \ \ \ \ \
\ \ 
\XX_3\equiv X_5 + i X_6 =
\cos  \g \ e^{ i \vp_3} \ , \la{relx} \ee  $$
\XX_1\equiv  X_1 + i X_2 = \sin   \g \ \cos
\psi \ e^{ i \vp_1} \ , \ \ \ \
\XX_2 \equiv X_3 + i X_4 =  \sin   \g \ \sin \psi \
e^{ i \vp_2} \ . $$
Then there is an obvious choice of the 3+3 Cartan generators  of
$SO(2,4) \times SO(6)$ corresponding to the 3+3 linear isometries
of the 
\adss metric 
\be\la{ads}
(ds^2)_{AdS_5}
= d \rho^2 - \cosh^2 \rho \ dt^2 + \sinh^2\rho \ (d \theta^2 +
 \cos^2 \theta \ d \phi^2_1 + \sin^2 \theta \ d\phi_2^2) \ , 
\ee
 \be\la{ses}
(ds^2)_{S^5}
= d\gamma^2 + \cos^2\gamma\ d\varphi_3^2 +\sin^2\gamma\
(d\psi^2 +
\cos^2\psi\ d\varphi_1^2+ \sin^2\psi\ d\varphi_2^2) \ , 
\ee
i.e. to the 
 translations in $AdS_5$ time $t$, in the two 
angles $\phi_a$   and in three angles of $S^5$ $\vp_i$:
\be \la{aas}
S_0 \equiv  S_{50} \equiv E = \sql \E\ , \ \ \ \  S_1\equiv
S_{12}= 
\sql \S_1 \ , \ \
\  S_2\equiv S_{34} =  \sql \S_2 \ , \ee
\be \la{jij}
J_1 \equiv  J_{12} = \sql \J_1 \ , \ \ \ \ \ 
J_2 \equiv  J_{34}  =\sql \J_2 \ , \ \ \ \ \ 
J_3 \equiv  J_{56}= \sql \J_3 \ . \ee 
We will be interested  in classical solutions that  have finite
values
of the target-space energy $E$ as well as of the spins
$S_a,J_i$. 
For a solution to  have a consistent semiclassical approximation, 
i.e. to correspond to an eigen-state of the Hamiltonian which
carries the corresponding quantum numbers (and thus 
being associated to a particular SYM operator with definite scaling
dimension)  
all other non-Cartan (i.e. non-commuting) 
components of the  symmetry generators \rf{cha},\rf{ss}  should
vanish \ci{ft1}. These correspond to ``highest-weight'' states; 
other members of the multiplet can be obtained by applying rotations 
to a ``highest-weight'' solution.

In the above  $R^{2,4}$
embedding representation of $AdS_5$  the charges of
the isometry group $SO(2,4)$   can be related to 
 the boundary  SYM theory 
  conformal group generators  as follows ($\m,\n=0,1,2,3$): 
  \be \la{geni}
S_{\m\n} = M_{\m\n} \ , \ \ \ \ \ \ 
 \ S_{\m 4} = \ha (K_\m - P_\m)\ , \ \ \ \ \ \ 
 \    S_{\m 5} = \ha (K_\m + P_\m) \  , \   \ \ \ \
  \ \ S_{54} = D\ . \ee
One can identify the  standard  spin with   $S_1 = S_{12} = M_{12}$,
the second  (conformal) spin  with  $S_2 = S_{34} = \ha (K_3- P_3)$,
 and finally the conformal energy 
 with the  rotation generator  in the $05$ plane, i.e. 
   with the
global $ AdS_5$ energy,  $E= S_{05} = \ha (K_0 + P_0)$.
\foot{The energy of a string state  in global $AdS_5$ space 
  with boundary $R \times
 S^3$  should be equal to the energy of the corresponding SYM  state 
 on $R\times S^3$ (which can  be mapped conformally to $R^4$). 
 Through radial quantization this  state may be
 associated to a local operator that creates it. 
 At the same time,  the $AdS_5$ energy or conformal Hamiltonian 
  generates an $SO(2)$ subgroup of $SO(2,4)$  while the dilatation
  operator (whose eigenvalues are scaling dimensions) 
  generates an $SO(1,1)$ subgroup of the conformal group.
   Their  eigenvalues happen to be  the same 
  since the two  representations (the unitary one 
  classified by $SO(4) \times SO(2)$  and the one 
  classified by  $SO(4) \times SO(1,1)$) 
   are related by a global $SO(2,4)$ 
  similarity transformation (see, e.g., \ci{dol}).
  Alternatively, after the
Euclidean continuation 
 $Y_0\to  i Y_{0E}$  (to allow for the  mapping 
 from  $R \times S^3$ to $R^4$)
 one may exchange   $Y_{0E}$ with $Y_4$ which 
exchanges the generator  $S_{54} = D$ with $ S_{05} = \ha ( P_0 +
K_0)= E$.
For all the solutions discussed below $S_{50}=E\not=0$ while 
$S_{54}=D=0$. 
  One could, in principle, apply a similar $Y_{0E} \to Y_4$
    transformation to 
 string solutions, getting  equivalent ones   (but more 
 complicated-looking, with radial direction of $AdS_5$ depending on
 $\tau$)
  that  would  have  
 non-zero values of the $SO(1,1)$  generator $S_{54}$. 
%  of the  $AdS_5$ symmetry group. 
% I am grateful to N. Beisert and G. Arutyunov for clarifying 
 % discussions of this issue.
  } 
  
\bigskip

Let us first consider  point-like string solutions, for which 
$Y_P=Y_P(\tau), \ X_M=X_M (\tau)$, i.e. massless
(cf. \rf{cv})  geodesics in \adss.
As follows from the second-order equations in 
 \rf{ade},\rf{sse},  in this case $\Lambda=\const$, 
$\tilde \Lambda=\const$, i.e. $Y_P$ and $X_M$  are given by 
trigonometric functions. The constraint \rf{cv} implies that the two 
frequences are related: $\Lambda = - \td \Lambda >  0$. 
Then generic massless geodesic  in \adss  
can be shown 
%(see also  \ci{tsec}) 
  to  be 
 of the two ``irreducible'' types
 (up to a global $SO(2,4) \times SO(6)$
transformation): (i) massless 
 geodesic that stays entirely within $AdS_5$; (ii) a   geodesic 
  that runs along the time direction  in $AdS_5$ 
  and  wraps a big 
 circle of $S^5$. 
In the latter case the angular motion in $S^5$ provides 
an effective (``Kaluza-Klein'') mass to a particle in $AdS_5$, 
i.e. the corresponding geodesic in $AdS_5$ is a massive one. 
Then we can choose the coordinates so that 
%This conclusion changes if a point-like 
%string is allowed to orbit in $S^5$: such  most general massless 
%\adss geodesic is equivalent, up to an  $SO(2,4)\times SO(6)$ 
%rotation,  to a  string being at rest at the center of $AdS_5$ 
%(moving only along time direction) and orbiting along big circle 
%of $S^5$, i.e. 
\be \la{bam}
Y_5+i Y_0 = e^{i \k \tau}\ , \ \
 \ \ \ 
X_5+ i X_6 = e^{i w \tau} \ , \ \ \ \ \ \k=w= \sqrt \Lambda  \ , \  \ \ \ 
Y_{1,2,3,4}=X_{1,2,3,4}=0 \ . \ee
Here  the only non-vanishing integrals of motion are 
$E=J_3=\sql  \k$, representing the energy and 
$SO(6)$ spin  of this BPS state, corresponding to tr$Z^{J_3}$
operator in SYM theory.
More generally, one may choose any massless geodesic 
in $AdS_5$ for which then $\eta^{PQ} \dot Y_P \dot Y_Q= - w^2$. 
The massless limit  $w \to 0$  corresponds to $J_3 \to 0$, 
i.e. the resulting state should be representing a 
 vacuum  in string theory or a  unit operator in SYM theory \ci{bmn}.

The former case, i.e. the  ``massless''  $w\to 0$  limit, 
 is actually subtle:  naively,  
 a massless geodesic in $AdS_5$ 
does not represent a semiclassical string state in the sense  it was 
defined  above.   
Indeed, for  a point-like string moving inside  $AdS_5$ we have 
$\eta^{PQ} \dot Y_P \dot Y_Q=0$, i.e. 
 $\ddot Y_P=0$. Thus in terms of the  embedding coordinates the massless 
 geodesic  is   a straight line 
\be\la{geo}
Y_P(\tau)  = A_P + B_P \tau  \ , \ \ \ \  \ \ \ \ 
 \eta^{PQ} B_P B_Q=\eta^{PQ} A_P B_Q=0\ , \ \ \ \
 \eta^{PQ} A_P A_Q=-1\ .
\ee 
Then the $SO(2,4)$ angular momentum tensor  \rf{cha} 
is $S_{PQ}= \sql (A_P B_Q- A_Q B_P)$  and can be shown to always have 
non-vanishing non-Cartan components. 
Indeed, by applying an $SO(2,4)$ rotation we may put the constant
vectors  $ A_P $ and $ B_P$ in a  canonical form:
$A_P=(0,0,0,0,0,1), \  B_P=({ p} ,0,0,{ p},0,0)$, 
i.e. 
\be \la{taks}
Y_5+ i Y_0 = 1 + i { p }\ \tau\ , \ \ \ \ \  \ \ \ \ 
 Y_3= { p } \ \tau\  , \ \ \ \ \ \ \ \  Y_{1,2,4}=0 \ .  \ee
Here  $ { p}$ is an arbitrary 
parameter  and 
$S_{50}=S_{53}=\sql { p}$.
An alternative choice of the parameters (related 
to the above one  by  an $SO(2,4)$ rotation with parameter 
 $u$) gives 
$
Y_5+ i Y_0 = {1 + u^2 \ov 2 u} + i {p\ov u} \tau ,  \ 
Y_1+ i Y_2 = {1 - u^2 \ov 2 u}  + i  {p\ov u} \tau , 
 \ Y_{3,4}=0$.\foot{In this case in addition to the Cartan  components 
$E=S_{50}= \sql {1 + u^2 \ov 2 u}p$ and 
$S_{12}= -\sql {1 - u^2 \ov 2 u}p$ we also have 
nonvanishing $S_{01}$ and $S_{25}$.}
 It  corresponds to
 the massless geodesic running parallel to the 
$R^{1,3}$ boundary in the Poincare coordinates where 
\ $ (ds^2)_{AdS_5}= {1 \ov z^2} ( dx^m dx_m + dz^2)$:
 $x_0=x_3=p \tau, \ z=u=\const$ (see also \ci{tsec}).
An expansion near this geodesic is used to define the light-cone
gauge in \ci{thorn}, i.e. it should represent a light-cone vacuum state.

\bigskip

Below   we would like to study  non-trivial ($\s$-dependent) 
solitonic solutions 
of classical closed string equations in \adss that have finite 
2-d energy and carry finite space-time energy and spins, 
i.e. 1+2 plus 3 commuting conserved charges of the $O(2,4) \times
O(6)$ isometry group. The conformal gauge constraints
 will then imply a
relation 
between the energy and the spins  ($a=1,2;\  i=1,2,3$) 
\be \la{spi}
\E=  \E (\S_a,\J_i; k_p) \ , \ \ \ \ \  {\rm i.e.}
\ \ \ \ \ \
E= \sql \E ({S_a\ov \sql},{J_i\ov \sql}; k_p) \ , \ee
where $k_p$ are ``topological'' numbers   determining
 particular type 
(e.g.,  shape) of the rotating solutions.
We will be interested in solutions that  have a regular
dependence of $E$ on 
on $\l$  in the large spin limit as in \rf{jojk}.
%i.e. there $E$ admits  an expansion 
%in integer powers of $\l$. 
A necessary condition for that appears to be to have 
large total angular momentum in $S^5$ direction.
That applies
to both rotating \ci{ft2}  and oscillating \ci{mina} 
 solutions. Note also  that rotating solutions in $S^5$ (but not in
 $AdS_5$) have a ``nearly-BPS'' interpretation \ci{matt} 
 in the  formal $\l\to 0$ limit.

\bigskip 

In general, coset space \sms are known to be integrable
\ci{lup,ZM}. To make this  
 formal integrability property explicit and useful 
 one needs to specify a class 
of solutions by choosing a  special  ansatz for string
coordinates. 
Before discussing particular 
rotating strings  in $S^5$ and 
 $AdS_5$  let us first make some general comments  on 
 the corresponding classical 
 sigma model and its conserved charges.

%%%%%%%%%%%%%%%%%%%%%%%%%%%%%%%%%%%%%%%%%%%%
\setcounter{equation}{0}
%%%%%%%%%%%%%%
 
%%%%%%%%%%%%%%%%%%%%%%%%%%%%%%%%%%%%%%%%%%%%
\setcounter{footnote}{0}
%%%%%%%%%%%%%%%%%%%%%%%%%%%%%%%%%%%
%%%%%%%%%%%%%%%%%%%%%%%%%%%%%%%%%
\section{$R_t \times S^5$  \sm : classical integrability \\ 
\ and conserved charges}
%%%%%%%%%%%%%%%%%%%%%%%%%%%%%%%%%%%

Let us  consider the  classical 
$S^5$ sigma model embedded into string theory  by adding 
extra time direction
$R_t$. This may be viewed as a special case of \adss  sigma model
where the string is placed at the center $\rho=0$ of $AdS_5$
while moving  in $S^5$.  

Introducing $\xi^\pm = \ha ( \tau \pm \s) $ and $ 
\d_\pm =\d_\t \pm \d_\s$  the corresponding equations
of motion  and conformal gauge constraints 
 can be written as (cf. \rf{sse})
\be\la{ssw}
\del_+ \del_-  X_M   + (\del_+ X_N \del_- X_N)  X_M =0 \ , \ \ \
\
\ \ \   X_M X_M =1  \ , \ee 
\be \la{coni} 
\del_+ X_M \d_+ X_M = (\d_+ t)^2  \ , \ \ \ \ \ \ \ \ \
\del_- X_M \d_- X_M  = (\d_- t)^2  \ ,  \ee
where $t$ satisfies $\del_+ \del_- t=0$,  which has  general
solution ($\k=\const$) 
\be \la{ggg}
t= \k \tau + h_+(\tau +\s) + h_-(\tau -\s) \ . \ee 
The  equations  \rf{ssw},\rf{coni} 
 are invariant under 2-d conformal transformations, 
$\xi^\pm \to  F_\pm  (\xi^\pm )$, so given a solution 
$X_M (\xi^+,\xi^-)$ one can find  another one as 
$\td X_M (\xi^+,\xi^-) = X_M (F_+ (\xi^+ ), F_- (\xi^- )).   $
One can also use this residual  conformal symmetry 
to  make the components of the  stress tensor 
$ \del_+ X_M \d_+ X_M$ and $\del_- X_M \d_- X_M$ equal to a
 constant, or, 
 what is equivalent in the
present case,  to  gauge away $h_\pm$ in \rf{ggg}, 
 putting  $t$ in the form $ t= \k \tau$.
When only 3 of $X_M$'s are non-zero (as
  in the case of the $O(3)$ invariant  \sm) 
  one can show \ci{lup} 
 that \rf{ssw} reduces to the 2-d sine-Gordon equation 
 \be 
 \del_+\del_- \alpha  + \sin \alpha =0 \ , \ \ \ \ \ \ \ \ 
 \cos \alpha = \del_+ X_M \del_- X_M \ . \ee 
 Similar reduced systems can be derived also from 
  other $O(n)$ invariant sigma models 
 \ci{eich}.\foot{A  relation to  
sine-Gordon  system appeared  previously 
in the context of strings 
moving in constant curvature spaces  in \ci{BN,veg}.}

The above  equations \rf{ssw}
admit  various  special solutions. One is the  ``flat-space'' 
or ``chiral'' 
solution (for which the Lagrange multiplier $\L$ in \rf{sse}
vanishes):
$X_M = f^+_M (\xi^+)$ or  $X_M = f^-_M (\xi^-)$ for 
particular   values of $M$. In contrast to flat-space case,  
 a linear combination of such solutions is no
longer a solution, so  
one may thus say that  \rf{ssw} 
describes scattering of left-moving and right-moving 
light-like  energy lumps \ci{lup}. 
 For chiral $X_M$ to  satisfy the string-theory  constraints \rf{coni}  we need 
 to make a special choice of $h_\pm$ in 
  $t$.
 %Since we consider the theory on a cylinder, 
% i.e. need also to impose the  periodicity conditions \rf{peri} 
% $h_+$ (and $f^+_M$) is then given  by Fourier series 
% of $\sin n (\tau + \s)$ and $\cos n (\tau + \s)$
 %and cannot contain a term linear in $\tau$, i.e. $t$ does  not
 %have a center-of-mass term, which precludes 
 %string-theory  interpretation of
 %such solutions. 

 Let  us now  review various  types of local and non-local
  conserved currents in
this \sm (see, e.g., \ci{eich,ogi}). 
One can define  a first-order linear system (Lax pair) 
\ci{lup} whose consistency 
is equivalent to  the equations \rf{ssw}:
\bea
\la{fof}
\d_+ R^{(\gl)} = {  (1-\gl^{-1}}) j_+ R^{(\gl)}  \ , \ \ \ &&\ \ \
\d_- R^{(\gl)} = {  (1-\gl ) } j_-   R^{(\gl)} \ , 
\\ R^{(\gl)}  (R^{(\gl)})^T&=& (R^{(\gl)})^T R^{(\gl)} 
=I \ , \nonumber \eea
where $R^{(\gl)} $ is an $so(6)$ matrix  and 
\be  \la{cur}
 (j_a)_{MN} = 2( X_M \d_a X_N - X_N \d_a   X_M)  \ . \ee  
One can then construct a new solution from a given one 
 as $X^{(\gl)}_M  = R^{(\gl)}_{MN}  X_N$. 
 Solving  \rf{fof} by the inverse scattering method 
 %on an infinite line appears to be unclear
 is subtle  due to complications related to the 
 choice of   boundary
 conditions \ci{eich,ZM} (e.g., on an infinite line, 
 for $j_a \to 0$ at spatial infinity the solution 
 $R^{(\gl)}$ does not have a plane-wave behaviour). 
 Still, \rf{fof}  may be used as  a  basis for analysing the
 integrability properties  of the \sm.

 One approach is to look at non-abelian (non-commuting)
non-local conserved charges related to  Yangians \ci{ber}. 
At the same time, it is important
also to study  an infinite family
of commuting local conserved charges 
whose existence is a manifestation of 
 integrability of the corresponding
 equations of motion. These may be  constructed 
 using the B\"acklund transformation.
 If $X_M$  is a given ``trial''  solution of
\rf{ssw}, 
 let is define 
its B\"acklund transform $X^{(\g)}_M$  as another solution  
satisfying  \ci{ogi,AS}
\be
\d_+ (X^{(\g)}_M  +  X_M) = \ha ( 1 + \g^{-2}) X^{(\g)}_N \d_+
X_N  \  (X^{(\g)}_M  - X_M)\ , \ee
\be \la{back}
\d_- (X^{(\g)}_M  -  X_M) = -\ha ( 1 + \g^{2}) X^{(\g)}_N \d_-
X_N  \  (X^{(\g)}_M  + X_M)\ , \ee
$$ 
 X^{(\g)}_M X^{(\g)}_M = 1 \ , \ \ \ \ \
X_M X_M =1 \ , \ \ \ \ \  X^{(\g)}_M X_M = { 1 - \g^2 \ov 1 +
\g^2} \ ,
\ \ \   \  X^{(0)}_M = X_M\  . $$
Here $\g$ is a spectral parameter. 
One  can write the solution of the equations \rf{back} 
in expansion in $\g$  
\be \la{kak}
X^{(\g)}_M =X_M +  \sum^\infty_{k=1} X_{(k) M} \g^k \ , \ \ \ \
\ \  \ \ 
X_{(1) M} = { 2 \d_+ X_M \ov \sqrt{ \d_+ X_N \d_+ X_N } }  \ ,
... \ . \ee 
One can define the generating function of {\it local}  commuting 
conserved {\it scalar} charges associated with the 
original solution $X_M$  by \ci{ogi,AS} 
\be \la{copm}
Q(\g) = \ha \g  \int^{2\pi}_0 { d \s \ov 2\pi} \ 
X^{(\g)}_M ( \del_+ X_M   + \g^2  \del_- X_M ) 
= \sum^\infty _{k=2} Q_k \g^k \ ,  \ee 
\be \la{vii} 
 Q_2 = \ha  \int^{2\pi}_0 { d \s \ov 2\pi}
  X_{(1) M} \d_+ X_M = \int^{2\pi}_0 { d \s \ov 2\pi}
\sqrt{ \d_+ X_N \d_+ X_N } = \int^{2\pi}_0 { d \s \ov 2\pi} 
\del_+ t  \ , \ee \be \la{nnn}
  Q_k = \ha  \int^{2\pi}_0 { d \s \ov 2\pi}
 \big( X_{(k-1) M} \d_+ X_M  + X_{(k-3) M} \d_+ X_M \big) 
 \ , \ \ \ \  k \geq 3 \ .    \ee 
Here in \rf{vii}   we used the constraint \rf{coni}.
Then $Q_2$ can be interpreted as the space-time energy:
since the general solution for $t$  is given by \rf{ggg},  
we conclude  that  $Q_2= \k= \E.$ 
For values of these charges on specific solutions see \ci{AS}.

One can also define 
an infinite number of conserved {\it non-local}  
$so(6)$ Lie algebra valued (i.e. {\it matrix}) 
currents and associated 
charges  as for any principal  or coset \sm 
\ci{lup,bre}.  Let us  follow \ci{bre} 
 and  replace $X_M$  by an orthogonal $O(6)$ 
(or unitary SU(4)) matrix 
\be g= e^{i \pi P} =  1 - 2 P  \ ,  \ee
where $P_{MN} \equiv  X_M X_N$ is a projector since $X_M X_M
=1$. 
Then $ g=g^{-1}= 1 - 2 P$  and
%($A_\pm = A_\tau \pm A_\s$) 
%\be A_a = S \d_a S^T\ , \ \ \ 
%U= \ha [(1 + \ell)^{-1} A_-  -  (1 - \ell)^{-1} A_+] \ ,  \ee 
%\be V= - \ha [(1 + \ell)^{-1} A_-  +  (1 - \ell)^{-1} A_+] \ ,
\ \ \
%\d_\tau U - \d_\s V + [U,V] =0 \ , \ee
\be \la{crr} 
\tr (\d^a g \d_a g^{-1}) = 8 \d^a X_M  \d_a  X_M \ , \ \ \ \ 
j_a\equiv g^{-1} \d_a  g = j_a(X) 
\ , \ \ \ \ \ \ \ \d^a j_a =0 \ ,  \ee
where the conservation of $j_a(X) $ given by \rf{cur} 
 follows from the equations 
of motion  \rf{sse},\rf{ssw}  for $X_M$. 
Defining $D_a =  \d_a + j_a $ we get 
$[\d_a, D_a]=0$. Then starting with the conserved current $j_a$
one
can construct an infinite sequence of conserved 
non-local currents $j^{n}_a$ 
using the following iterative procedure.
Given a conserved current $j^{(n)}_a$ we define a
matrix  function $\chi^{(n)}$
and use it to construct the next conserved current
\be 
j^{(n)}_a = \ep_{ab} \d^b \chi^{(n)}\ , \ \ \ \ \ \ 
j^{(n+1)}_a = D_a  \chi^{(n)}\ ,\ \ \ \ 
j^{(1)}_a \equiv j_a \ , \ \ \  \chi^{(0)}=1 \ . \ee 
This  leads to an infinite set of conserved charges 
\be 
\Q^{(n)} = \int^{2\pi}_0 d \s \ j^{(n)}_\t ( \tau, \s) 
\ . \ee 
For example, 
\be \la{pop}
\Q^{(1)}_{MN} 
  = \int^{2\pi}_0 d \s\  j_{\t MN} (\t, \s)
= 2  \int^{2\pi}_0 d \s\  (X_M \d_\tau  X_N - X_N \d_\tau  X_M) 
\ , \ee 
 is proportional to the  $O(6)$
 angular momentum 
$J_{MN}$ \rf{ss}, 
and 
\bea \la{ttt} 
\d_\s \chi^{(1)} &=& j_\t \ , \ \ \ \ \  \ \ \ \ 
 \chi^{(1)}_{MN}  (\t, \s) = \int d \s'\  j_{\t MN}  (\t, \s') \
 ,  \\
  \la{uuy}
 \Q^{(2)}_{MN}  &=& 
 \int^{2\pi}_0 d \s'\  [ \d_\t +  j_\t ( \tau, \s') ]_{_{MK}}
 \chi^{(1)}_{KN}  ( \t, \s') \ . \eea 
These relations can be consistently  defined on an infinite
spatial ($\s$) line 
but not on a  circle as in the closed string 
case  we are interested in: 
for $X_M$ periodic,  $j_a$ \rf{cur} is also   periodic, but its
integral 
in \rf{ttt} may not be, and thus $\Q^{(2)}$ may not be 
well-defined
(see also related comments in \ci{ben,nap}).
There are, however, particular classes of solutions (like
circular solutions discussed below) for  which  these charges 
may be well-defined.

Let us mention also  that as 
 in other \sms  with a   current satisfying 
$j_a = g^{-1} \d_a  g , \ \d^a j_a =0$ we can construct  a
 set of chiral currents -- 
 symmetric  higher  spin  2-d local    currents 
 which are scalars under $O(6)$
  \ci{poly,moun}
\be \la{uuur}
T_{+...+} = \tr j_+^n \ , \ \   \ \
\del_- T_{+...+}=0 \ , \ \ \ \ 
  T_{-...-} = \tr j_-^n \ , \ \ \  \ \
\del_+   T_{-...-}=0 \ . \ee 
The special case of spin 2 currents 
  $T_{++}$ and $ T_{--}$ are the components of the \sm 
 stress tensor  proportional to $\d_\pm X_M \d_\pm
 X_M$. 
There are also other   examples 
of local chiral currents built out of totally symmetric invariant
tensors associated with the corresponding Lie algebra \ci{moun}.
%Thus a series of nonlocal  conserved charges  can 
%be viewed as a ``generalisation'' of the angular momentum
%her spin chiral currents   -- as a
%``generalisation'' of the stress tensor. 
 
 Similar non-local  and local  currents  can be defined 
\ci{ben} also  for  the full 
\adss  supercoset string \sm of \ci{Met}.\foot{That the \adss 
superstring should be integrable being a conformal extension 
of a bosonic coset \sm was suggested 
 previously  in \ci{thorn,wad} (see also \ci{ald}).}

\medskip 

To conclude, the  above  sigma model admits an 
infinite set of conserved
charges  which is usually interpreted as 
 implying its  integrability \ci{lup}.  
 The   use 
 of this integrability of a 2-d system  for classifying 
 finite energy solutions on a 2-d 
 cylinder  is not immediately  clear however.
 In  one dimension a 
system is integrable if it has the same number of commuting
integrals of
motion as the  number of its degrees of freedom. 
In 2-d where one has  infinite  number of degrees of freedom it
is usually assumed that having infinite   set 
of conserved  quantities  implies integrability.
%Commuting charges 
%  may be used at least to classify the solutions.
A more practical  definition of integrability could  be  a
prescription 
of how to construct explicitly a generic solution with required
properties. Formal solution-generating techniques
(see, e.g., \ci{bab})  are  not guaranteed 
a priori to produce  finite energy solutions
(cf. \ci{pie}). 

As we shall see  below, one can 
 understand  the  integrability of the  $R_t \times S^5$
  \sm in a very explicit 
  way    by reducing 
it \ci{afrt,art}  on a special ``rotating string''  ansatz \ci{ft2} 
to a  well-known 1-d  integrable system --
Neumann  system  \ci{bab,babi} or its generalization
 -- Neumann-Rosochatius (NR)  system \ci{pere,gag}.
% While the Neumann  or NR 1-d systems have a
%finite number  of independent  commuting integrals,  
%there is an  infinite number of commuting conserved charges 
%in the original 2-d \sm and  these
%may play an important role in the context of the 
%AdS/CFT duality  (see \ci{AS,mz3} for details). 

%%%%%%%%%%%%%%%%%%%%%%%%%%%%%%%%%%%%%%%%%%%%
\setcounter{equation}{0}
%%%%%%%%%%%%%%
 
%%%%%%%%%%%%%%%%%%%%%%%%%%%%%%%%%%%%%%%%%%%%
\setcounter{footnote}{0}
%%%%%%%%%%%%%%%%%%%%%%%%%%%%%%%%%%%
%%%%%%%%%%%%%%%%%%%%%%%%%%%%%%%%%%%%%%%%%%%%%%%%%%%
\section{Reduction of  $R_t \times S^5$ sigma-model \\ 
\ to 
1-d Neumann system}
%%%%%%%%%%%%%%%%%%%%%%%%%%%%%%%%%%%%%%%%%%%%%%
%%%%%%%%%%%%%%%%%%%%%%%%%%%%%%%%%%%%%%%%%%%%
\subsection{Rotating string ansatz}
%%%%%%%%%%%%%%%%%%%%%%%%%%%%%%%%%%%%%%%

Let us  consider a  string located at the center 
of spatial part of  $AdS_5$  with time coordinate 
being proportional to the worldsheet time, i.e. with 
the $AdS_5$  coordinates in \rf{relx} given by  
\be \la{uuoo}
Y_5+iY_0=e^{i t }\ , \ \ \ \ \ \ Y_1,...,Y_4=0 \ , \ \ \ \ 
t= \k \tau \ . \ee 
The  general case when the string can  be extended and rotate in
both
$AdS_5$ and $S^5$ will be discussed below in section 6. 
Since the  $S^5$ metric \rf{ses} 
has three commuting translational
isometries in $\vp_i$ which give rise to three global 
commuting integrals of motion  $J_i$ \rf{jij} 
 to get solutions  with non-zero $J_i$ it is natural 
 to choose  the following ``rotating string'' 
 ansatz for the $S^5$ coordinates $X_M$ in \rf{relx} 
 \ci{ft2,afrt,art}
\bea
\XX_1\equiv X_1+iX_2=z_1(\sigma)\ e^{iw_1\tau}\, , ~~~\ \ \ &&\ \
\ \
\XX_2\equiv X_3+iX_4=z_2(\sigma)\ e^{iw_2\tau}\, , ~~~ \nonumber\\
\la{emb}
\XX_3\equiv X_5+iX_6&=&z_3(\sigma)\ e^{iw_3\tau}\, .
\eea
Here $w_i$ may be interpreted as frequences of rotation in the
three 
orthogonal planes. The  functions $z_i$ may be real \ci{afrt} or, in
general,  
complex \ci{art}  and  should satisfy $X_M X_M=1$, i.e.  
\be \la{sat}
z_i =  r_i   e^{ i \a_i}\ , \ \ \ \ \ \ \ \ \ \ss r^2_i =1 \ .
\ee 
Thus the shape of a rotating string is not changed in time 
(i.e. the string is rigid)
 and it always belongs to a 2-sphere.
%so that the radial functions $r_i$ should belong to a 2-sphere.

The closed string periodicity condition \rf{peri} implies 
\be\la{pep}  r_i (\s + 2 \pi) = r_i (\s ) \ , 
 \ \ \ \ \ \ \ \ 
\a_i(\s+ 2 \pi ) =  \a_i + 2 \pi  m_i  \ , \   
  \ \ \ \ \ m_i =
0,\pm 1,\pm 2, ... \ . \ee 
Comparing \rf{emb} to \rf{relx} we conclude that 
 the $S^5$ angles 
 $\vp_i$ may depend on both $\tau$ and $\sigma$, 
 \be \vp_i = w_i \tau + \a_i (\s)\ , \ee
 with the  integers $m_i$ in \rf{pep} 
 thus playing   the role of ``winding numbers'' in the 
 Cartan  directions $\vp_i$. 
 
The space-time energy $E$ of the string in \rf{aas} 
 and the spins \rf{jij} forming  a  Cartan subalgebra of  $O(6)$
 here  are given by 
\bea \la{enn}
E=\sql \E  \ , \ \ \ \ \ \ \ \ \                                     
     \E=\k 
\ , 
\eea
\be
\la{spins}
J_i\equiv\sqrt{\lambda}\ {\J}_i\ , \ \ \ \ \ \ \ \
\J_i =  w_i
\int_0^{2\pi}\frac{d\sigma}{2\pi}\ 
 r_i^2(\sigma) \, , \ \ \ \ { \J_i \ov w_i} = 1 \ .
\ee
All  other components of the conserved angular momentum tensor 
$J_{MN}$  \rf{ss} vanish
  automatically    if all $w_i$ are 
different \ci{afrt}, but their vanishing should be checked 
if two of the three frequences are equal.

%%%%%%%%%%%%%%%%%%%%%%%%%%%%%%%%%%%%%%%%%%%%
\subsection{Integrals of motion  and constraints }
%%%%%%%%%%%%%%%%%%%%%%%%%%%%%%%%%%%%%%%

In general, starting with 
\be 
\XX_i(\t,\s) = r_i(\t,\s)\ e^{i\vp(\t,\s)}\ ,  \ee
one finds that  the Lagrangian \rf{SSL} reduces to 
\be
\label{lah}
L_S=\frac{1}{2}\sum^3_{i=1} \big[  {\dot r}_i^2
- r'^2_i + r^2_i ( {\dot \vp}_i^2 
- \vp'^2_i  ) \big] +  \frac{1}{2}
\Lambda(\sum^3_{i=1}  r^2_i-1) \, .
\ee
One can  then check that the ``rotating string'' 
ansatz \rf{emb}, i.e. 
\be \la{gig}
r_i = r_i (\s) \ , \ \ \ \ \ \ \ \  \
\vp_i = w_i \t + \a_i (\s) \ 
\ee
is indeed consistent with the equations of motion
following from \rf{lah}.
Note that because of the formal $\tau \leftrightarrow \s$
symmetry 
of the 2-d equations of motion  another special solution is 
given by the ``pulsating string'' ansatz: 
$r_i = r_i (\t) ,  \  \vp_i = m_i \s + \b_i (\t)$, where 
$m_i$ are now integer winding numbers. Then $r_i(\t)$ is a
solution of a
similar Neumann system discussed below (see also 
section 8).

Substituting  \rf{emb} into   \eq{SSL} 
 we get the following effective  1-d ``mechanical''  system
for a particle on a  5-d sphere 
\be
\label{L}
\rL=\frac{1}{2}\sum^3_{i=1} ( z'_i z'^*_i -  w_i^2  z_i
z^*_i)- \frac{1}{2}
\Lambda(\sum^3_{i=1}  z_i z^*_i-1) \, , 
\ee
with $\s$ playing the role of time
(we changed the sign of $L$).
If we set  $z_k = x_k + i  x_{k+3}$ ($k=1,2,3$),
 this is recognized 
as a special case of the 
well-known integrable system -- the 
standard $n=6$ {\it  Neumann } model \ci{pere,bab,
babi} describing a harmonic oscillator on a 5-sphere
\be \label{Lx}
\rL_N =\frac{1}{2}\sum^6_{M=1} ( x'^2_M -  w_M^2 x^2_M )-
\frac{1}{2}
\Lambda(\sum^6_{M=1}  x^2_M -1) \, .
\ee
 Here three
of the six frequencies are equal to the other three, 
$ w_{k+3}=w_k$, $k=1,2,3$. 
This implies integrability of the  model \rf{L} 
and determines its integrals  of motion. 
Indeed, the  Neumann system \rf{L} has the following 
six  commuting  integrals of motion (see, e.g., \ci{pere,babi}):
\bea \la{inti} 
F_M=x_M^2+\sum^6_{M\neq N}\frac{(x_M x'_N-x_N x'_M)^2}
{w_M^2-w_N^2} \,  , \ \ \ \ \ \ \ \ \ \ 
\sum^6_{M=1}  F_M=1 \ .
\eea
Since in the present case 3 of the 6 frequences are equal 
one needs to consider 3 non-singular combinations of $F_M$ 
which then give the 3 integrals of \rf{L}:
\be \la{inih}
I_i = F_i + F_{i+3} \ , \ \ \ \ \ \ i=1,2,3  \ , \ \ \ \ \ \ \ 
\ss I_i =1 \ .  \ee 
More explicitly, \rf{L} can be written as 
\be
\label{La}
\rL=\frac{1}{2}\sum^3_{i=1} ( r'^2_i  + r^2_i \a_i'^2  -  w_i^2 
r^2_i)- 
\frac{1}{2}
\Lambda(\sum^3_{i=1}  r^2_i-1) \, , 
\ee
implying that 
\be\la{ank}
\a'_i = { v_i \ov r^2_i } \ ,\ \ \ \ \ \ \ \ 
\ \ v_i= {\rm const} \ , 
\ee
where $v_i$ are three  integrals of motion, which complement 
the two independent integrals in 
\rf{inih}.  
Eliminating  $\a'_i$ 
  from  \rf{La} (changing the sign of the corresponding term to
  reproduce the same equations for $r_i$)
   we then get the following effective  Lagrangian
  for the  radial coordinates 
\be
\label{Laa}
\rL=\frac{1}{2}\sum^3_{i=1} \big(  r'^2_i   -   w_i^2  r^2_i -
   { v_i^2 \ov
r_i^2} \big)-  \frac{1}{2}
\Lambda(\sum^3_{i=1}  r^2_i-1) \,  . 
\ee
When the new integration constants $v_i$ vanish, i.e.
$\a_i$ are constant, 
we go back to the case of the 
$n=3$ Neumann model studied in \ci{afrt}.
 For non-zero $v_i$ \ci{art} 
 the Lagrangian \rf{Laa}
 describes the so called {\it  Neumann-Rosochatius} (NR) 
 integrable system (see, e.g., \ci{gag}). As explained above, 
  its integrability  follows from 
 the fact that it is  a special case
 of the $n=6$  Neumann system, with the integrals of motion
 \rf{inih}
 taking the following explicit form:
\be\la{nti}
I_i =r^2_i + \sum^3_{j\not=i} { 1 \ov w^2_i - w^2_j} 
\bigg[ (r_i r'_j - r_j r'_i)^2  + { v^2_i \ov r^2_i} r^2_j 
+ { v^2_j \ov r^2_j} r^2_i  \bigg]  \ . 
\ee
This gives two  additional (besides $v_i$)  independent  integrals of
motion.
%(which may be 
%expressed in terms of  two constants  $b_a$, $a=1,2$).
  
The  conformal gauge constraints \rf{cv},\rf{cvv} or \rf{coni}
now  become  
%\be \la{cv}
%\dot{X}_M\dot{X}_M+ X_M ' X_M '
%=\sum_{i=1}^3 ( r'^2_i    + r^2_i \a'^2_i   + w_i^2  r^2_i )  
%=\kappa^2   \, , \ee
%\be \la{cvv}
%\dot{X}_M{X'}_M= 2\sum_{i=1}^3 w_i r^2_i \a'_i =0   \ . \ee
\bea \la{cve}
\kappa^2=\sum_{i=1}^3( r'^2_i      + w_i^2 r^2_i  +  
 { v_i^2 \ov r^2_i} )    \ , \\
 \la{cvve}
\ \sum_{i=1}^3 w_i v_i  =0   \ .
\eea
As a consequence of \rf{cvve}
only   two  of the  three integrals of motion   $v_i$ are 
  independent of $w_i$.  
 Note also that \rf{cve},\rf{cvve} imply 
 %\be \la{imp}
$\kappa^2=\sum_{i=1}^3\big[ r'^2_i  + (w_i r_i  \pm   { v_i \ov
r_i} )^2
\big]    ,$ 
%\ee
 so that the space-time energy $E$ or $\kappa$  is minimized if 
 $r_i=\const$ and $ ( w_i r_i  \pm   { v_i \ov r_i})^2$ take 
 minimal value, i.e. if 
 $w_i^2 r_i  -  { v_i^2 \ov r^3_i} =0$. Since all three 
 $w_i/v_i$
 cannot be positive, this does not mean that $\kappa$ should 
 vanish.
   We shall return to the
 discussion of such solutions below at the end of section 5.1.

\bigskip 
To summarize, 
we are  interested in finding  periodic finite-energy
solitonic solutions of the  $O(6)$ sigma model defined
on a  2-cylinder  that carry three  global  charges
$J_i$.
As discussed  in
\ci{afrt} (see also below),  the periodicity condition \rf{pep}
on
$r_i$ 
  implies  that the two  integrals of motion $b_a$ (two 
  appropriate  independent
  combinations of $I_i$ in \rf{nti}) 
 can be traded for two integers $n_a$  labeling different 
  types of solutions. 
Imposing the periodicity condition \rf{pep} on $\a_i$  
gives, in view of \rf{ank}, the following constraint:
\be\la{okl}
v_i \int^{2\pi}_0 { d\s \ov r^2_i(\s)} = 2 \pi m_i  \ . \ee
It implies that $v_i$ should be expressible in terms of the 
integers $m_i$, frequencies $w_i$ and the ``radial'' integrals
$b_a$
or $n_a$ (note also  that since the integral in \rf{okl} 
is of a positive function, $m_i=0$ implies $v_i=0$). 
As a result, the moduli  space of solutions 
  will thus be parametrized  by 
  $(w_1,w_2,w_3; n_1,n_2; m_1,m_2,m_3)$.
  The constraint \rf{cvve}  will give one 
  relation between these
3+2+3 parameters. As a consequence, trading $w_i$ for the 
angular momenta using \rf{spins}, the energy of the solutions 
as determined by \rf{enn} and the conformal gauge constraint 
\rf{cve} will be 
a  function of the $SO(6)$ spins and the 
 ``topological'' numbers $n_a$ and $m_i$ (cf. \rf{spi}) 
\be \la{ennp} 
\E=\E(\J_i; n_a,  m_i)\ , \ \ \ \ \ \ \ {\rm i.e.} \ \  \ \ \ \ \ 
E= \sql\ \E ( { J_i \ov \sql}; n_a,  m_i)  \ . \ee 
The constraint \rf{cvve} will provide one  additional relation 
between $J_i$ and $n_a,m_i$. 
Our aim will be  to study the relation \rf{ennp} for various 
types of  solutions and in various limits.

%%%%%%%%%%%%%%%%%%%%%%%%%%%%%%%%%%%%%
\subsection{Special case of $n=3$ Neumann system}
%%%%%%%%%%%%%%%%%%%%%%%%%%%%%%%%%%%%%%%%%

In the special case of $v_i=0$  (when the angles 
$\a_i$ are constant, i.e. $\vp_i$ in \rf{gig} 
depend only on $\tau$)
the NR system \rf{Laa} 
reduces to the $n=3$ Neumann system with the two 
independent integrals
in \rf{nti} and only one non-trivial constraint \rf{cve}, 
which expresses the fact that $\k$ is related  to the 1-d
Hamiltonian of the Neumann system, 
\be \la{joj}
H=\ha \sum^3_{i=1 } (r'^2_i + w_i^2 r_i^2)= \ha \k^2 \ . \ee
Note that  this  Hamiltonian is related to the 
3 integrals of motion in
\rf{nti} by 
$ H=\frac{1}{2}\sum^3_{i=1} w_i^2I_i $.
Any two of these three  integrals
are enough to integrate this
dynamical system.
% since  the motion occurs on a surface of constant integrals.
In order to find the relevant closed string  solutions
 we need also to impose
the periodicity condition \rf{pep}, 
i.e. we are  interested in  the subsector of periodic  solutions of 
the Neumann model.

Since $r_i$ belong to a 2-sphere \rf{sat}, the corresponding equations  can be 
 expressed in terms of the two $S^2$ angles, namely, 
  $\g$ and $\psi$ in 
 \rf{relx}. However, in the general case 
it is convenient to  use another parametrization of $S^2$ --
to replace $r_i$ by the two ``ellipsoidal''
coordinates $\zeta_a$  which are the roots of the equation 
$ \ss { r^2_i \ov \zeta - w^2_i} =0$: 
\bea\label{xizi}
r_1&=&\sqrt{\frac{(\zeta_1-w_1^2)(\zeta_2-w_1^2)}{w_{21}^2
w_{31}^2}}\ , \ \ \ \ \ \ \ 
r_2=\sqrt{\frac{(w_2^2-\zeta_1)(\zeta_2-w_2^2)}{w_{21}^2
w_{32}^2}}\, , \\
r_3&=&\sqrt{\frac{(w_3^2-\zeta_1)(w_3^2-\zeta_2)}{w_{31}^2
w_{32}^2}} \, ,
\ \ \ \ \ \ \ \ \ \ w_{ij}^2\equiv w_i^2-w_j^2\ . 
 \eea
Expressing the integrals of motion \eq{inti}
in terms of $\zeta_a$ one finds a system of two 1-st
order equations 
\bea
\la{sep}
(\frac{d\zeta_1}{d\sigma})^2=
-4\frac{P(\zeta_1)}{(\zeta_2-\zeta_1)^2}\, , ~~~~~~
(\frac{d\zeta_2}{d\sigma})^2=
-4\frac{P(\zeta_2)}{(\zeta_2-\zeta_1)^2}\, . 
\eea
%which is easily separable. 
The function $P(\zeta)$ is the following  5-th order
polynomial
\bea \la{poll}
P(\zeta)=(\zeta-w_1^2)(\zeta-w_2^2)(\zeta
-w_3^2)(\zeta -b_1)
(\zeta -b_2)\ .
\eea
The parameters
 $b_1,b_2$ here are the
 two constants of motion which can be expressed in
  terms of  the integrals $I_i$ in \eq{nti}  by solving 
\bea
\nonumber
b_1+b_2&=&(w_2^2+w_3^2)I_1+(w_1^2+w_3^2)I_2+(w_1^2+w_2^2)I_3\,
, \\
b_1b_2&=&w_2^2w_3^2I_1+w_1^2w_3^2I_2+w_1^2w_2^2I_3\, .
\la{bby} \eea
The Neumann system Hamiltonian \rf{joj}
is then 
$
H=\frac{1}{2}\big(w_1^2+w_2^2+w_3^2-b_1-b_2\big) = \ha
\k^2 .$
The  polynomial $P(\zeta)$ in
\eq{poll}
 can be interpreted  as
defining  a hyperelliptic curve of genus 2 defined by the equation 
$
s^2+P(\zeta)=0 , $ 
with $s$ and $\z$ being two complex coordinates of $C^2$. 
The formal solution of the system \rf{sep} 
is then given in terms of the 
related  $\theta$-functions \ci{mam,afrt}. 

Thus, 
the most general three-spin string solutions
 are naturally associated  with special genus 2 
   hyperelliptic curves \ci{afrt}. 
   The simpler two-spin case (e.g., $w_3=0$)  is
   associated with an elliptic curve and the
   corresponding relation between the energy and the spins 
   then involves  elliptic functions (see \ci{ft1,ft4,bfst}). 
   Elliptic integrals appear also in 
   the one-spin case \ci{vega,gkp}.
      
The system \eq{sep} allows one to achieve the full
separation of the
variables:
dividing one equation in (\ref{sep}) by the other
 one can integrate, e.g.,
$\zeta_2$ in terms of $\zeta_1$ and then obtain a
closed equation
for $\zeta_1$ as a function of $\sigma$.
In finding  solutions
we  need also to take into account the periodicity
conditions on $r_i$    now
viewed as conditions on $\z_1,\z_2$.
The spins $\J_i $ in
\eq{spins} expressed in terms of $\z_1,\z_2$
satisfy   \ci{afrt}
\bea
\la{spii}
\ss w_i (w_i -\J_i ) = 
\int_{0}^{2\pi}\frac{d\sigma}{2\pi} \
(\zeta_1+\zeta_2) \ ,\\
\la{spiii}
\ss {\J_i\ov w_i^3}={1\ov w_1^2w_2^2w_3^2}\int_{0}^{2\pi}
\frac{d\sigma}{2\pi} \
\zeta_1\zeta_2\, \ , \ \ \ \ \ \ \ \ \  
 \ss {\J_i\ov w_i} =1 \ . 
\eea
To find  the energy \eq{enn} as a function  of the spins $\J_i$
we need to express
 the frequencies $w_i$ and the Neumann
 integrals of motion or  $b_a$ in \rf{bby}
  in terms of $\J_i$.
 After finding a periodic solution of
 (\ref{sep}), this
  reduces to  the problem of computing the two
  independent
integrals on the r.h.s. of (\ref{spii}) and (\ref{spiii}).

\medskip 

Let us briefly mention that  the 
 case of the NR system \rf{Laa} 
 with $v_i\not=0$ can be treated similarly
\ci{art}. We can again introduce 
 the ellipsoidal coordinates $(\zeta_1,\zeta_2)$, 
 and expressing the integrals of motion (\ref{nti}) in terms of 
$\zeta_a$
 we  end up with the same system \rf{sep} where now 
\bea
P(\zeta)&=&(\zeta-b_1)(\zeta-b_2)(\zeta-w_1^2)
(\zeta-w_2^2)(\zeta-w_3^2)
+ v_1^2 (\zeta-w_2^2)^2(\zeta-w_3^2)^2 \nonumber \\
  &+&\ v_2^2
(\zeta-w_1^2)^2(\zeta-w_3^2)^2
+v_3^2 (\zeta-w_1^2)^2(\zeta-w_2^2)^2\ . 
\eea
The  Hamiltonian
of the NR system reduces  to 
$
H=\frac{1}{2}\big[\sum_{i=1}^3(w_i^2+v_i^2)-b_1-b_2\big]  .
$
As in the pure Neumann case, $P(\zeta)$ is the fifth order 
polynomial which again defines a hyperelliptic curve
$s^2+P(\zeta)=0$.  
The general solution of  eqs. (\ref{sep})
can be again given in terms of theta-functions associated with the
Jacobian of the hyperelliptic curve.  An example 
of a solution is provided by the  $v_3=0$ case 
where  $\zeta=w_3^2$ is a root of $P(\zeta)$
and then the  NR system  can be solved in terms of
the  elliptic functions \ci{art}.

%\bigskip 
%%%%%%%%%%%%%%%%%%%%%%%%%%%%%%%%%%%%%%%%%%

\subsection{Types of solutions  and 
rotating strings in flat space}

%%%%%%%%%%%%%%%%%%%%%%%%%%%%%%%%%%%%%%%%%%%%%%

Let us  consider for simplicity the case with  $v_i=0$
described by   
the $n=3$ Neumann system  (general  solutions of the NR system 
have similar structure).
The five parameters $w_i $ (or $\J_i$)   and $b_a$ 
of the solutions of the Neumann system 
may be viewed as
coordinates on the  moduli space of periodic finite-energy 
 solitons.
%Because of the closed string periodicity condition, 
 The values of $b_a$ will not be arbitrary:
 such  solutions will be classified
by  two    integer ``winding number''  parameters $n_a$
which  will  be  related to  $w_i$ and $b_a$ through 
the periodicity condition.
 In general, there will   be several different
solutions for given values of $\J_i$,
i.e. the energy of the string $E$ will be a function not only of
$\J_i$ but also of  $n_a$:
there will be a  discrete series of  solutions with energies 
starting from some minimal value. 
Solutions with the same spins 
$\J_i$  will be  distinguished 
by values of higher conserved commuting charges.

Depending on the values of these parameters
(i.e. location in the moduli space) one may find
different geometric types (or shapes)  of the resulting
rotating string solutions.
The shape of the string does not change with time and  
 the string may be  {\it ``folded''}
(with topology of an interval)
 or  {\it ``circular''}
(with topology of a circle).
 A folded string may then 
  be  {\it ``straight''} as in the one- and
 two-spin examples
 considered in \ci{gkp} and \ci{ft4}, 
 or {\it ``bent''}  (at one or several points)
  as in the general three-spin case \ci{afrt}.
 A ``circular'' string  may  have the form
 of a round circle as in the two-spin and three-spin
 solutions   of \ci{ft2,ft3}  or may have a  more general
 ``bent circle'' shape as in the three-spin solutions
 in \ci{afrt}.

It is useful   to review  how these different string shapes
appear in the case of a closed string rotating in flat
$R^{1,5}$ Minkowski space.
%\foot{As follows from \rf{sse}, 
%those  flat-space  solutions that have $\dot X_M^2 - X'^2_M =0$
%are also the solutions of the $R \times S^5$ model
%with $\L=0$,see also below.}
 In the conformal  gauge, string coordinates are then given by
solutions of free 2-d wave equation, i.e. by combinations 
of $e^{i n(\tau \pm  \s)}$, subject  to the standard
 constraints \rf{coni}.
%$\dot X^2_M + X'^2_M = 0, \ \dot X X'=0$.
For a closed string rotating in the two
orthogonal spatial 2-planes  and  moving along
the 5-th spatial direction we find (cf. \eq{emb}; here $t=\k \tau$ as
in \rf{uuoo})
\be \la{flat}
X_1 + i X_2 = \ r_1(\s) \ e^{ i w_1 \tau } \ ,
\ \ \ \   X_3 + i X_4 = \ r_2(\s) \ e^{ i w_2 \tau} \ ,
\ \ \ \  X_5 = p_5 \tau  \ , \ee
\be \la{taak}
w_1= n_1 \ , \ \  w_2 = n_2 \ , \ \ \
 r_1  = a_1   \sin (n_1 \s) \ ,  \ \ \ \ \ \
r_2  = a_2  \sin [n_2 (\s + \s_0) ]  \ .   \ee
Here $\s_0$ is an arbitrary integration constant, and $n_a$
are arbitrary integers. The 
 conformal gauge constraint implies that
 $   \k^2 = p_5^2 +  n^2_1 a_1^2 + n^2_2 a_2^2 .$
 Then the  energy,  the two spins and the 5-th component of
 the linear momentum are (here the tension parameter is 
 $ \sql \to {1 \ov \a'}$)
\be
\la{eess}
 E= {\k  \ov \a'} \ , \ \ \
\ \  J_1 = {n_1 a^2_1 \ov 2 \a'}  \ , \ \ \
\    J_2 =  {n_2 a^2_2 \ov 2 \a'}\ ,   \ \ \
P_5 ={p_5  \ov \a'} \ , \ee
i.e.
\be\la{rrr}
 E= \sqrt { P^2_5  + { 2 \ov \a'} ( n_1 J_1 + n_2 J_2 ) } \
 . \ee
To get  the two-spin states on the leading  Regge
trajectory
(having minimal
energy for given  values of the {\it two} non-zero spins)
 one  is to choose $n_1=n_2=1$.
 % with $\s_0= {\pi \ov 2}$.
The shape of the string depends on the values
of $\s_0$ and $n_1,n_2$.
If $\s_0\ov \pi $ is irrational then the string always has a
``circular'' (loop-like) shape. In general, the ``circular''  
string
will not be lying in one plane, i.e. will have one
or several bends.
For rational values of $\s_0\ov \pi$ the string can be
either circular or folded,  depending  on
the values of $n_1,n_2$.

Let  $\s_0=0$. If  $n_1=n_2$ the string is folded and straight, i.e. have
no bends.
Indeed, then
$X_1+ i X_2$ is proportional to $X_3 + i X_4$
and thus one may put the string in  a single  2-plane
by  a global O(4) rotation.
If both $n_1$ and $n_2$ are either even or odd and
different
then the string is folded and has several bends
(in the 13 and 24 planes). For example, if $n_2 = 3 n_1$
then the folded string  is wound $n_1$ times and
has two bends (for $a_1=a_2$ we have 
 $r_2 = r_1 (3-4r_1^2)$).
Next, let  us choose  $\s_0 ={\pi\ov 2 n_2}$. Then 
for   $n_1=n_2$ the string is an ellipsoid,
 becoming a round circle in the
special case of $a_1=a_2$ (i.e. $\J_1=\J_2$)   \ci{ft2}.
The string is also circular if $n_1$ is even and $n_2$ is
odd. If, however, $n_1$ is odd and $n_2$ is even
the string is folded, e.g., if $n_2 = 2 n_1$
then the folded string  is wound $n_1$ times and
has a single bend at one  point.

The structure of spinning   string  soliton  solutions
in   curved $S^5$ case is analogous.
The equations of motion of the Neumann system are
linearized
on the Jacobian of the hyperelliptic curve.
The image of the string in the Jacobian whose real
connected part
is identified with the Liouville torus can wind around two
non-trivial cycles with the winding numbers $n_1$ and $n_2$
respectively  \ci{afrt}.
The size and the shape of the Liouville torus
are governed by the moduli $(w_i,b_a)$. Specifying the
winding numbers
$n_1,n_2$,  two of the five parameters $(w_i,b_a)$
are then  uniquely determined by
the  periodicity conditions. The actual rigid shape of the
physical string 
   lying on two-sphere 
  will depend
on the numbers $n_1,n_2$ and on the remaining moduli
parameters (relative values of $b_a$ and $w^2_i$): it  
 may be of
(bent) folded type or of (deformed)  circular type.
Various examples of 
 folded and  circular 3-spin string solutions 
 and their energies  were discussed in \ci{ft2,afrt}.  
 In most 3-spin cases finding explicit relation for 
 the energy \rf{ennp} is complicated, but one can always 
 develop   large $\J_i$ perturbation theory \ci{afrt}. 
 We shall discuss some  examples of such   solutions below.

Finally, let  us note that while the Neumann 
 or NR 1-d systems have a small 
 finite number of commuting integrals,  
there are infinite  commuting conserved charges 
in the original 2-d \sm and  the corresponding integrable spin
chain on the SYM side. These are expressed in terms
 of the NR integrals 
in the present case, see \ci{AS} for details.

%%%%%%%%%%%%%%%%%%%%%%%%%%%%%%%%%%%%%%%%%%%%%
\setcounter{equation}{0}
%%%%%%%%%%%%%%

%%%%%%%%%%%%%%%%%%%%%%%%%%%%%%%%%%%%%%%%%%%%
\setcounter{footnote}{0}
%%%%%%%%%%%%%%%%%%%%%%%%%%%%%%%%%%%
%%%%%%%%%%%%%%%%%%%%%%%%%%%%%%%%%%%%%%%%%
\section{Simplest circular solutions in $R_t \times S^5$:  $\L=\const$}
%%%%%%%%%%%%%%%%%%%%%%%%%%%%%%%%%%%%%%%

A simple  special class of solutions of the system 
\rf{L} or \rf{Laa}
is found by demanding  that the Lagrange multiplier 
$\Lambda$ in \rf{sse} is constant,  
i.e. $\dot X^2_M - X'^2_M=\const$.  
In this case the radii $r_i$ turn out to be constant 
  (and  $n_a=0$, i.e. there are no bends).
  This  represents 
  an interesting new class  of  circular
  3-spin solutions \ci{art} which  includes as a 
  special case the circular  solution 
   of  \ci{ft2} where two out of three  spins 
   were equal. 
   
%%%%%%%%%%%%%%%%%%%%%%%%%%%%%%%%%%
\subsection{Constant radii solution} 
%%%%%%%%%%%%%%%%%%%%%%%%%%%%%%%%%%%%%%%%%%%%%%

Let us 
 start with the Lagrangian 
  \rf{L} written in terms of 3 complex coordinates
$z_i$. Then  the equations of motion are 
\bea \la{jk} 
z''_i   +  m^2_i z_i = 0  \ , \ \ \ \ \ \ 
m^2_i &\equiv& w^2_i + \L \  , \ \ \ \ \ \ \ \ss |z_i|^2 =1 \ , \\  
 \L&=&\ss (|z'_i|^2   -  w^2_i | z_i|^2)\ . \eea
Eq. \rf{jk} 
can be easily integrated if one assumes that $\L=\const$, 
\be \la{uou} 
z_i = a_i e^{i m_i \s}  + b_i e^{-i m_i \s}\ ,  \ee
where $a_i,b_i$ are complex coefficients.
The periodicity condition $z_i(\s + 2 \pi) = z_i(\s)$ 
implies that $m_i$ must be integer. 
It is easy to show \ci{art} 
that modulo the global  $SU(3)\in SO(6) $ 
invariance the solution of \rf{uou} that satisfies both $\L=\const$
and 
$\ss |z_i|^2=1$  should have  
$b_i=0$ ({\it or} $a_i=0$), i.e. should look like 
($m_i$ may be positive or negative and $a_i$ may be made 
real by $U(1)$ rotations) 
\be \la{iii}
z_i = a_i e^{i m_i \s} \ , \ \ \ \ \ \ \ 
\ \ \ \  \ss a^2_i=1 \ . \ee 
It may seem that one may  get a new solution 
if two of the windings $m_i$ are equal  while the third  is 
zero, i.e. $z_1 = a \cos m \s , \ 
  z_2 = a \sin m \s ,  \ z_3=\sqrt{1- a^2}$
  (which is, in fact, the circular solution of \ci{ft2}),  
  but this configuration 
   can be transformed into the form \rf{iii} by a global 
  $SU(2)$ rotation.            
%%%%%%%%%%%%%%%%%%%%%%%%%%%%%
        
 One can also   rederive  \rf{iii}
  by starting with 
\rf{Laa},\rf{ank}.           
The potential $  w_i r^2_i + { v^2_i \ov r^2_i} $ 
in \rf{Laa} has a minimum, and that suggests 
that $r_i=$const may be a solution. 
The equations of motion that follow from \rf{Laa}
\bea
\label{Eqofm}
r_i''=-w_i^2 r_i &+& {v_i^2 \ov r^3_i}   -  \L  r_i \  , \\
 \la{lamm}
\L =  \sum_{j=1}^3
 \Big(r'^2_j-w_j^2r_j^2 + {v_j^2 \ov r_j^2} \Big) \, &,& \ \ \ \ \ 
 \ \ \ \ \ \ \ 
\sum_{j=1}^3 r^2_j =1 \ 
\eea
are indeed solved by 
\be \la{kop}
r_i(\s) =a_i =\const\ , \ \ \ \ \ \ \ 
w^2_i -  {v^2_i \ov a_i^4}  = \nn^2 = \const\ , 
\ \ \ \ \Lambda= - \nn^2 \ , 
\ee
where $\n$ is an arbitrary constant 
(which may be positive or negative).
Eq. \rf{kop} then implies 
$
a_i^2 =  {| v_i | \ov \sqrt{ w_i^2 - \nn^2 } } $, 
$ \a_i'= { v_i \ov a_i^2 }= { v_i \ov  | { v_i } | } 
\sqrt{ w_i^2 - \nn^2 } \equiv  m_i$, 
i.e.
$   \a_i = \a_{0i} + m_i \s $, where  $m_i$ must be 
integer to satisfy the periodicity 
condition \rf{pep} and $\a_{0i}$ may 
 be set to  zero  by independent $SO(2)$ rotations.  Then 
\be\la{kol}
w_i^2 = m^2_i + \nn^2 \ , \ \ \ \ \ \ \  \ \ \ \ 
 v_i = a^2_i m_i  \ , \ \ \ \ \ \ \sum_{i=1}^3  a^2_i=1 \ . 
\ee 
The constraints  \rf{cve},\rf{cvve} give 
$
\k^2 =  2 \ss  a^2_i w_i^2 - \nn^2
  ,$ and $  
\sum_{i=1}^3  a^2_i  w_i m_i =0$ .  
 As a result, we get the following relations for the 
  energy and spins \ci{art} 
 (cf. \rf{enn},\rf{spins})
 \bea \la{onk}
 \E^2 = 2 \ss  \sqrt{m^2_i + \nn^2}\   \J_i  - \nn^2 \ , \\
 \la{kpp}
 \ss {\J_i \ov  \sqrt{m^2_i + \nn^2} } =1 \  ,\\
 \la{uiy}
 \ss m_i \J_i =0 \ .  \eea
 We shall assume 
for definiteness that all $w_i$ and thus all $\J_i$ 
are non-negative. Then \rf{uiy} implies that  one 
of the three $m_i$'s
 must have the  opposite sign to the  other two.
One can check directly that the only non-vanishing components of
the $SO(6)$  angular momentum tensor
$J_{MN}$ \rf{ss}  on this solution are indeed the Cartan ones $J_i$
in \rf{jij}.

The special case of 
 $\n^2=0$ (or $\Lambda=0$)   corresponds to 
a solution for the string in flat space 
which can be  embedded into  $S^5$ by 
choosing the free radial parameters of a 
 circular string to satisfy the condition 
$\ss a^2_i =1$.  Indeed, as follows from \rf{kol} for
 $\n^2=0$ we find that all frequencies must be integer
  $w_i= |m_i|$. We may choose,  
 e.g., $m_1 < 0, \ m_2> 0,  \ m_3 > 0 ,$
so that 
 the solution is a combination of the 
left and right moving waves in different directions 
(we use complex combinations of the coordinates in
\rf{emb}, cf. \rf{flat},\rf{taak})
\be\la{iot}  \XX_1 =a_1 e^{ i m_1 (\s - \tau)} \ , \ \ 
 \XX_2 =a_2 e^{ i m_2 (\s + \tau)} \  , \ \ \ 
 \XX_3 =a_3 e^{ i m_3 (\s + \tau)} \  , \  \ \ \ss a^2_i=1 \ . 
\ee
Here we get from \rf{onk}--\rf{uiy}
\be \la{yyr}
 \E^2 = 2 \ss  |m_i|  \J_i   \ , \  \ \ \ \ \ \ 
  \ss {\J_i \ov |m_i|}
 =1 \ , \ \ \ \ \ \ \ \ 
 \ss m_i \J_i =0 \  .
   \ee
This corresponds to a very special point 
in the moduli space of solutions. 
For fixed $m_i$, we get two  constraints on $\J_i$,
 and the energy is
given by the standard flat-space Regge relation (cf. \rf{rrr}). 
Then 
$|m_1 | \J_1 = m_2 \J_2 + m_3 \J_3$ (where $\J_2$ and $\J_3$ are
related via $\ss {\J_i \ov |m_i|}
 =1$) 
  and thus 
$\E^2 = 4  |m_1|  \J_1$. 
 The energy of 
  this ``flat'' solution thus  does {\it not} have 
 a regular \rf{jojk}  expansion  in integer   powers 
 of ${1\ov \J^2}={\l \ov J^2}, \ \J=\ss \J_i$.
 This will  no longer be so in the genuinely ``curved'' 
  $\n\not=0$ 
 case where we will have indeed a  regular expansion 
 for the energy in $1\ov \J^2$,   as in the case of the circular
 solution of \ci{ft2}. This then 
   opens up  a  possibility of 
 direct comparison with 
 perturbative  anomalous dimensions in SYM theory.

 %%%%%%%%%%%%%%%%%%%%%%%%%%%%%%%%%%%%%%%%%%
 \subsection{Energy as  function of spins}
 %%%%%%%%%%%%%%%%%%%%%%%%%%%%%%%%%%%%%%%%%%%%
 
In general, to  express $\E$ in terms of $\J_i$ and $m_i$ 
one is  first  to solve the condition 
\rf{kpp} in terms of $\n$, determining $\nu$  as a function of $\J_i$
and
$m_i$  and then substitute the result into \rf{onk}.
The condition \rf{uiy} may be imposed  at the very end, 
implying that for given spins $\J_i$ the 
solution exists only for a special choice of 
the integers $m_i$. 
Expanding in  large total spin $\J=\ss \J_i$ 
  as in \ci{ft2,afrt} ane finds \ci{art} 
that $ 
\n^2=\J^2  - \ss m^2_i  {\J_i\over \J} 
+...$ and thus 
\be \la{epo}
\E = \J  + {1\over 2\J } \ss m^2_i  {\J_i\over \J} 
+...\ . \ee
 Thus, as in  other  examples in \ci{ft2,ft4,afrt}, 
here  the energy   admits a regular expansion in
 $1\ov \J^2 $=${\lambda\ov J^2}$ as in \rf{jojk} 
\be \la{epon}
E = J\big(1   + {\l\over 2J^2 } \ss m^2_i  {J_i\over J} 
+...\big)= J + {\l\over 2J } \ss m^2_i  {J_i\over J} 
+...\ ,  
\ee
where $m_i$ should satisfy  the constraint $\ss m_i J_i=0$. 

Let us now look at some special cases. 
In the one-spin case $(0,0,J_3)$, i.e.   $ \J_1=\J_2=0, \ a_1=a_2=0$,  
 we have  $w^2_3= \n^2$, i.e. $m_3=0$
and $J_3= w_3$, and then $\E= \J_3$. 
This  is simply the point-like BMN geodesic case:
 there is no $\s$-dependence. 
 
In the two-spin case $(J_1,J_1,0)$, i.e.  $\J_3=0, \ a_3=0$, 
the equation  \rf{kpp} for $\n^2 $ becomes a quartic equation.
Its simple explicit  solution is found in the equal-spin 
 case when 
$\J_1=\J_2$, i.e.  when 
\be\la{two}
  a_1= a_2 = {1\ov \sqrt {2}}\ , 
   \ \ \ \ m_2=-m_1\equiv m  > 0 \ , \ee 
%\n^2 =\J^2 -m^2 \ , \ \ \ \ \  \J\equiv \J_1 + \J_2 = 2 \J_1 \ , \ee
so that 
\be \la{uio}
\E=\sqrt{ \J^2+ m^2 } \ , \ \ \ \ \ \ \ \ \ \ \ 
\  \J\equiv \J_1 + \J_2 = 2 \J_2 \ ,  \ee
i.e.
\be\la{hih}
E = J \sqrt{ 1 + m^2 {  \l  \ov J^2} } \ . \ee
We get 
 \be \la{hoh}
 \XX_1 = {1\ov \sqrt {2}} e^{ i w \tau - i m \s} \ , \ \ \ \ \ \ \ \ \
 \XX_2 = {1\ov \sqrt {2}} e^{ i w \tau  + i m \s} \ ,  \ \ \ \ \ 
 w= \sqrt{ \nu^2 + m^2 } \ .  \ee
 This solution is thus equivalent to the circular 2-spin solution of
 \ci{ft2} -- it is related to it  by an $SO(4)$ rotation:
$
 \XX'_1 = {1\ov \sqrt {2}}(\XX_1 + \XX_2)  ,  \ \ 
 \XX'_2 ={1\ov \sqrt {2}}( -\XX_1 + \XX_2 ).$
In the general case of two unequal spins we can again  solve \rf{kpp} 
in the limit of large 
$\J_1,\J_2$ (for fixed $m_1,m_2$), 
getting the special case of \rf{epo} with $m_1 \J_1+ m_2\J_2=0$, 
$\J_3=0$. 

Another  special case  is $(J_2,J_2,J_3)$ when two out 
of three non-vanishing spins are  equal, e.g., $ \J_1=\J_2$.   
Setting 
\be \la{spec}
m_3=0\ , \ \ \ \ m_1=-m_2=m\ , \ \ \ \ \ \ 
a_3 = a< 1 \ , \ \ \ \  a_1=a_2= \sqrt{1-a^2} \ , \ee 
\be\la{kik}  \J_3= a^2 \nu\ , \ \ \ \ \ \ \ \ \ \ 
 \J_1=\J_2= \ha (1-a^2) \sqrt{m^2 + \nu^2}\ ,  \ee 
we thus  find  from \rf{epo}
\be \la{ipo}
\E = \J  + { m^2 \J_2 \over \J^2 } 
+...\ , \ \ \ \ \  {\rm i.e.} \ \ \ \ \ 
E= J + { m^2 \l  J_2 \over J^2 } + ... \ , \ \ \ \ \ \ 
 J= 2 J_2 + J_3 \ .  
\ee
This solution is equivalent to the circular 3-spin solution with two equal spins 
in \ci{ft2,ft3} (the two backgrounds are related by  a global 
rotation in $X_2,X_3$ directions 
converting $e^{i m \s}$ into $\cos m \s$ and $\sin m \s$).
The corresponding 
operator in the gauge theory 
$\tr ( X^{J_1} Y^{J_1} Z^{J_3}) + ....$  (belonging to 
 the SO(6)  representation 
with Dynkin indices $[J_2+J_3, 0, J_2-J_3]$ for $J_2 > J_3$)
which has  the 1-loop anomalous dimension equal to \rf{ipo} 
does indeed exist as was found in \ci{mz3}.

More generally, we may consider a 3-spin solution $(J_2,J_2,J_3)$ 
with $m_3\not=0$, so that  $(m_1 + m_2) J_2 + m_3 J_3=0$. 
Then \rf{epon} gives 
\be \la{ipoj}
\E = \J  +  {\J_2 \over 2\J^2 } 
\big[m^2_1 + m^2_2  + (m_1 + m_2)^2 { \J_2\ov \J_3} \big] + ...
 \ , \ee
which generalizes \rf{ipo} to the case when $m_1+ m_2 \not=0$.
The  energy is minimal in the latter case. 
This suggests that the band of such states in the same 
representation  $[J_2+J_3, 0, J_2-J_3]$ (if 
 $| {m_3 \ov m_1 + m_2}| > 1$)  but with higher energy than \rf{ipo} 
 should also be found on the SYM side. 
 
\medskip

To summarize, the constant-radii solutions of the NR system 
represent a simple generalization of the circular 2-spin and 3-spin
solutions of \ci{ft2} which have regular expansion of the 
energy in powers
of  $\l\ov J^2$. Therefore,  it should be possible to match, as in
\ci{mz2,afrt,bfst,mz3},  
the coefficient of the $O(\l)$ term in  \rf{epo}
 with  the SYM
 anomalous dimensions determined by the Hamiltonian of the 
integrable $SU(2,2|4)$ spin chain 
  \ci{mz1,BM} in the corresponding   3-spin subsector
  of states.

%%%%%%%%%%%%%%%%%%%%%%%%%%%%%%%%%%%%%%
\subsection{Quadratic fluctuations  near circular solutions }
%%%%%%%%%%%%%%%%%%%%%%%%%%%%%%%%%%%%%%%%%%%%%%%%

The remarkable simplicity  of the circular solutions discussed above 
makes it easy to  find the quadratic fluctuation 
action and to compute the corresponding spectrum of string fluctuations.
This in turn allows one 
to  analyse the stability of the solution
  and to find the string 1-loop 
correction to the ground-state energy, 
 in the same way as this was done
in \ci{ft3}
for  a particular 3-spin circular 
solution with two equal spins \rf{kik}.
In spite of the $\s$-dependence of the  solution, 
 the   quadratic action turns out to  have 
  constant coefficients, just like 
 in the BMN case \ci{bmn,meet}
 when one expands near the point-like geodesic in $S^5$
 \ci{gkp,ft1}. 
 Sending $J\to \infty$ for fixed ${\l \ov J^2} \ll 1 $ 
 suppresses higher loop corrections to  masses 
 of  excited string states. As a result, 
 as  in the ``plane-wave'' BMN case, 
 the string fluctuation  spectrum can be found {\it exactly}. 
 
To illustrate this, here we shall consider  the 
bosonic part of the quadratic fluctuation action following \ci{art}.
 The fermionic part of the spectrum 
can be easily  found in the same way as was done  
(in a special case \rf{kik}) 
in a \ci{ft3}.
In contrast to the BMN case,  here we are expanding near a
non-supersymmetric solution, and the resulting world-sheet string action
(in the static or light-cone type gauge) 
will not have a  world-sheet supersymmetry. 
There  remains an interesting question if a ``nearly-BPS'' 
property  of similar rotating string solutions
 in the $\l\to 0$ limit observed in \ci{matt} 
imposes certain constraints on the world-sheet action.

It is straightforward to find the quadratic fluctuation Lagrangian 
by  expanding near the solution \rf{iii} or \rf{kop}--\rf{kol}
following  
\ci{ft3,art}. 
Using 3 complex combinations of coordinates in \rf{emb} and
expanding
($\XX_i \to \XX_i + \td \XX_i$) 
the sigma model action \rf{SL} near the classical solution \rf{iii}, i.e. 
\bea \la{soll}
\XX_i = a_i e^{i w_i \tau + i m_i \s} \ , \ \ \ \ \ \ \ \ \ \ 
w^2_i =\sqrt{ m^2_i + \n^2 } \ ,
\la{csop}  \ \ \ 
\ss a^2_i=1 \ , \ \ \ \ \ \ \ \ \ \ \   \ss a^2_i w_i m_i =0 \ , 
\eea
we  find the following Lagrangian for the quadratic fluctuations 
(see \ci{ft3}) 
\be \la{flu}
\td L = - \ha \del_a \td \XX_i \del^a \td \XX^*_i  + \ha \L 
\td \XX_i  \td \XX^*_i \ , \ee
where  $\L= -\nu^2$ (see \rf{kop}) and 
$\td \XX_i $ are subject to the constraint\foot{The  
imposition of the conformal gauge 
  constraints on the fluctuations is not necessary 
in order to determine the non-trivial part of the fluctuation
spectrum \ci{ft2,ft3} (solving the constraints 
in terms of fluctuation of $t$ leads to equivalent result \ci{ft3}). 
In addition to $S^5$ fluctuations there
 are also  $AdS_5$ fluctuations: one massless and 
  4 massive ones with 
  mass $\k$ coming from the classical value of the Lagrange
  multiplier $\td \Lambda$   \ci{ft2,ft3}.} 
\be \la{koi} 
\ss ( \XX_i \td \XX^*_i + \XX^*_i \td \XX_i ) = 0 \ . \ee 
To solve this  constraint we set 
\be \la{sett}
\td \XX_i = e^{i w_i \tau  + i m_i \s}  (g_i + i f_i)   \ , \ee
where $g_i$ and $f_i$ are real functions of $\tau$ and $\s$. Then  
 \rf{koi} reduces to 
\be
\la{oyy}  \ss a_i  g_i  = 0 \ . \ee
Using \rf{sett} the  Lagrangian \rf{flu}  becomes (after integrating by parts, 
cf. \ci{ft3})
\be \la{fle}
\td L = 
 \ss \bigg[ \ha (\dot f^2_i   + \dot g^2_i - f'^2_i - g'^2_i )
-  2 w_i f_i \dot g_i  + 2 m_i f_i g'_i \bigg] \ . \ee
To solve the linear relation 
\rf{oyy}  we may apply a global $O(3)$ rotation to $g_i$, 
\ $\bar g_i = M_{ij} (a) g_j$,  which  preserves the kinetic terms in
\rf{fle}  and 
transforms $\ss a_i  g_i $ into $\bar g_1$; then we may  set the latter 
 to zero in the resulting Lagrangian \rf{fle}. 
 Equivalently, we may solve \rf{oyy} directly 
 for $g_1$ and substitute it   into \rf{fle}.
The result (after diagonalization) 
  is a special case of the following
 2-d Lagrangian (summation over $p,q$ is assumed) 
\be\la{lll}
  L =  \ha \dot  x_p^2    -\ha  x'^2_p
 +  F_{pq}   x_p   \dot x_q
 -   H_{pq}   x_p     x'_q \ , \ee
where $x_p=(f_1,f_2,f_3,g_2,g_3)$  and  $F_{pq}$ and $H_{pq}$ are 
constant antisymmetric matrices depending on $a_i,w_i,m_i$. 
Eq. \rf{lll}  can  be written also as (ignoring total derivative)
\be \la{spac}
 L =  \ha (\dot  x_p  + F_{pq} x_q )^2    - \ha ( x'_p   +
H_{pq} x_q)^2  -   (F_{pq}F_{qk} - H_{pq}H_{qk}) x_p x_k \ ,   
    \ee
i.e. it represents a   massive scalar 2-d theory coupled to a
constant 2-d  gauge field
(which can be ``rotated away''  at the expense of
 making  the mass term $\tau$ and $\s$
dependent).
The   Lagrangian \rf{lll}      can  be also interpreted
as a {\it light-cone gauge}   ($u=\tau$)  Lagrangian  for the
bosonic string sigma model
$L= - ( \eta^{ab} g_{mn} + \epsilon^{ab} B_{mn} ) \del_a x^m
\del_b x^n $  in  (in general, non-conformal)  {\it  plane-wave}
background  with the following metric
and antisymmetric 2-form field 
\be\la{uyq}
ds^2 = 2 dudv   + 2 F_{pq} x_p dx_q du  +  dx_p dx_p  \ ,
 \ \  \ \  \ \ \ \
B_2  =  2 H_{pq} x_p  d x_q \wedge  d u  \  . \ee
By analogy with the   BMN case, 
 one may say that the geometry 
``seen'' in the large $\J$ limit by  the circular 
rotating string is a generalized plane-wave background.
The resulting  quadratic string excitation spectrum 
for such an action can  be found 
in a more or less explicit way (as in   \ci{papd}).

For example, let us consider  the 2-spin case
 where  $m_3=0$ and 
\be\la{oll}
a_1^2 + a^2_2 =1   \ , \  \ \  a_3=0\ ,
\ \ \ 
a^2_1 m_1 w_1  +a^2_2 m_2 w_2  = 0 \ , \ \  \ \ \ 
w^2_1 - m^2_1 = w^2_2 - m^2_2 = \n^2 \ . \ee 
We shall  assume that $w_i >0, \ m_1 < 0, \ m_2 > 0$.
In this case $f_3,g_3$ decouple (they have mass $\nu$, cf. \rf{flu}) 
  and we get 
  the following Lagrangian for the remaining three  $x_s$-fluctuations 
 $f_1,f_2$  and (rescaled) $g_2$
 \bea \td L &=&  \ha (\dot f^2_1   + \dot f^2_2 
 + \dot {g}^2_2 - f'^2_1  -f'^2_2 -    g'^2_2 ) \nonumber \\
 \la{uuu}
  + \ 
 2 (  a_2 w_1 f_1  &-&  a_1 w_2 f_2 )\dot g_2 
 -  2 (  a_2  m_1 f_1  - a_1 m_2 f_2 ) g'_2  
 \ . \eea
 To find the spectrum  of characteristic frequencies
corresponding to this  action    we  note that 
since $f_i$ and $g_i$ must 
be periodic in $\s$   one can expand the solution of the
quadratic fluctuation equations  in modes
\be 
x_s = \sum_{n=-\infty}^{\infty} \sum_{k=1}^8  A^{(k)}_{sn}
e^{i( \ome_{n,k} \tau  \ + \  n \s)}  \ \ , \ee 
where $k$ labels different frequencies  for a given value of $n$
(we shall   suppress the index $k$ below).
Plugging this into the classical equations that
follow from \rf{uuu}   one finds  the following
equation for the 4 non-trivial characteristic frequences 
(it expresses the vanishing of determinant of the characteristic matrix)
 \be \la{bbbb}
 (\ome^2 - n^2)^2  - 4  a^2_2 ( w_1 \ome - m_1 n)^2 
 - 4 a^2_1  ( w_2 \ome - m_2 n)^2 =0 \ .  \ee 
 The stability condition is 
 that all 4  roots should be  real. 
 The solutions are obviously real for $n=0$ so an instability 
 may appear only for $n=\pm 1, ...$. 
  In the special case of the equal-spin  circular
 solution of \ci{ft2}, i.e.   \rf{two},\rf{uio},  we find  
$
 (\ome^2 - n^2)^2  - 4 w^2  \ome^2  - 4 m^2  n^2  =0 , $
 i.e. \ci{ft2} 
 \be \la{ooo}
 \ome^2_{\pm} = n^2 + 2 \nu^2 +  2 m^2  \pm 2 \sqrt{ ( \nu^2 +
 m^2)^2 
 + n^2 ( \nu^2 + 2 m^2) } \ , \ee
which implies instability  when $ n^2 - 4 m^2 < 0$, 
i.e. for 
$n= \pm 1, ...,\pm (2m-1)$ \ci{ft3}. 
This instability  is present also for generic  2-spin solutions 
with $a_1\not= a_2, \ m_1 \not=-m_2$. 

In spite of the  instability it is  useful to work out the
 spectrum of frequences 
 in the limit of large spins (i.e. large  $\nu$, cf. \rf{kik})
   since the resulting energies may  be
    compared to SYM theory.
      The large $\nu$ expansion of \rf{ooo}  gives 
 (for the lower-energy modes)
 \be \la{kkk}
 \ome_{-} = \pm { 1\ov 2\nu} n \sqrt{n^2-4m^2} 
 + O( { 1\ov \nu^3}) 
 \ ,  \ee 
 and so the 
 contribution to the energy of rotating string 
  from (a pair of) such modes
 is (here $\k^2 = \nu^2 + 2 m^2$, $  J =J_1+ J_2 = \sql 
  \sqrt{ \nu^2 + m^2}$)
 \be \la{huh}
 \Delta E_n = { 2 |\ome_{-}| \ov \kappa} 
 = { 1\ov  \nu^2} n \sqrt{n^2-4m^2}  + O( { 1\ov \nu^4}) 
= { \l\ov  J^2} n \sqrt{n^2-4m^2}  +O( { \l^2 \ov J^4})
\ . \ee
This expression was indeed reproduced \ci{mz2} on the SYM side
(for $m=1$) as the anomalous dimension of excited string states 
corresponding to a particular Bethe root distribution of 
 a Heisenberg  spin chain related to the dilatation 
 operator in the two R-charge sector. 
 
In the general $(m_1, m_2)$ case, there are modes that 
have $\ome \sim {1 \ov \nu}$ and modes for which 
$\ome^2 \to 4
\n^2$ at large $\nu$ (see \ci{ft3}). 
Expanding \rf{bbbb} at 
large $\nu$ assuming
$\ome = O( {1\ov \nu})$ we find the following generalization of 
\rf{kkk}
\be \la{kuk}
 \ome_{-} =  { 1\ov 2\nu}
n \bigg[ 2a_2^2 m_1   +  2  a_1^2 m_2  
\pm   \sqrt{n^2-  4   a^2_1 a^2_2 (m_1 -m_2)^2  } \bigg]
 + O( { 1\ov \nu^3}) 
 \ ,  \ee 
 where $a^2_1+ a^2_2=1$. 
 Eq. \rf{kuk}  reduces to \rf{kkk} in the equal-spin case when 
 $a^2_1=a^2_2=\ha, \ m_1=-m_2$. 
 Recalling  that we have the constraint $m_1 J_1 + m_2 J_2=0$ 
 where $J_i = a^2_i \sqrt{m^2_i + \nu^2}$,  
 one concludes  about  the existence of 
 unstable  modes  with $n^2  < 4|m_1m_2|$ \ci{art}.
Again, one  should be able to reproduce the analog of \rf{huh} in the case
of \rf{kuk} on the gauge theory side.

\bigskip 

It is  straightforward to find the 
 generalisation of \rf{uuu},\rf{bbbb}  
to the 3-spin case, i.e. when $a_3$ is non-zero. 
The resulting spectrum is  similar to the spectrum in the 
$(J_1,J_2=J_3)$ case in \ci{ft3}. 
The generalization of the eq.\rf{bbbb} to the 3-spin case is \ci{art}
\bea
(\ome^2 - n^2)^4 &-& (\ome^2 - n^2)^2 \big[(a^2_2+ a^2_3) \Om^2_1 
+ (a^2_2+ a^2_3) \Om^2_2 +(a^2_1+ a^2_2) \Om^2_3 \big]
\nonumber\\
 \la{aaa} 
&+& \ a^2_3 \Om^2_1 \Om^2_2 + a^2_2 \Om^2_1 \Om^2_3 + 
a^2_1 \Om^2_2 \Om^2_3 =0 \ , \eea
where 
$ \Om_i \equiv 2 (w_i \ome - m_i n) \  $
and $a_i$ and $w_i$ can be expressed in terms of $\nu$ and $m_i$ using 
\rf{soll}. 
Setting $\Om_3=0,\ a_3=0$  leads us  back to \rf{bbbb}.
Eq. \rf{aaa}  gives 8 characteristic frequences, 
4 of which scale as 
${1 \ov \nu}\bar \ome$ in the large $\nu$ (large $\J$) limit.  
In general, there is 
a range of parameters for which the solution is stable 
\ci{ft3,art}, i.e.  
  $\bar \ome$'s  are real. 
 
For example, for the  choice of the parameters in 
 \rf{spec} when two of the spins are equal, 
 we find \ci{ft3} ($\ome \to  {1 \ov \nu} \bar \ome$) 
 \be 
 \bar \ome^2 = { 1 \ov 4} n^2  \bigg[  {n^2}   + 2(3  a^2 -1)m^2 
 \pm 2 m  \sqrt{ (3 a^2-1)^2 m^2  +  
 4a^2 ({n^2} -m^2 ) }\bigg] \ . 
 \ee 
 Note that the limit $m=0$ corresponds to the point-particle  
 (BMN) case when $\ome=  \sqrt{ \nu^2  +  n^2}$. 
 The condition of stability, i.e. $\ome^2 \geq 0$ 
 is obtained  by demanding that 
 $(p^2 - 4 ) ( p^2 - 4 a^2 ) \geq 0$ and 
 $  (3 a^2-1)^2  +  4a^2 (p^2- 1)    \geq 0$, where
 $p\equiv { n\ov m}$. For $m=1$ the  stability condition is satisfied 
 if  $a^2 \geq { 1 \ov 4}$  \ci{ft3}.
 Similar stability 
 conditions on $a$ (or $\cos \g_0$ in the notation of \ci{ft3}) 
  are found for other values of $m$ \ci{ft3,art}.
 
 %%%%%%%%%%%%%%%%%%%%%%%%%%%%%%%%%%%%%%%%%%%%%%%%%%%%%%%%%%
 \subsection{1-loop string correction to the classical energy}
 %%%%%%%%%%%%%%%%%%%%%%%%%%%%%%%%%%%%%%%%%%%%%%%%%%%%%%%
 
 As was shown in \ci{ft3}, for the  stable 3-spin solution  \rf{spec}
 one can compute
 the 1-loop correction to the classical energy \rf{ipo}   by summing over
 all (bosonic and fermionic) fluctuation frequences.
  As in the static gauge,
     here  $t=\k \tau$  and so   
the space-time energy and the 2-d energy (sum of $\ha \w_n$ for all
oscillator  frequencies) are related by \ci{ft1,ft3}
$E = { 1 \ov \k} {\rm E}_{\rm 2-d}$. Thus 
the 1-loop correction is given by the  standard sum
of the oscillator frequencies
\be    E = { 1 \ov \k} {\rm E}_{2-d}
= {1\ov 2\kappa} \big(\ \sum_{n\in Z} \w_n^B -
\sum_{r\in Z+\ha} \w_r^F\ \big)\ \ ,    \ee
where $\w_n^B=\sum_{k=1}^8 \ome_{n,k}^B$ and
 $\w_r^F= \sum_{k=1}^8 \ome_{r,k}^F$
 and  the index $k$ labels the characteristic frequencies.
 Here we need also to include contributions of $AdS_5$ 
 fluctuations with masses equal to $\k$ \ci{ft3}. 
As expected on the basis of conformal invariance of the \adss
string theory, 
this   expression is found to be UV finite \ci{ft3}.
The 1-loop correction vanishes  in the
``point-particle'' limit  when $m=0$ in  \rf{spec}, 
in agreement with the
non-renormalization of the energy of the corresponding    BPS state
dual to a gauge theory  operator with protected conformal
dimension \ci{bmn}. 

As was found in \ci{ft3}, the leading term in $E_1$  
in the large 
$\k \approx  \nu \approx  \J$ limit  is given for $m=1$
 by 
\bea\la{yyy}
E_1 &=&{1\ov \k^2} d_1 + O({1\ov \k^3}) \ ,\\
\la{eee}
d_1 = - \ha \big[  5 a^2 &+& 4 - \sqrt{ 3(4a^2  -1)}  
- 4 \sqrt{ 3a^2 + 1} \big]
\ .  \eea
We are interested in the limit  when $J_i \to \infty$ with 
${J_i\ov J}  $ held fixed (here $J=\ss J_i= J_1 + 2 J_2$).
Since  at large $\k$ we have 
$  {1\ov \,\k^2} =  {\l\ov J^2} + ...$, 
and (see \rf{kik})
$
a = a({J_2\ov J})  \approx 1 - {J_2\ov J } \geq { 1 \ov 2} $
we get (cf. \rf{oppu})
\be 
 E_1 ={\l\ov  J^2} d_1({J_2\ov J}) + ... \ .\ee
For  $J\gg J_2$ we find  $d_1({J_2\ov J})
  \approx 1 - { 7J_2 \ov J}$.
Combining this  with the classical result  for the  energy
\rf{ipo}, we obtain \ci{ft3} 
\be
\la{geg}
E = J + {\l \ov J^2} \big[ J_2 + d_1({J_2\ov J})  +
... \big] + ...
= J + {\l \ov J} \big[ {J_2\ov J} + { 1 \ov J} d_1({J_2\ov J})  +
...\big] + ...\ , \ee
i.e.
 \be \la{oneo}
E= J + {\l \ov J} \bigg[ { J_2\ov J}  + { 1 \ov J} \big[1- {7 J_2 \ov J}+  
O( ({J_2\ov J})^2)\big] + 
 O({1 \ov J^2})  \bigg] +  O( {\l^2 \ov J^3} ) 
\ .\ee
We conclude that  the leading order $J$ term 
in the clssical  energy  is not modified 
by the 1-loop correction,
 and that  the 1-loop contribution to the first 
classical correction term ${\l \ov J}$ 
 is subleading in the ${1\ov J}$ expansion, 
  in agreement with \rf{iipy}.
  
It is natural to conjecture that all higher-loop sigma
model superstring corrections are also subleading at large $J$.
As in the BMN case (see \ci{ft1} and sect. 3.2 in \ci{tsec}), 
the underlying   reasons  for this  should be 
that  (i) the 2-d energy of this  2-d  UV finite 
QFT on a compact space (cylinder)
should admit a regular inverse-mass expansion, 
and 
(ii) the space-time supersymmetry of the superstring action
``spontaneously''
broken by the solution  should 
imply  some kind of ``asymptotic supersymmetry''. 
%(which may be related to the observation in \ci{matt}). 
That would mean that in the limit when 
 $J $ is sent to  infinity for fixed ${\l\ov J^2}$ 
the classical expression for the ground-state energy \rf{epon},\rf{ipo}
{\it and} the energies of excited string states obtained from quadratic
fluctuations  are {\it exact}, just like in the BMN case. 

Similar conclusions should apply for all multispin  string solutions 
that have energy admitting a regular expansion in 
 ${\l\ov J^2}$ as in 
\rf{jojk}. If there is indeed a 
 general relation between the regularity of the classical
expression of the energy and suppression of quantum 
corrections to it in
the $J \to \infty$ limit remains to be understood. 

As discussed in section 1, 
it  should be  then  be possible to compare the classical 
energy with the 
SYM anomalous dimension also computed in the limit of large $J$. 
Such a comparison was indeed successfully performed for the 2-spin circular
\ci{mz2,afrt,bfst} and folded
 \ci{mz2,ft4,bfst} string solutions and  the 3-spin circular solution 
 of \ci{ft2} with two equal spins \ci{mz3}. 
 
Another  interesting open problem  is to compare the string 
1-loop $1\ov J$ correction to 
the leading $\l \ov J$ term in 
\rf{oneo} with  the corresponding $1\ov 
J$ correction to the thermodynamic 
limit of the Bethe ansatz expressions
 for the anomalous dimension
\ci{mz3} 
on the gauge theory side.

%Comments:    2 types of frequences;  asymptotic susy 
%? why suppression. 

%%%%%%%%%%%%%%%%%%%%%%%%%%%%%%%%%%%%%%%%%%%%%
\setcounter{equation}{0}
%%%%%%%%%%%%%%

%%%%%%%%%%%%%%%%%%%%%%%%%%%%%%%%%%%%%%%%%%%%
\setcounter{footnote}{0}
%%%%%%%%%%%%%%%%%%%%%%%%%%%%%%%%%%%
%%%%%%%%%%%%%%%%%%%%%%%%%%%%%%%%%%%%%%%%%
\section{Rotating strings in  \adss
}
%%%%%%%%%%%%%%%%%%%%%%%%%%%%%%%%%%%%%%%

Let us now generalize the discussion of section 4
to the case when the string can rotate in both $AdS_5$ 
and $S^5$. 
For that we need to supplement the 
$S^5$ rotating string ansatz \rf{emb}  by a 
similar $AdS_5$ one \ci{ft2,afrt,art}:
$$
\Y_0\equiv Y_5+i Y_0= \zz_0(\s ) e^{i \w_0\tau }\ , \ 
 $$
\be \la{adr}
\Y_1\equiv Y_1+i Y_2= \zz_1(\s ) e^{i \w_1\tau }\ ,
\ \ \ \ \  \ \ \Y_2\equiv  Y_3+i Y_4= \zz_2(\s ) e^{i \w_2\tau } \ . 
\ee 
Here  the  functions 
$\zz_r=(\zz_0,\zz_1,\zz_2)$ are, in general, 
 complex, and  because of
the condition
$\eta_{MN}Y^M Y^N=-1$, their real  radial parts 
 lie on a  hyperboloid ($\eta_{rs}=(-1,1,1)$, cf. \rf{sat}) 
\be\la{dos} 
\zz_r = \rr_r e^{i \b_r} \ , \ \ \ \ \ \ 
\eta^{rs}\rr_r\rr_s \equiv -\rr_0^2+ \rr_1^2 + \rr^2_2  
=-1\ . 
\ee
In sections 4 and 5 we had $\rr_0=1, \ \rr_1=\rr_2=0, \ \b_r=0.$
To satisfy the  closed string periodicity conditions  we need, as in
\rf{pep},  
\be \la{tqr}
\rr_r (\s + 2 \pi)= \rr_r (\s) \ , \ \ \ \ \ \ 
\b_r (\s + 2 \pi)= \b_r (\s)  + 2 \pi k_r \ , \ee
where $k_r$ are integers. Comparing \rf{adr}  to \rf{relx} we conclude that 
the $AdS_5$ time $t$ and the angular coordinates $\p_1,\p_2$
are related to $\b_r$ by 
\be \la{tte}
t = \w_0  \tau  + \b_0 (\s) \ , \ \ \ \ \ 
\p_1 = \w_1  \tau  + \b_1 (\s) \ , \ \ \ \ \
\p_2 = \w_2 \tau  + \b_2 (\s) \ . \ee 
We shall require  the time coordinate $t$
to  be single-valued, i.e. ignore windings in time direction  
and will also rename $\w_0$ into $\k$, i.e. 
\be  k_0=0 \ , \ \ \ \ \ \ \ \  \   \w_0 \equiv \k   \ . \ee 
The three $O(2,4)$ Cartan generators (spins) in \rf{aas} here 
are ($S_0=E, \ \w_r=(\w_0,\w_1,\w_2)$)
\be \la{chak}
S_r=\sqrt{\lambda } \w_r \int_0^{2\pi } {d\sigma\over 2\pi }\ 
\rr_r^2(\sigma )\equiv \sqrt{\lambda }\ \S_r \ . 
\ee
In view  of (\ref{dos}), they satisfy the relation
\be\la{hjh}
\sum_{s,r} \eta^{sr} {\S_r\ov \w_s} =-1 \ , \ \ \ {\rm i.e.} \ \ \ 
{\E\over \kappa } - {\S_1 \over \w_1}  - {\S_2 \over \w_2}=1 \ . 
\ee
Substituting the above rotational ansatz into the $AdS_5$ Lagrangian 
(and changing overall sign) 
we find the analog of the 1-d  Lagrangian \rf{L} in 
the $S^5$ case \ci{art} (we assume sum over repeated indices $r,s$)
\be\la{tl}
\td \rL = \ha \eta^{rs} (\zz_r'{\zz_s^*}'-\w_r^2  \zz_r\zz_s^* )-
\ha \tilde \Lambda (\eta^{rs} \zz_r\zz_s^*   + 1) \ . 
\ee
Like its $S^5$ counterpart \rf{L}, this 1-d 
 Lagrangian is a special case of an $n=6$ Neumann system, 
now with  signature $(-++++-)$, and thus represents again 
an integrable system, 
being related  to  a special 
euclidean-signature Neumann  model by an analytic continuation.  
 The  reduction of the total \adss Lagrangian on the rotation ansatz 
 is then given by the sum of \rf{L}  and \rf{tl}. 
From \rf{tl} we find as in \rf{ank}
\be\la{ankk}
\b'_r  = { u_r\ov \rr^2_r}\ , \ \ \ \ \ \  \ \ \ u_r=\const \ , \ee
 so that the effective Lagrangian for the radial coordinates becomes  
\be\label{la}
\tilde \rL= \ha \eta^{rs} ({\rr_r'}{\rr_s'}
- \w_r^2 \rr_s\rr_s - {u_r u_s\over
\rr_r \rr_s}  )  - 
  \ha \tilde \Lambda (\eta^{rs} \rr_r\rr_s    + 1) \ . 
\ee
Thus   (\ref{la}) describes a 
Neumann-Rosochatius integrable system with 
indefinite signature, i.e. 
with $\delta_{ij}$ replaced by $\eta_{rs}$ (cf. \rf{Laa}). 

We should also require the periodicity condition analogous to 
\rf{okl}: $
u_r \int^{2\pi}_0 { d\s \ov \rr^2_r(\s)} = 2 \pi k_r $.
Then $k_0$ implies that  we should set $u_0=0$
as a consequence of single-valuedness of the $AdS_5$ time.

While  the  equations for $r_i$  and $\rr_r$ following from 
\rf{La} and \rf{la} are decoupled, the variables of the two NR 
systems are mixed in the conformal gauge constraints 
\rf{cv},\rf{cvv} which now take the form (generalizing 
\rf{cve},\rf{cvve} where we had $\rr_0=1, \ u_r=0, \ \rr_a=0$)
\bea \la{cvi}
 \rr'^2_0   + \k^2 \rr^2_0     
 = \sum_{a=1}^2 ( \rr'^2_a   + \w_a^2 \rr^2_a  +  
 { u_a^2 \ov \rr^2_a} )   &+&    \sum_{i=1}^3( r'^2_i   + w_i^2 r^2_i  +  
 { v_i^2 \ov r^2_i} )    \ ,\\ 
\la{cvvi}
 \sum_{a=1}^2 \w_a  u_a   &+&  \sum_{i=1}^3 w_i v_i  =0   \ .
\eea
Here 
$  \rr_0^2 - \sum_{a=1}^2 \rr_a^2 =1  ,$ and  $
\sum_{i=1}^3 r^2_i =1 $ and we used that $u_0=0$. 
One  can then repeat the discussion of sections  4.2, 4.3
and 4.4 in the
present case, classifying   general solutions 
of the resulting NR system.
One again finds folded and circular solutions,
 and the two-spin folded
solution exists only if the string is bent \ci{afrt}.

%%%%%%%%%%%%%%%%%%%%%%%%%%%%%%%%%%%%%%%%%%%%%%%%
\subsection{Simple circular strings in $AdS_5$ }
%%%%%%%%%%%%%%%%%%%%%%%%%%%%%%%%%%%%%%%%%%%%%

Let us first  assume that the string 
is not rotating in  $S^5$ (i.e. $w_i,v_i=0, \ r_i=\const$) 
and consider the $AdS_5$ analog of the simplest circular solution 
of section 5 by demanding $\td \L=\const$. 
The discussion is exactly the same as in section 5 with 
few  signs reversed. 
As in section 5.1, finding solutions with $\td \L=\const$ 
 turns out to be equivalent 
to looking for constant radii ($\rr_r=\const$)  solutions.
Then (cf. \rf{kop},\rf{kol}) 
\be \la{hop}
\rr_r=\ab_r=\const \ , \ \ \ \ \  \beta_{a} =k_{a}\s \ , \ \ 
\ \ k_0=0\ , \ \ \ \  u_0=0 \ , \ \ \ \
u_a = \ab^2_a k_a \ , \ee
 \be \ \w^2_0\equiv \k^2=  \td \Lambda \ , \ \ \ \ \ \ \ 
\w^2_a = k^2_a  + \k^2  \ , \ \ \ \ \ \ a=1,2 \ .  
\ee
The energy  as a function of spins is then obtained by solving 
the system  that follows from the definition
of the charges \rf{chak}
 and  the  constraints \rf{cvi},\rf{cvvi}
 with $\kappa$ as a parameter (cf.
\rf{kpp}--\rf{uiy})
\be
{\E\over\kappa } - {\S_1 \over \sqrt{ k^2_1 + \k^2}}  - 
{\S_2\over  \sqrt{k^2_2 + \k^2} } = 1\ , 
\label{cuat}
\ee
\be\la{nnb}
\k \E - \ha \k^2  = \sqrt{ k^2_1 + \k^2 }\ 
 \S_1  +  \sqrt{k^2_2 + \k^2}\  \S_2   
\ , \ \ \ \ \ \  k_1 \S_1  +  k_2 \S_2=0\ . 
\ee 
This implies 
$
{\S_1 k^2_1 \over \sqrt{k^2_1 + \k^2}}  +  
{\S_2 k^2_2 \over  \sqrt{k^2_2 + \k^2} } = \ha \k^2 .$
Considering the limit of large spins  $\S_i  \gg 1$, 
with  $k_a$ being fixed 
we conclude that
 $\k = ( 2 k_1^2 \S_1 + 2 k_2^2 \S_2)^{1/3} + ...$ 
 and then  
\be
\E= \S_1 + \S_2 + {3\over 4} (2k_1^2\S_1+2k_2^2\S_2)^{1/3}+...
\ ,   \ee
or,  in view of  $k_1 \S_1 = -k_2 \S_2$
(treating $S_1,S_2$ and $k_1$ as independent) 
\be\la{pqw}
\E= \S + {3\over 4} \big( 2k^2_1 \S  { \S_1\ov \S_2} \big)^{1/3}
+...\ , \ \ \ \ \ \ \ \  \S\equiv \S_1 + \S_2 \ .   
\ee
Using \rf{chak} this can be rewritten  as 
\be\la{qqw}
E= S + {3\over 4} (\l S)^{1/3}  \big( 2k_1^2 {S_1\over S_2}\big)^{1/3}
+ ...\ .    
\ee
The case of  $k_1=-k_2=k$ when the two spins are equal 
$\S_1=\S_2=\ha \S$  is that of the the circular 
solution found in  \ci{ft2}  for which we get 
\be 
E=  S + {3\over 4}(2k^2 \l S)^{1/3} + ...\ . 
\ee 
As was shown in \ci{ft2}, this $k_1=-k_2$ solution is 
stable only for  small enough $\S$
(namely, $\S \leq {5\ov 8} \sqrt{7\ov 2}$ for $k=1$). 
% An obvious question is if these more general solutions
%are more stable. 

The ``non-perturbative'' scaling of the subleading 
term  in \rf{qqw} with $\l$ precludes  one  from direct comparison
of the above energies  to
anomalous dimensions of the 
corresponding \ci{ft2} SYM operators which 
(in euclidean version) have the following structure \ci{ft2}
tr$(\bar \Phi (D_1 + i D_2)^{S_1} 
(D_3 + i D_4)^{S_2}\Phi) + ... $.
%This is different from  what   was found in the $S^5$ 
%rotation case. 
This is unfortunate, since such 
 operators are of more ``realistic'' 
type similar to the ones relevant for 
 high-energy scattering in 
 non-supersymmetric gaure theories -- they 
 contain many covariant
derivatives 
instead 
of many scalars and thus may appear  in less supersymmetric 
gauge theories without 
adjoint scalars. 

It turns out that one needs a  large $J$ spin in $S^5$
directions to have a regular \rf{jojk} 
 expansion of the energy.
Indeed, the situation changes 
 when we consider 
``hybrid'' solutions where the circular string rotates in both  
$AdS_5$ and $S_5$ directions.

%%%%%%%%%%%%%%%%%%%%%%%%%%%%%%%%%%%%%%%%%%%%%%%%
\subsection{Constant radii 
 circular strings in $AdS_5\times S^5$ }
%%%%%%%%%%%%%%%%%%%%%%%%%%%%%%%%%%%%%%%%%%%%%

It is straightforward to  combine the  solutions 
of sections 5.1 and 6.1 to write down the most general 
circular constant-radii solution in $AdS_5\times S^5$ 
\ci{art}.
It is  parametrized by the  frequences ($a=1,2; \ i=1,2,3$)
\be \la{fff}  \w_0=\k \ , \ \ \ \ \ 
\w^2_a = k_a^2 + \k^2 \ , \ \ \ \  \ \ 
w^2_i = m^2_i + \nu^2 \ ,  \ \ \ \ \ \k^2= \td \L \ , \ \ \ 
\nu^2 = -\L\ ,  \ee 
related to the 
energy $\E$ and 2+3 spins $\S_a $ and 
$\J_i$   and topological numbers $k_a$ and $m_i$.
These will be  
related by \rf{kpp} and \rf{hjh} as well as by the conformal
gauge constraints \rf{cvi} and \rf{cvvi}. Explicitly, 
we get the following generalization of both \rf{onk}--\rf{uiy}
and \rf{cuat},\rf{nnb}
\bea\la{kqp}
 \ss {\J_i \ov  \sqrt{m^2_i + \nn^2} } =1 \  , \ \ \ && \ \ \
 {\E\over\kappa } - 
 \sum_{a=1}^2 {\S_a \over \sqrt{ k^2_a + \k^2}} = 1 \ , \\
\la{oiy}
 2\k \E -  2 \sum^2_{a=1} \sqrt{ k^2_a + \k^2 }\ 
 \S_a  - \k^2   &=&  2 \ss  \sqrt{m^2_i + \nn^2}\   \J_i  - \nn^2 \ , \\
 \la{seo}
\sum_{a=1}^2   k_a \S_a  + \ss m_i \J_i &=&0\ . 
\eea
Here $\k$ and $\nu$ (or the two Lagrange multipliers in \rf{fff}) 
 are parameters that need to be solved for 
in order to find $\E$ as a function of the spins 
$\S_a,J_i$ and windings $k_a,m_i$. 
The solution exists only for such integers 
$k_a$ and $m_i$ that satisfy \rf{seo}.

If all spins are of the same order and large 
$\S_a \sim \J_i \gg 1$ we find 
\bea 
\k &=& \J  + { 1 \ov 2 \J^2 } ( \ss  m^2_i \J_i+2 \sa  k^2_a
\S_a)  + O({1 \ov \J^2}) \ , \ \ \ \ \ \ 
   \J\equiv  \ss  \J_i \ ,\nonumber  \\ \la{kool}
\nu&=&\J  -   { 1 \ov 2 \J^2 }  \ss  m^2_i \J_i  
  + O({1 \ov \J^2}) \ ,\eea
and thus ($\S\equiv  \sa  \S_a$) 
\be \la{eenn}
\E = \J + \S + { 1 \ov 2 \J^2 } ( \ss  m^2_i \J_i+ \sa  k^2_a \S_a) 
 + O({1 \ov \J^3}) \  , \ee
 or  \ci{art}
 \be \la{ennl}
E = J + S + { \l \ov 2 J^2 } ( \ss  m^2_i J_i+ \sa  k^2_a S_a) 
 + O({\l^2 \ov J^3}) \  . \ee
This expression is a direct generalization of  \rf{epon}
in the $\S_a=0$ case. 
The energy is minimal  if $m^2_i$ and $k_a^2$ 
have minimal possible values (0 or 1). 
We may also 
look at a different  limit when $ \J \gg \S \gg 1$.
In this case we get a ``BMN-type'' (single $J$ rotation)
asymptotics with the leading term still given by 
\rf{ennl}, i.e. $\Delta E \sim { \l \ov 2 J^2 } S$.

As an  example, let us consider the simplest hybrid  solution
when only one of  each types of the spin is  non-zero, 
i.e. $\J_1=\J, \ \S_1=\S, \  \S_2=\J_2=\J_3=0$.
Then $\rr_0^2 - \rr^2_1=1,\  \rr_3=0$ 
and $r_1=1,\ r_2=r_3=0$, i.e.  (cf. \rf{relx}) 
\be \la{joji}
\YY_0 = \cosh \r_0 \ e^{i\k \tau} \ , \ \ \ \ 
\YY_1 = \sinh \r_0 \ e^{i\w \tau + i k \s}\ , \ \ \ 
\XX_1= e^{i w \tau + i m \s} \ , \ee
where $\rr_0= \cosh \r_0$ 
determines  the fixed radial coordinate in
$AdS_5$ at which the string is located while it is spread 
and rotating in $\phi_1$ (it is positioned  at  $\theta={\pi\ov 2}$
and $\phi_2=0$ in $S^3$ of  $AdS_5$). 
Also, the string is a rotating circle along $\vp_1$ in
$S^5$  located at $\vp_2=\vp_3=0, \ \g={\pi\ov 2}, \ \psi=0$. 
Its energy  for $\J \sim \S \gg 1$ is then
%\foot{Here 
%$m \J = - k \S$
%and $\J = \sqrt{ m^2 + \nu^2}$, 
%$2 \k \E - \k^2 = 2 \sqrt{ 
%k^2 + \k^2} \S + \J^2 + m^2$, i.e. 
%\ $\E = \k + { \k \S \ov \sqrt{
% k^2 + \k^2}} $.}
%\foot{The BMN limit 
%here corresponds to $ J \gg S$,
% or $ |k| \gg |m|$.}
\be \la{spl}
E= J + S + { \l \ov 2 J^2} ( m^2 J + k^2 S) + ... 
   \ . \ee 
%or, for example, for $k=-m$ when  $\J=\S$ 
%we get $
%\E= 2\J  + { m^2 \ov  \J}  + ... \ . $
One can easily analyse the fluctuations near 
 this solution as was done in in section 5.3 \ci{art}.
 We find 1 massless and 4 massive (mass $\nu$) fluctuations in $S^5$
 directions; in addition to 2 massive (mass $\k$) 
 decoupled $AdS_5$ fluctuations 
 there are also 3 coupled ones with a Lagrangian similar to \rf{uuu}.
 Then  the equation \rf{bbbb} for
  the characteristic frequences becomes
  \be 
  (\Omega^2 -n^2)^2 + 4 \rr^2_1 ( \k \Omega)^2  - 4 \rr^2_0 
  ( \om_1 \Omega - k n)^2 =0 \ ,  \ee
 and one concludes that this $(S,J)$   solution is 
 always {stable}. Indeed,  setting 
  $ \rr_0=a, \ \rr_1=\sqrt{a^2-1}$ we get 
 \be \la{kuek}
 \Omega_{-} =  { 1\ov 2\k}
n \bigg[  2  a^2  k 
\pm   \sqrt{n^2 +   4   a^2 (a^2-1)   k^2  } \bigg]
 + O( { 1\ov \k^3}) 
 \ ,  \ee 
 so that for any  $a=\cosh \r_0\geq 1$ 
  there are no unstable modes.
  
  The conclusion is that  for a 
regular large-spin expansion of the energy  one needs to have at
least one (large)  component of spin in $S^5$ direction. 
This turns out to be true also in the case of other
(folded and circular) 
spinning string solutions with more complicated 
$\s$-dependence.

%%%%%%%%%%%%%%%%%%%%%%%%%%%%%%%%%%%%%%%%%%%%%
\setcounter{equation}{0}
%%%%%%%%%%%%%%

%%%%%%%%%%%%%%%%%%%%%%%%%%%%%%%%%%%%%%%%%%%%
\setcounter{footnote}{0}
%%%%%%%%%%%%%%%%%%%%%%%%%%%%%%%%%%%
%%%%%%%%%%%%%%%%%%%%%%%%%%%%%%%%%%%%%%%%%
\section{``Inhomogeneous'' two-spin   solutions in \adss }
%%%%%%%%%%%%%%%%%%%%%%%%%%%%%%%%%%%%%%%%%%%%%%%

%%%%%%%%%%%%%%%%%%%%%%%%%%%%%%%%%%%%%%%%%%%%%%%%%%
\subsection{Rotating ansatz  in terms of angles}
%%%%%%%%%%%%%%%%%%%%%%%%%%%%%%%%%%%%%%

If we set $k_a$ and $m_i$ or $u_r$ in \rf{la} and $v_i$
in \rf{Laa}  to zero 
(i.e. assume that the angles $\p_a$ and $\vp_i$ do not
 depend on
$\s$),
the \adss NR system reduces to the sum of the two $n=3$ 
Neumann systems. Then  rotating strings  carrying 
2+3 charges $(S_1,S_2; J_1,J_2,J_3)$ and the energy $E$ 
  are described by the following ansatz in terms of angles in 
  \rf{relx} \ci{ft2}
\be\la{anss}
t= \kappa \ta \ , \quad \phi_a = \vo_a \ta\ , \ \ \
 \vp_i= \wup_i \ta \ ,\ \ \ \  \rho(\sigma)=
  \rho(\sigma + 2 \pi)\ . \ee
The  remaining angles may depend only on $\s$, i.e. 
   $\theta=
\theta(\sigma)$,
$\gamma=\gamma(\sigma)$ and 
 $\psi=\psi(\sigma)$  and may be periodic modulo 
$2\pi$ shift, e.g., 
$\psi (\sigma + 2 \pi)= \psi(\s) + 2 \pi n $. If $n=0$ we get 
{\it folded}  solutions, if $n\not=0$ we get {\it circular} 
solutions \ci{afrt}.

The conserved charges in \rf{aas},\rf{jij} then have the following
explicit form 
\[\label{tti}\arraycolsep0pt
\begin{array}{rclcrcl}
\mathcal{S}_1 \earel{=} 
\vo_1 \int^{2 \pi}_0 \frac{ 
d \sigma}{2 \pi} \, \sinh^2 \rho\,  \cos^2
\theta\,,
 &\quad&\ 
\mathcal{J}_1
\earel{=}
\wup_1  \int^{2 \pi}_0 \frac{d 
\sigma}{2 \pi} \, \sin^2 \gamma \, 
\cos^2 \psi \,,
\\[15pt]
\mathcal{S}_2
\earel{=} 
\vo_2  \int^{2 \pi}_0 \frac{d
 \sigma}{2 \pi} \, \sinh^2 \rho\,  \sin^2
\theta\, ,
 &\quad&\ 
\mathcal{J}_2
\earel
{=}
\wup_2  \int^{2 \pi}_0 
 \frac{d \sigma }{2 \pi} \,  \sin^2 \gamma \, 
\sin^2 \psi\, , 
\\[15pt]
\mathcal{E}
\earel
{=}
\kappa \int^{2 \pi}_0 \frac{d \sigma}{2 \pi} \, \cosh^2 \rho\, ,
 &\quad&\ 
\mathcal{J}_3
\earel
{=}
\wup_3 \int^{2 \pi}_0  \frac{d \sigma}{2 \pi} \,  \cos^2 \gamma\, .  
\end{array}\]
The  \sm 
 equations for the $\s$-dependent angles 
  $(\rho,\theta)$  
\[\label{rt}
\begin{array}{c}
\rho'' -  \sinh \rho\, \cosh \rho \, 
(\kappa^2 + \theta '^2 - \vo_1^2 \cos^2 \theta - 
\vo^2_2 \sin^2 \theta )   =0
\,,
\\[6pt]
(\sinh^2 \rho \ \theta')'   -  
(\vo_1^2 - \vo^2_2) \sinh^2 \rho\,  \sin \theta \, \cos \theta  =0 \,,
\end{array}
\]
 and  $(\gamma,\psi)$ 
\[\label{rta}
\begin{array}{c}
\gamma'' -  \sin \gamma\, \cos \gamma\, 
(\wup_3^2 + \psi'^2 - \wup_1^2 \cos^2 \psi - \wup^2_2 \sin^2 \psi )  
=0  \,,
\\[6pt]
(\sin^2 \gamma \ \psi')'   -  
(\wup_1^2 - \wup^2_2) \sin^2 \gamma \, \sin \psi \, \cos \psi  =0 \,,
\end{array}
\]
are decoupled from each other. 
As explained above and in   \cite{afrt}, the resulting system of
equations 
is completely integrable, being equivalent 
 to a combination  of the two
Neumann dynamical systems.
 As a result, there are  2+2 
``hidden'' integrals of motion, reducing the 
general problem to solution 
of two independent 
systems of   two coupled  first-order  equations, 
with parameters related through  the 
only one  nontrivial conformal gauge constraint 
\<\label{conf}
 \rho'^2  - \kappa^2 \cosh^2 \rho   + \sinh^2\rho \ ( \theta'^2 +
 \vo_1^2 \cos^2 \theta  + \vo^2_2\sin^2 \theta ) \qquad \nonumber 
 \\
+ \  \gamma'^2 +  \wup_3^2 \cos^2\gamma  + \sin^2\gamma\ (\psi'^2 +
\wup_1^2 \cos^2\psi + \wup^2_2 \sin^2\psi) = 0 \ . 
\>
Note that the  two metrics in \rf{ads},\rf{ses} 
 are related  by the obvious 
  analytic continuation 
and change of the overall sign,  which is equivalent,  for  the
present rotational ansatz \rf{anss},    to 
\[\label{con}
\rho \lra  i \gamma \ , \ \ \ \  \ \ \theta \lra \psi \ , \ 
\ \ \ \ \  \kappa \lra -\wup_3  \ ,\ \ \ \ \ \  
\vo_1 \lra  -\wup_1\ ,\ \ \ \ \ \  
\vo_2 \lra  -\wup_2 
\  .  \]
 This transformation  maps  the system 
\rf{rt} into the system \rf{rta} and 
also preserves the constraint \rf{conf}. Thus it 
formally maps   solutions into solutions \ci{afrt,bfst}.
Under  \rf{con}
the conserved charges \rf{tti}
transform as follows
%\foot{In order for some  two physical solutions 
%to be related by this analytic continuation prescription 
%at least one of them  should 
% have  a non-vanishing $J_3$ spin
%(which transforms into the energy of the solution).} 
%
\[\label{chaa}
S_1 \lra J_1 \ , \ \ \ \ \ 
S_2 \lra J_2 \ , \ \ \ \ \ 
E \lra  - J_3  \ . \]
This corresponds to interchanging different $SO(2)$ generators 
of the symmetry group $SO(2,4)\times SO(6)$.
One can   find  also other similar  transformations that 
map solutions into solutions by combining \rf{con}
 with discrete  $SO(2,4) \times SO(6)$ isometries
like  interchanging the angular coordinates (see below).
  
%that do not induce other components  of the 
%rotation generators except  the above Cartan ones 
%(e.g.,  interchanging the angular coordinates 
%induces interchanging of the charges in \rf{tti}, 
%etc.).  
%Below  we shall consider such an example.

%%%%%%%%%%%%%%%%%%%%%%
\subsection{Folded two-spin solutions: $(S,J)$ and $(J_1,J_2)$ }
%%%%%%%%%%%%%%%%%%%%%%%%%%%%%%

Let us now review  the  two non-trivial two-spin folded string 
solutions which  are,  in fact,  related by 
the above analytic continuation.

The first is the ``$(S,J)$'' solution  \cite{ft1} 
\[ \label{sj}  
 \kappa,\vo_1, \wup_3 \not=0 \ , \ \ \ \
\rho= \rho(\sigma) \ , \ \ \ \ \theta=0\ ,\ \ \ \
\gamma= 0 \ , \ \ \ \ \psi=0 \ ,  \]
where the string is stretched in
the radial direction $\rho$ of $AdS_5$  and 
 rotates ($\vo_1$) in $AdS_5$ about  its center of mass. The  latter 
 in turn moves ($\wup_3$) along a large circle of $S^5$.
 In the limit of point-like string ($\S=0$) this  becomes 
 a massless geodesic in $S^5$ as in \ci{bmn,gkp}. 
 In the case of $w_3=0$ this becomes the folded string rotating 
 in $AdS_5$ \ci{vega,gkp}.
The gauge constraint \rf{conf} and the 
integrals of motion \rf{tti} here 
become 
\[\la{rrho}
\begin{array}{c}
   \rho'^2  - \kappa^2\, \cosh^2 \rho   + \vo^2_1\,
  \sinh^2\rho =- \wup_3^2   \ , 
\ \ \ \ \ \ \ \J \equiv \J_3 =  \wup_3 \ , 
\\[6pt]\displaystyle
\S\equiv \S_1 =  \vo_1   \int^{2 \pi}_0 
{\frac{ d \sigma}{2 \pi}} \ \sinh^2 \rho\  
  \ , \  \ \ \ \ \ \ \ 
 \E= \kappa  \int^{2 \pi}_0 {\frac{ d \sigma }{2 \pi}} \ \cosh^2 \rho 
 \ . 
\end{array} \]
For the second  ``$(J_1,J_2)$'' solution one has \ci{ft2,ft4} 
\[ \label{jj}  
  \kappa,\wup_1, \wup_2 \not=0 \ , \ \ \ \
\rho= 0 \ , \ \ \ \ \theta=0 \ ,\ \ \ \
\gamma= {\frac{ \pi}{2}} \ , \ \ \ \ \psi=\psi(\sigma)  \ .   \]
Here   the string is 
located at the center of $AdS_5$ while it 
is stretched ($\psi$) along a great circle of $S^5$
and rotates ($\wup_2$)
about its center of mass which moves ($\wup_1$)
along an orthogonal great circle of $S^5$.
The gauge constraint \rf{conf} and the integrals of motion 
\rf{tti} here  are 
\[\la{psii}
\begin{array}{c}
  \psi'^2 +
\wup_1^2 \cos^2\psi + \wup^2_2 \sin^2\psi  =\k^2 \ ,  
\ \ \ \ \ \  \E=  \kappa  \ ,\
\
\\[6pt]\displaystyle
\J_1=  \wup_1   \int^{2 \pi}_0 
{ \frac{d \sigma }{2 \pi}} \    \cos^2
\psi 
 \ , \  \ \ \ \ 
\J_2 =  \wup_2    \int^{2 \pi}_0 { \frac{d \sigma}{2 \pi}} \   
 \sin^2
\psi
  \ .  
\end{array}\]
In view of \rf{con},\rf{chaa} we  conclude  that these two solutions are related
by the following 
analytic continuation:
\[\label{caa}
\begin{array}{c}
\rho \to  i \psi \ , \
\ \ \ \ \  \kappa\to  -\wup_1  \ ,\ \ \ \ \ \  
\vo_1  \to  -\wup_2 \  , \ \ \ \ \  \wup_3 \to - \kappa \ , 
\\[6pt]
   E \to -J_1 \ , \ \ \ \ \ \ \ 
S \to  J_2  \ , \ \ \ \  \ \ \ 
 J \to  - E \ .
\end{array}\]
Here we are  assuming that $\r(\s + 2\pi) = \r(\s)$ {\it and} 
$\psi(\s + 2\pi) = \psi(\s)$, i.e. $\psi$ does 
not have a winding
number. This choice corresponds to a folded string solution
(a two-spin generalization of the solution of \ci{vega,gkp}).

The first-order equations in \rf{rrho} and \rf{psii} 
are first integrals of the sinh-Gordon and sine-Gordon 
equations for $\rho$ and $\psi$, respectively, 
which are the only non-trivial equations of the total  Neumann 
systems one is to solve in the present two  cases: here  the 
related hyperelliptic curve (see section 4.3) 
reduces to an elliptic one. Indeed, their solutions can be readily 
expressed in terms of elliptic functions (see below).  

One can  also directly relate \ci{bfst} the systems of 
equations  expressing the  periodicity
condition and the  respective energies and spins. 
In the first case \cite{ft1} we get, 
introducing a modular parameter 
 $\q <0$ related to the maximal value of the radial $AdS_5$ coordinate
 $\r_0$ 
\bea  \label{qui}
\q\equiv  - \sinh^2 \rho_0&=&
 \frac{\k^2- \wup_3^2}{\kappa^2-\vo_1^2} <0 
\ ,  \\  \la{pui}
\sqrt{\k^2-\wup_3^2}=
\frac{2\,\sqrt{-\q}}{\pi}\,\ellK(\q)\ ,
\qquad \ \  
 {\mathcal{E}} &=& \k+\frac{ \k}{\omega_1}\,
\mathcal{S}
=\frac{2\k\,\sqrt{-\q}}{\pi\sqrt{\k^2-\wup_3^2}}\,\ellE(\q)\  ,
\nonumber\eea
where 
$\ellK(\q)$ and $\ellE(\q)$ are the standard complete elliptic integrals
of the first and the second kind.\foot{They 
are defined by 
$\ellK(\q)= \int^{\pi\ov 2}_0 { d \a \ov \sqrt{1 - \q \sin^2 \a}} $ and 
$\ellE(\q)= \int^{\pi\ov 2}_0 { d \a  \sqrt{1 - \q \sin^2 \a}} $
and are 
related to  hypergeometric functions 
 $ {}_2F_1(\half,\half;1,\q)=
\frac{2}{\pi}\,\ellK(\q) ,\  $ $ 
{}_2F_1(-\half,\half;1;\q)
=\frac{2}{\pi}\,\ellE(\q) $.
Let us note also that the elliptic integral of the third kind is 
defined  by
$\Pi(m^2,\q)=
\int_0^{\pi/2}\frac{d\a}{(1-m^2 \sin^2 \a)
\sqrt{1-\q\sin^2 \a}}.$}
Solving  for $\omega_1$ and $\k$ in terms of $\mathcal{J}$ and $\q$
we  get the system of two equations 
for the energy  as a function of the spins $\E= \E(\S,\J) $ \ci{bfst} 
\bea\label{a5}
\lrbrk{\frac{\mathcal{J}}{\ellK(\q)}}^2
-\lrbrk{\frac{\mathcal{E}}{\ellE(\q)}}^2=\frac{4}{\pi^2}\,\q\ ,
\\ \la{aa5}
\lrbrk{\frac{\mathcal{S}}{\ellK(\q)-\ellE(\q)}}^2
-\lrbrk{\frac{\mathcal{J}}{\ellK(\q)}}^2=\frac{4}{\pi^2}\,(1-\q) \ ,  
\eea
where the parameters $\q $ is negative for a physical folded solution.
The second of these two  parametric equations 
  determines $\q$ in terms of $\mathcal{S}$ and
$\mathcal{J}$, while 
the first  one then gives the energy as a function of the spins.

Similarly, for the $(J_1,J_2)$ solution \rf{jj} one finds 
\cite{ft4} (we assume $\vo^2_2 > \vo^2_1$; here $\psi_0$ is the maximal
value of $\psi$)
\<\label{kui}
\q \equiv \sin^2 \psi_0 = \frac{\kappa^2 - w^2_1 }
{w_2^2 - w_1^2} >0 \ , 
\qquad
1=\frac{\mathcal{J}_1}{w_1}+\frac{\mathcal{J}_2}{w_2},\qquad
\mathcal{E}=\kappa\ ,
\nln
\mathcal{J}_1=\frac{2 w_1}{\pi \sqrt{w_2^2-w_1^2}}\,\ellE(\q)\ , 
\qquad \ \ \ \ \
\sqrt{w_2^2-w_1^2} =\frac{2}{\pi}\,\ellK(\q)\ .
\>
The solution of the equation in \rf{psii} for $\psi$ 
can be written as follows
\be\la{elip}
 \cos  \psi(\s)  = r_1 (\s)= \dn ( A \s, \q) \ , \  \ \ \ \ 
 \sin   \psi(\s)  = r_2 (\s)= \sqrt q\  \sn ( A \s, \q)  \ ,\ee
  where $r_3(\s)=0$ ($\g={\pi\ov 2}$, cf.\rf{relx}), 
    $A\equiv  {2 \ov \pi} \ellK(\q)$ and 
  $\dn$ and $\sn$ are the standard elliptic 
  functions.\foot{The Jacobi 
  elliptic function $\sn (u,\q)$ is defined by 
  $ u= \int^{\sn (u,\q)}_0  { dy \ov \sqrt{ (1 - y^2) ( 1- \q y^2)} } $. 
  Equivalently, if $\sn\ u = \sin \phi$ then 
  $u= \int^{\phi}_0  { d\a \ov \sqrt{  1- \q \sin^2\a} } $.
  One has also $\dn^2 (u,\q) +\  \q\  \sn^2 (u,\q)=1$ and 
  $\cn^2 (u,\q) +  \sn^2 (u,\q)=1$. 
  These three functions (given by ratios of theta-functions)
  are meromorphic. Also, $\sn (-u,\q)= - \sn (u,q)$ and 
  $\sn (u+ 2I,q)= -  \sn (u,q)$, where the half-period is 
  $I= \ellK(\q)$.}
Here 
we finish with the system of two equations determining 
$\mathcal{E}= \mathcal{E} ( \mathcal{J}_1, \mathcal{J}_2)$
\bea\label{s5}
\lrbrk{\frac{\mathcal{E}}{\ellK(\q)}}^2
-\lrbrk{\frac{\mathcal{J}_1}{\ellE(\q)}}^2=\frac{4}{\pi^2}\,\q\ , \\
\la{ss5}
\lrbrk{\frac{\mathcal{J}_2}{\ellK(\q)-\ellE(\q)}}^2
-\lrbrk{\frac{\mathcal{J}_1}{\ellE(\q)}}^2=\frac{4}{\pi^2}\ ,
\eea
where $\q >0$. 
A manifestation of the analytic 
continuation relation \rf{caa} between the  two   solutions 
is  the  equivalence \ci{bfst} of the  systems 
\rf{a5},\rf{aa5}   and \rf{s5},\rf{ss5} under the 
substitution 
\[\label{gmap}
\mathcal{E}\mapsto - \mathcal{J}_1\ ,\qquad\ \ 
\mathcal{S}\mapsto  \mathcal{J}_2\ ,\qquad\ \ 
\mathcal{J}\mapsto - \mathcal{E} \ , 
\]
{\it and} the analytic continuation from $\q >0$ to $\q<0$ in  the
elliptic integrals.

%\bigskip
%%%%%%%%%%%%%%%%%%%%%%%%%%%%%%%%%%%%%%%%%%%%%%%%%%%%

\subsection{Energy as function of  spins}
%%%%%%%%%%%%%%%%%%%%%%%%%%%%%%%%%%%%%%%%%%%%%%%%%%%%%%%%%%%

Depending on a  region of  parameter space  
(or values of the integrals of motion) one finds different functional 
form of dependence of the energy on the two spins. 
A direct comparison with gauge theory 
 is possible in the case 
 when  the two spins are 
  large compared to $\sql$,  i.e.~$ \S \gg 1, \  \J \gg 1$
  in the $(S,J)$ case and 
 $\J_1 \gg 1 , \  \J_2 \gg 1$  in the $(J_1,J_2)$ case.
We can then expand 
the energies, e.g., in powers of the total $S^5$ spin $\J$.
This  amounts to an  expansion 
in (inverse) 
powers of $\J\equiv \J_3= {J_3\ov \sql} $ in the $(S,J)$ case and 
 of 
$ \J\equiv \J_1+\J_2={1 \ov \sql}( J_1+J_2)
 $ in the $(J_1,J_2)$ case, respectively, 
\[\label{jjs}
E= S + J  + \ { \frac{\lambda}{ J}}\ \td \ep_1 ( {\frac{S}{J}}) + 
 \ { \frac{\lambda^2}{ J^3}}\  \td \ep_2 ( {\frac{S}{ J }})  + \ldots  
 \ ,\ \ \ \  \ \ \ \ \ \ 
  J\equiv J_3, \ \ J, S \gg \sql  \ ,  \]
\[\label{jjss}
 E= J  
+\  { \frac{\lambda}{ J}} \  \ep_1 ( {\frac{J_2}{J}}) + 
 \ { \frac{\lambda^2}{J^3 }}\  \ep_2 ( {\frac{J_2}{J}})  + \ldots   \ , \
 \ \ \ \ 
 \ \ \ \  J\equiv J_1 + J_2,\  \ J_1,J_2 \gg \sql   \ . \]
The coefficient functions  $\td \ep_n$ and $ \ep_n$ 
in \rf{jjs} and \rf{jjss} (the analogs of $c_n$ in \rf{jojk})
 can be  related, given 
that the two solutions are related by the analytic continuation \rf{gmap}. 
By expanding in large spins 
 one finds a simple  relation between the leading-order (``one-loop'')
corrections for the energies of the two solutions \ci{bfst}:
\[ \label{rell}
  \td \ep_1 (y) =  -    \ep_1 (- y  ) \ .\]
The same relation is obtained  also on the gauge-theory side \ci{bfst}.

Eq. \rf{rell}
follows also directly from  \rf{pui} and 
\rf{kui} or the systems \rf{a5},\rf{aa5} and \rf{s5},\rf{ss5}. 
 In the $(J_1,J_2)$ case, 
expanding the parameter $\q$ for large $\J$
 as (with $\J$ being $\J_1 + \J_2$)
\[ \label{xex} 
\q = \q_0 + {\frac{ \q_1 }{\J^2 }} + { \frac{\q_2}{ \J^4} } + \ldots \ ,
 \] 
one finds  that 
$\q_0$ is given by  the  solution of the transcendental equation 
\[\label{xxx}
\frac{\ellE(\q_0)}{\ellK(\q_0)}= 1-  \frac{J_2}{J} \ , \qquad\ \ \  
\q_0= \q_0 ({\frac{J_2}{ J}}) \ .  \]
The rest of the expansion coefficients
in $\q$ and the energy \rf{jjss} are then determined simply by 
 linear algebra. In particular, one finds \ci{bfst}
\be
\label{eep}
\epsilon_1 = \frac{2}{\pi^2} \,\ellK(\q_0)
\big[ {\ellE(\q_0)-(1-\q_0)\ellK(\q_0)}\big]  \ . \ee
In the $(S,J)$ case,  using the same expansion \rf{xex} for the
corresponding 
parameter $\q$ in \rf{qui}  where 
now $\J=\J_3$ we find that  $\q_0$  satisfies 
\[\label{yyra}
\frac{\ellE(\q_0)}{\ellK(\q_0)}=1+\frac{{S}}{{J}}\ , \ \ \ \ \ \ \ \ 
\ \ \   \q_0 = \q_0 ({\frac{S}{ J}}) \ ,  
\]
and  also  
\be \la{yee}
\td \epsilon_1=
-\frac{2}{\pi^2} \,
\ellK(\q_0)\big[\ellE(\q_0)-(1-\q_0)\ellK(\q_0)\big]\  .
\ee
Comparing \rf{xxx},\rf{eep} to  \rf{yyra},\rf{yee}  and
observing 
that to the leading order \rf{caa} implies $J_2 \to S, \ J\to - J$, 
we indeed  observe  the relation \rf{rell}, or 
$(\td \epsilon_1)_{\q_0}= - (\epsilon_1)_{-\q_0}$. 

\bigskip 

%%%%%%%%%%%%%%%%%%%%%%%%%%%%%%
Let us now comment on  dependence of the energy on the spins  in 
other regions of the parameter space. 
Let us start with the $(S,J)$ solution.
In the limit of short strings with $\J \ll 1, \ \S \ll 1$ one finds 
\ci{ft1}
\be \la{pxx}
E= \sqrt{J^2 + 2 \sql S } + ... \ . \ee
This limit   probes  a small-curvature region of
$AdS_5$   where   $\r\approx 0$, and where  the energy spectrum
should thus be approximately the same as in flat space.
Indeed, \rf{pxx} is  the standard  relativistic  expression
for the  energy of a string in flat space
 moving   
 %along a $\vp$-direction
%of $S^5$ circle 
with momentum $\J$   and rotating in 
a 2-plane %in \ads\  
with spin $\S$.
If the  boost energy  is smaller than the rotation one,
i.e. if $ \J^2 \ll \S$, then
we get  the flat-space Regge trajectory relation 
$E \approx \sqrt{2\sql S}   + { J^2 \ov 2 \sqrt{2 \sql S} }.$
Such  expression (with a non-analytic dependence on $\l$) 
cannot be directly compared to SYM theory before one 
computes all quantum string \sm $1\ov \sql$ 
corrections (and resums them to have a regular $\l\to 0$ limit).

For short strings with  $\J \gg 1$ (and thus with $\J \gg  \S$) 
\be \la{koip}
E=   J   + S   +  {\l   S \ov 2 J^2 } + ... \ . \ee
This corresponds to the BMN limit  with $S$ playing the role of the
string excitation number \ci{ft1}.\foot{The BMN case corresponds to 
expanding near a point-like string moving along big circle of $S^5$.
In the limit $J \to \infty,\  \frac{\lambda}{J^2}$=fixed one 
may drop all but quadratic fluctuation terms in the string action
(which becomes then equivalent to the plane-wave \ci{papa} action 
\ci{meet} in the 
light-cone gauge). The energies of fluctuations above the BPS 
ground state $E=J$ are then determined by the string 
fluctuation masses given  
by  ${\rm m}^2= \frac{1}{\J^2}= \frac{\lambda}{J^2}$.}
One may also consider a ``near BMN'' limit ${S\ov J}\ll 1$ of this 
two-spin solution \ci{ft1,bfst}
\[\la{shoo}
E=J+S \sqrt{1+\frac{\lambda}{J^2}}
-\frac{\lambda\,S^2}{2J^3}\cdot \frac{1+{\lambda\ov 2J^2}}{1+{\lambda\ov J^2}}
+\ldots \ , \ \ \ \ \  \ \  \ \ \ \  S \ll J \ .  
\]
This represents  the near BMN limit for a total of $S$ excitations of
the oscillation  modes with $n=\pm 1$. 
Thus solving non-linear  classical 
\sm equations  gives the same  semiclassical spectrum as  expanding
the \sm  action  near 
a point-like geodesic  and then quantizing the small-fluctuation Lagrangian. 
 We see that there is an overlap
 between the leading-order (large $\sql$)
  quantum spectrum   obtained
 by expanding near $S^5$-boosted  point-like string state
 with {\it no}  rotation in \ads and  a
 classical spectrum  obtained by expanding
 near a highly boosted  {\it and} rotating
 string solution.
This supports   the  suggestion  \ci{gkp,ft1,tsec}
 that  parts of semiclassical \adss string spectrum
  can be captured by expanding  near  different classical
 string solutions.

Other asymptotic expressions are found  when $S$ is large.
In this case the string 
can become very long and 
approach the boundary of $AdS_5$, i.e.~$\rho_0\to\infty$. 
For $\J \ll \ln \S, \ \S \gg 1$ one finds
$\E \approx  \S +{1\ov\pi}\ln\S \ +\  
{\pi\J^2\ov 2\ln\S }$,  i.e. \ci{ft1} 
\be\la{longi}
E \approx   S  +{\sql \ov\pi}\ln {S \ov \sql} \ + \
{\pi J ^2\ov 2 \sqrt{\l} \ln { S \ov \sql}  } \ .
\ee
In the limit of $J=0$  this reduces
to  the remarkable $\ln S$ behaviour found 
in \ci{gkp} for the single-spin  $AdS_5$ rotating 
string   solution in  $AdS_5$. 
Having only large $AdS_5$ spin thus does not lead to an analytic dependence of
the energy on $\l$ {\it and}, not surprisingly, 
 is not enough to suppress quantum string \sm 
corrections.
 Indeed, the 1-loop string correction shifts  the  coefficient of the 
$\ln S$  term by a constant \ci{ft1}, and, in general, 
the classical $\sql$ coefficient should thus be   replaced by an 
{\it ``interpolating''} function\foot{The 
one-loop coefficient computed in  \ci{ft1}
is $a_1 \approx - {3\ov 2\pi} \ln 2 .$
 We use this opportunity to 
correct  factor of  1/2 misprints   in eqs. (6.6) and (6.9) 
in \ci{ft1}.}
\be \la{nter}
E =  S  + f(\l) \ln {S } + ... \ , \ \ \ \ \ 
f(\l)_{_{\l \gg 1}} = {\sql \ov\pi} + a_1 + { a_2\ov  \sql } + ... \ . \ee
The AdS/CFT correspondence implies that after a resummation 
 $f(\l)$ should admit a regular 
weak-coupling expansion $f(\l)_{\l \ll  1} = q_1 \l  + q_2 \l^2  + ...$, 
with \rf{nter} reproducing the anomalous dimension 
of the corresponding 
gauge-theory  operators like tr$( \bar \Phi  D^S \Phi)+...$
(see \ci{kot}).

In the intermediate case where 
 $   \ln {\S\ov \J } \ll  \J \ll  \S $  we get \ci{ft1} 
 \[ \la{iiiy} 
E=S+J+\frac{\lambda}{2 \pi^2 J}\, \ln^2 \frac{S}{J} + \ldots
\,,   \,.
\]
In contrast to the large $\J$ limit  of the
short string (small $\S$) case \rf{shoo}  here  the third correction term
is not related to the BMN-type   spectrum:
 there the 
 boost is large and  string oscillations are small, while
in the long-string case the spin $\S$ is always larger than the boost
parameter $\J$.
Eq. \rf{iiiy}  appears to be analytic in $\l$ and, assuming that 
string loop  corrections to the coefficient of the 
$\ln^2\frac{S}{J} $ term are suppressed in the limit $S \gg 1, \ J \gg 1$,
one could  hope to  relate the 
 ${\l \ov 2\pi^2 J }\ln^2  {S\ov J } $
 term  to the 1-loop  anomalous  dimension  of the
 gauge-theory operators   with large  spin and large
  $R$-charge. Indeed, \rf{iiiy} may be viewed as a special case 
  of \rf{jjs}, where $\td \epsilon_1 \approx 
   \frac{1 }{2 \pi^2 }\, \ln^2
  \frac{S}{J}$. This asymptotics is 
   indeed observed on the gauge theory side  as a special case of the general 
   relation between the string theory and the 
   gauge theory results 
   for the function $ \td \epsilon_1 (\frac{S}{J})$ 
      established in  \ci{bfst}. 
 One concludes, in particular,  
  that  the coefficient of the 
    $\ln S$ term  in the anomalous dimensions of the  corresponding
 $\cN=4$ SYM  operators  with large spin {\it and}  large R-charge $J$
 is indeed  suppressed  also at weak 't Hooft coupling.
 
 \bigskip 
 
 A similar analysis  can be repeated for the $(J_1,J_2)$ 
 solution. 
The energy of a short string rotating in $S^5$
with $\J_1 \gg1 , \ \J_1 \gg \J_2$ is given by a 
BMN type expression (cf. \rf{koip})
\[
E=J+ \frac{\lambda\,J_2}{2J^2}+\ldots \,,\qquad \ \  
J_2 \ll J_1 \ , \ \ \  J=J_1 + J_2    \, . 
\]
The full expression in the near BMN limit is (cf.  \rf{shoo})
\[\la{oko} 
E=
J_1 +J_2 \sqrt{1+\frac{\lambda}{J_1^2}}
-\frac{\lambda\,J_2^2}{2J_1^3}\cdot
\frac{1+{3\lambda\ov 2J_1^2}}{1+{\lambda\ov J_1^2}}
+\ldots \ , \ \ \ \ \ \    J_2 \ll J_1 \ .  
\]
To compare to the BMN case 
we may set $J_1=J$  and then $J_2$ represents the number 
of excitations.

Making string longer corresponds to increasing the spin $J_2$. 
For example, at 
$J_1=J_2$ we get \ci{ft4}
\[\label{half}
E=J+ c_1 \frac{\lambda}{J}+\ldots
 \ , \ \ \  c_1 \equiv \ep_1 (\ha ) = 0.356\ldots 
 \ , \ \ \ \ \  J_2=J_1=\ha J \ . 
\]
When $J_2\to J=J_1+J_2$,  
i.e. $J_1$ becomes small, 
 the string extends over half a great $S^5$  circle and \ci{bfst}
\[
E=J+\frac{2\,\lambda}{\pi^2\,J(1-J_2/J)}+\ldots=J+\frac{2\,\lambda}{\pi^2\,J_1}+\ldots
 \ , \ \ \ \ \ \    J_2\approx J \ .  
\]
 The point where $J_2=J$ can be viewed as a transition point: 
 one half of the string can be  unfolded to give a  circular string
discussed below. 
Alternatively, the case of  $J_1=0$  can be studied  by starting 
with a  single-spin folded rotating $S^5$ solution  with its 
center of mass at rest at  a   pole of $S^5$  \ci{gkp}.  In this case 
for   $\J=\J_2\gg 1 $  
one finds \ci{gkp}
\be\la{jkl}
E= J + {2 \ov \pi} \sql + ... \  , \ \ \ \ \ \   \ \ \   { \sql \ov J} \ll 1 
\ .
 \ee
Like in the single-spin $AdS_5$ case 
(cf.\rf{longi}) here the expansion of the energy is not analytic in $\l$, 
and one expects that quantum \sm corrections should 
promote the subleading $\sql $ term into a  nontrivial 
function $h(\l)= 
{2 \ov \pi} \sql + k_1 + {k_2 \ov \sql} + ....$.
% precluding a 
%direct comparison  with perturbative gauge theory.

%%%%%%%%%%%%%%%%%%%%%%%%%%%%%%%%
\subsection{Circular two-spin solution}
%%%%%%%%%%%%%%%%%%%%%%%%%%%%%%

In addition to the ``homogeneous'' circular two-spin  solutions 
discussed in section 6 there are also different 
circular two-spin  solutions of the Neumann system \rf{Laa} with $v_i=0$
  that generalize the ``round circle'' $J_1=J_2$ solution of 
\ci{ft2} to the case of $J_1 \not=J_2$ \ci{afrt}. 
The circular string solution is given  by the same ansatz \rf{jj}
 as for the folded string but now 
$\psi(\sigma)$ is  assumed to be periodic modulo
$2\pi$
\[ \la{pss}
\psi(\sigma+2\pi)=\psi(\sigma)+2\pi k \ .
\]
In what follows we shall set the winding number $k$ to be 1. 
In  general, in spherical
coordinates $(\gamma,\psi)$ the equations of motion \rf{rta}
 describing
this   type of  string are $\gamma=\frac{\pi}{2}$ and $\psi'' +
\ha w_{21}^2 \sin 2\psi =0$, \ $ w_{21}^2= w^2_2 - w^2_1$. 
Integrating  once, we get
$\psi'^2 =w_{21}^2(\q^{-1}  - \sin^2\psi )$, where $\q
$ is  an
integration constant.  If $\q >1$, then $\q^{-1} =\sin^2\psi_0$ and this
solution describes a {\it folded} string  extending from
$-\psi_0 $ to $\psi_0$.
If instead  $\q<1$, then there is no turning point where $\psi' =0$, 
and the solution  describes a circular string
 extending all the way around the equator $\g={\pi\over 2}$
with $\psi $ from $0$ to $2\pi $:
instead of folding back onto itself, the
string wraps completely around a great circle of $S^5$.
 In the limit  $\q\to 0 $, this
solution approaches the circular string with $J_1=J_2$. 
Thus the parameter $\q$ provides an interpolation
between the circular and the folded string
 configurations.
 Note that 
after a rescaling 
$\psi\to \frac{1}{2}\psi$ the equation for $\psi$
 describes a plane motion of a 
 pendulum in a gravitational field.
 Clearly, the rotation 
of the pendulum requires more  energy than the oscillatory motion 
and this explains why the energy of the circular string is bigger 
than that of the folded one.

The radial coordinates in \rf{emb} in the circular 
 case are given by 
(cf. \rf{elip}) 
\be\la{lip}
 \cos  \psi(\s)  = r_1 (\s)= \sn ( A \s, \q) \ , \  \ \ \ \ 
 \sin   \psi(\s)  = r_2 (\s)=   \cn ( A \s, \q)  \ ,\ee
  where again $r_3=0$ and   $A\equiv  {2 \ov \pi} \ellK(\q)$. 
  
The set of equations for the energy and spins 
of   this solution is (cf. \rf{kui}) \ci{afrt}
\bea\la{crcl}
\mathcal{J}_2 &=& \frac{w_2}{\q}\lrbrk{1-
\frac{\ellE(\q)}{\ellK(\q)}}\ , \qquad
\mathcal{J}_1 = \frac{w_1}{\q}\lrbrk{\q-1
+\frac{\ellE(\q)}{\ellK(\q)}},
\\
\mathcal{E}^2&=& w_1^2+\frac{1}{\q}(w_2^2-w_1^2)\    ,\qquad
\ellK(\q)= \frac{\pi}{2}\sqrt{\frac{1}{\q}( w_2^2-w_1^2    )}\ . 
\eea
Solving  for $w_1,w_2$ we get a system of two equations
for $\E=\E(\J_1,\J_2)$ 
similar to the one in \rf{s5},\rf{ss5} 
\< \la{yyo}
\lrbrk{\frac{\mathcal{E}}{\ellK(\q)}}^2-
\lrbrk{\frac{\q\mathcal{J}_1}{(1-\q)\ellK(\q)-\ellE(\q)}}^2
=\frac{4}{\pi^2},\\ \la{yyoo}
\lrbrk{\frac{\q\mathcal{J}_2}{\ellK(\q)-\ellE(\q)}}^2-
\lrbrk{\frac{\q\mathcal{J}_1}{(1-\q)\ellK(\q)-\ellE(\q)}}^2
=\frac{4}{\pi^2}\,\q\ .
\>
Note that the  ansatz for the circular solution is symmetric
 under 
$\mathcal{J}_1\leftrightarrow\mathcal{J}_2$,\foot{The direct
limit  $J_2=0$  is 
not, however, well-defined for a circular $(J_1,J_2)$ case.} 
and this 
symmetry may be seen  by applying a 
modular transformation to   the elliptic integrals \ci{afrt,bfst}:
$
\ellK(\q)=\sqrt{1-\q'}\,\,\ellK(\q'),\ \
\ellE(\q)=\frac{\ellE(q')}{\sqrt{1-\q'}},\ 
1-\q= \frac{1}{1-\q'}.$ 

In the limit when both spins are large 
we can expand $\q$  and the energy   in powers of $1\ov \mathcal{J}^2$, 
i.e.  $\q=\q_0+{\q_1\ov \mathcal{J}^2} +...$, 
$\J=\J_1+\J_2,$ and 
  (cf.  \rf{xex},\rf{xxx},\rf{eep})
\bea \la{hio} 
E&=& J + {\l \ov J} \ \hat \epsilon_1 (\frac{J_2}{J})  + .... \ , \\
\la{hyh}
\hat\epsilon_1=\frac{2}{\pi^2}
\ellK(\q_0)\ellE(\q_0)\ ,&& \qquad 
\frac{J_2}{J}=\frac{1}{\q_0}
\big[1-\frac{\ellE(\q_0)}{\ellK(\q_0)}\big] \ .
\eea
The same relations
 for $\hat \epsilon_1 (\frac{J_2}{J})$ 
 in \rf{hyh}  were  reproduced  for  the corresponding 
  1-loop anomalous dimension 
on the gauge theory side \ci{mz2,afrt,bfst}.

\bigskip

One may also construct  other similar 
 solutions, e.g.,   by combining a 
folded string solution  in $AdS_5$ with folded or circular $(J_1,J_2)$ 
 $S^5$ solution. 
 %(instead of just having the string 
 %being point-like in $S^5$).
   In this case the energy 
will be given by a system of { three} 
 parametric equations involving 
$\E,\S,\J_1,\J_2$. 
 Such $(S,J_1,J_2)$ solutions 
may  be related   by an analytic continuation to special 
$(J_1,J_2,J_3)$ solutions. 

\bigskip

To conclude, as  we have seen on    the examples discussed above,  
to have  regular \rf{jojk}  dependence of the string 
energy on ${\l\ov J^2} \ll 1$ we need  at least one large
 ``center-of-mass''  momentum  in $S^5$. 
In such  case quantum sigma model  corrections are expected to be 
suppressed in the limit $J \gg 1$, and thus 
the classical string energy should represent 
the exact gauge-theory anomalous 
dimension computed in the limit $ J \gg 1  $, 
to all orders in perturbative expansion in $\l$.
This was  explicitly verified 
(for the leading ``one-loop'' ${\l\ov J}$ term) 
 for several types of such ``regular''
spinning string solutions  \ci{mz2,afrt,bfst,mz3}. 

Having  large spins in $AdS_5$ only or only one  large 
spin (with center of mass being at rest) in $S^5$ 
appears to be  not enough
for the energy to have an expansion in even  powers of 
${\sql \ov J}$ (note that the circular $(S,J)$ solution 
of the NR system  
discussed in section  6.2 represents an exception from 
this rule, cf.\rf{spl}).
 In these latter cases 
quantum \sm corrections are not expected to be suppressed in the large 
spin limit and thus the classical $\sql$-coefficients
of the leading terms in the expansion of the energy  
should become promoted by the string quantum $1\ov (\sql)^n$
 corrections 
to  non-trivial ``interpolating'' 
functions of $\l$. The latter  should be resummed 
before one may  try to compare to  perturbative gauge-theory results. 
%This remains beyond the presently available techniques
Comparing string theory to gauge theory 
at a {\it quantitative}  level in such  cases  remains a  challenge.

 %%%%%%%%%%%%%%%%%%%%%%%%%%%%%%%%%%%%%%%%%%%%
\setcounter{equation}{0}

%%%%%%%%%%%%%%%%%%%%%%%%%%%%%%%%%%%%%%%%%%%%
\setcounter{footnote}{0}
%%%%%%%%%%%%%%%%%%%%%%%%%%%%%%%%%%%
%%%%%%%%%%%%%%%%%%%%%%%%%%%%%%%%%%%%%%%%%%%%%
\section{Open questions and  generalizations}
%%%%%%%%%%%%%%%%%%%%%%%%%%%%%%%%%%%%%%%%%%

The  above discussion  of particular string solutions 
with ``regular'' expansion  of the energy $E$ in powers of 
$\l \ov J^2$  raises several questions. 

Since the direct comparison with  gauge theory  at  present
can be done  only in the $J\to \infty$ limit  and in 
expansion in $ {\l \ov J^2} \ll 1$ it would be interesting to classify 
all possible solutions with such property \rf{jojk}
of the energy. 
One may also  try to derive the general expression  for 
the leading order coefficient 
 in $E= J + {\l \ov J} c_1  + ...$, i.e.  
for $c_1$ as  a functional on a  space of such solutions. 
An interesting work in this direction is \ci{mik} which utilizes the 
observation \ci{matt} that the induced world-sheet  metric  
of rotating strings with  large $J$ becomes degenerate, and  that one 
can then  develop a perturbative expansion near such world sheet.  
Deriving equations for the functional $c_1$
(with the expressions in 
\rf{ennl}, \rf{eep},\rf{yee}  and  \rf{hyh} as special solutions)
may help to establish the 
correspondence with spin chain energy eigenstates 
in a more universal way than the presently 
known  procedure based on 
association of a particular Bethe 
root distribution with  a particular string solution 
  \ci{mz2,bfst,mz3}.

It remains also to prove   that for all solutions 
with ``regular'' expansion of the energy in $ {\l \ov J^2} \ll 1$
quantum superstring \sm corrections are indeed suppressed 
by extra ${1\ov J^n}$ factors as in \rf{oppu}. The underlying 
supersymmetry of the 
\adss string theory is certainly important for that  conclusion,  
and a   possible role  of asymptotic 
supersymmetry at  finite $J$ and $\l\to 0$
 observed in \ci{matt}  for   
 simple $S^5$ rotating solutions 
remains to be clarified. 
The string/gauge theory 
matching for  a  pulsating solution in \ci{mz3} 
(when the $\l\to 0$ limit does not give a BPS state) 
seems to indicate 
that suppression of quantum \sm corrections may occur  even 
under more general conditions.

More generally, as discussed in section 1, 
the full expression for the classical energy 
of a ``regular'' solution 
$E= \sql \E ({ J\ov \sql},...)$ 
should be 
representing the exact dimensions of the corresponding gauge-theory
operators  computed in the large $J$ limit.
 Here we assume that like the energy 
on the string side, the anomalous dimension on the gauge theory side 
should admit a regular double expansion in ${\l\ov J^2} < 1 $   and
${1\ov J}\to 0$, i.e. should have the form 
\rf{ity},\rf{jz}.
 This remains to be proved in general
 for multi R-charge/spin  SYM  operators.
 The full  expression for the classical energy 
 should be  a solution of some differential 
  equations  or an equivalent system of 
 algebraic or  transcendental  equations involving 
 moduli parameters  of the string  solutions 
  (cf. \rf{kqp}-\rf{seo} or \rf{a5},\rf{aa5} and 
  \rf{yyo},\rf{yyoo}, see also \ci{art}). Thus,  
   like in the examples discussed above, the 
    full expression for the energy $E$   
     should  be  effectively 
   determined by its leading-order term $ c_1$. 
That suggests a possibility to  derive (in large $J$ limit)
the exact in $\l$  expressions for the corresponding 
anomalous dimensions in SYM theory. 
By analogy with  how the simple 
square root expression $\sqrt{1 + {\l \ov J^2} n^2} $ 
of  the near-BPS BMN case  was reproduced in \ci{zan}, 
one may  expect that in the $J \to \infty$ limit 
there  may  then be a relation between the  values of anomalous
dimensions (or, in fact, the expressions for the dilatation 
operator restricted to a particular subsector of states) 
at different orders in expansion in $\l$. 

Let us add  that while the full expression for  the classical
string energy comes 
from the conformal gauge constraint (and  looks like a
``relativistic'' expression $E^2 = J^2 + 2\sql c_1 + ...$)
the 1-loop anomalous dimensions on the SYM side are obtained 
by solving the quantum spin chain Hamiltonian eigenvalue  problem 
($\D-J = a_1 {\l \ov J} + ...$), which looks like a first term in 
a ``non-relativistic'' expansion. It would be 
important to understand how the perturbative series on the SYM 
side can be summed up, i.e. how the 1-loop expression for anomalous 
dimension can be promoted to the full ``relativistic'' expression 
without order-by-order analysis of modification of the 
dilatation operator (interpreted as a generalised ``non-local'' 
 spin chain Hamiltonian, cf. \ci{beik,bei3}).

Another  interesting problem is to  compare 
subleading terms 
in $1/J$ expansion, as was done in the BMN case in
\ci{par,cal}.\foot{Since the classical energy 
of multi-spin (say $(J_1,J_2)$) solutions
  reproduces  the 
near-BMN spectrum in the limit $J_1 \gg J_2$ 
(see \rf{oko}), 
 computing the 1-loop  \sm correction to the  energy 
would the  effectively determine the $1/J$ (2-loop, in BMN case)
correction to the BMN spectrum and thus could be compared 
to the result of \ci{par,cal}.}
This will  involve computing one-loop correction 
to the classical string energy \rf{oppu} and comparing it with 
the subleading  correction to  the ``thermodynamic'' limit 
of the 1-loop  Bethe   energies. 
Note  that  the 1-loop (order $\l$) SYM 
result for the anomalous dimension for any value of $J$, i.e. 
$q_1(J)$ in \rf{ity}, should  represent the  sum of all string \sm
loop corrections to the leading  $\l \ov J$ term ($c_1$ coefficient)
 in \rf{jojk}. 
The subleading $1/J$ terms should be 
governed by the same integrable structures on the two sides of the
duality. 
 For any value of $J$, one certainly expects that, 
in view of  the  conformal invariance of 
the \adss string theory (absence of mass generation),
the  classical integrability of 
the \adss  \sm \ci{ben} should have a direct 
extension to the quantum level. 
On the $\N=4$ SYM side, there are strong indications
 that the one-
and two-loop integrability of the dilatation operator 
extends to all 
loop orders  \ci{beik,bei3,plef}.

It is  important   to understand  if the 
precise check of the string-theory / gauge-theory 
correspondence in the large spin sector of  states may be 
 extended to other semiclassical string  states with  large oscillation
 numbers. An indication that this is indeed 
 the case comes from  recent work \ci{mz3}. 
 As was noticed earlier in \ci{mina}, 
 the circular string oscillating in $S^5$ (but not in $AdS_5$) 
 has energy that admits a regular expansion in $\l\ov \rN^2$, where 
 $\rN$ is the oscillation level number. The leading term in 
 this expansion was matched in \ci{mz2,mz3} onto  a particular 
 eigenvalue of the corresponding \ci{mz1} $SO(6)$ 
 spin chain Hamiltonian.
  
This raises the question of generalization of the rotation ansatz 
\rf{emb},\rf{adr} of the previous sections to include 
the possibility of string oscillations, i.e. 
of changing of string shape 
in time. It is not a priori clear which  should be the most general 
rotation/oscillation 
ansatz for the $\s$ and $\tau$ dependence of the \adss coordinates
 consistent with the full 2-d classical \sm equations of
motion, but for each consistent ansatz one should expect that 
the 2-d \sm  should  again reduce to an 
integrable 1-d system, whose solutions (and thus their energies)
 could  be found in a relatively explicit way.
\foot{An alternative to 
 direct procedure of finding  classical solutions may be the 
  semiclassical 
 quantization method used in \ci{vesa,mina,mz3}.} 
 
An example is provided  
by  a ``2d-dual'' version of the  rotation ansatz
\rf{emb} with  $\tau$ and $\sigma$ 
 interchanged (but   keeping  
 the $AdS_5$ time as $t=\k \tau$), i.e. \ci{art} 
\be\la{tat}
\XX_i = z_i(\tau)\ e^{i m_i \s} = 
r_i(\tau) e^{ i \a_i (\t) + i m_i \s} 
  \ , \ \ \ \ \ \  \ \ \ \ \ \ \ \ss r^2_i(\t) =1 \ . \ee
 In this case the radial directions depend on $\t$ instead of $\s$ 
  and the ``frequencies''  $m_i$ must take integer values 
 in order  to satisfy the closed string periodicity 
 condition. 
  This ansatz  describes an ``oscillating'' or ``pulsating''
  $S^5$  string configurations, 
   special cases of which (with motion in both $AdS_5$ and
   $S^5$) 
   were discussed previously in \ci{vesa,gkp,mina,larse,mz3}.
    Since the sigma model  Lagrangian \rf{SSL} is 
   formally invariant under $\s \leftrightarrow \t$, the resulting  
1-d effective Lagrangian will have essentially 
the same form as \rf{L},\rf{La} 
 \be
\label{Le}
L=\frac{1}{2}\sum^3_{i=1} ( \dot z_i \dot z^*_i -  m_i^2  z_i
z^*_i) +  \frac{1}{2} \Lambda(\sum^3_{i=1}  z_i z^*_i-1) \, .
\ee
Solving for $\dot \a_i$ as in \rf{ank} we get  
$r^2_i \dot \a_i = \J_i$=const, where  the counterparts of 
the integration constants $v_i$  are now
 the angular momenta in \rf{spins}.
 Then  we end up with the following analogue of 
 \rf{Laa}
  \be
\label{Lae}
L=\frac{1}{2}\sum^3_{i=1} \big(  \dot r^2_i   -   m_i^2  r^2_i -
   { \J_i^2 \ov r_i^2}
\big)  +   \frac{1}{2} \Lambda(\sum^3_{i=1}  r^2_i-1) \,  . 
\ee
Thus pulsating solutions (carrying also 3 spins $\J_i$) 
are again  described by  a special 
Neumann-Rosochatius integrable system \ci{art}. 
Since  the  corresponding conformal
gauge constraints are  also $\t \leftrightarrow \s$ symmetric,  
they take the form similar  to \rf{cv},\rf{cvv} or
 \rf{cve},\rf{cvve}: 
$
\kappa^2=\sum_{i=1}^3( \dot r^2_i  + 
m_i^2 r^2_i  +  { \J_i^2 \ov r^2_i} )      $ and 
%\ \ \ \ \ \ \ \ \ \ \
$\ \sum_{i=1}^3  m_i \J_i  =0   .$
 One may then look for   periodic  solutions of the above NR
 system \rf{Lae} subject to the above constraints constraint, 
  i.e. having finite 
 1-d energy. 
 The resulting class of pulsating string
 solutions  deserves a  detailed study.
In the simplest (``elliptic'') case reducing to a sine-Gordon 
type system we may follow \ci{dhn,vesa,mina} and  introduce, as
  for any periodic solitonic solution,   an 
   oscillation ``level number'' $\rN $.
%  This  may be achieved  by   considering a
% semiclassical (WKB) quantization of the action \rf{Lae}. 
 In the case of the $S^5$ pulsating solution in \ci{mina,mz3} 
  the expansion of the energy at large level $\rN\gg 1$ appears to be 
  regular in $\l\ov \rN^2$  and, moreover, 
  the leading $\l\ov \rN$ term in $E$ 
  can  be matched onto a particular  
  anomalous dimension on the SYM side   \ci{mz2,mz3}.
 
  \bigskip 
  
One would certainly like to go beyond comparison of particular
 string states to particular  SYM operators and to 
 establish a more general  relation between
  the string  \sm   and the dilatation operator
  on the SYM side, implied by the emergence of similar integrable 
  structures on the two sides  \ci{AS,mz3}. The spin chain Hamiltonian 
  may be  associated  (in a thermodynamic limit) 
  to an effective  coset \sm with the same global symmetries.  
   Then one may hope to relate this \sm 
     to ($J \to \infty$ limit of)
     the string \sm  
   by a kind of
   % non-local
   duality transformation (for a recent progress see \ci{kru}).\foot{
   %The spatial direction of  the spin 
  % chain may be interepreted as
 % a ``momentum'' direction from the point of view 
  % of the string sigma model.
   Note also  that there is
    a known relation between a
    discrete version of the Neumann  model  and 
spin chains \ci{gorrr}, 
but discretising $\s$ in the string \sm seems to imply a departure from the
planar limit on the dual SYM side.}

%%%%%%%%%%%%%%%%%%%%%%%%%%%%%%%%%%%%%%%%%%%%%%%%
  
Finally, one would like also 
 to extend the successes of checking 
the gauge/string 
duality in the non-supersymmetric 
semiclassical sectors of states from  $\cN=4$ SYM theory 
to less supersymmetric gauge theories.
 As was already mentioned in the introduction, 
 evidence of integrable  
structures 
in the high-energy  (near-conformal) limit of QCD 
appeared in \ci{lip,bra,beli,belm,lipk}, and  so the spin chain relation 
 of the the 1-loop dilatation operator of $\N=4$ SYM theory 
\ci{mz1,BM,bel} should   have  generalizations to other 
$\cN=1,2$ supersymmetric theories (and not only in  twist 
2 sector \ci{wwu}).
On the string side, while finding similar
classical rotating string 
solutions 
in other less supersymmetric conformal $AdS_5 \times M^5$ 
(and non-conformal, see, e.g., \ci{vega,other})  backgrounds 
in type IIB theory or its orbifolds 
is,  in principle,  straightforward, it is not clear if the string 
\sm corrections to the leading terms in the classical 
energy  are again suppressed in the $J\to \infty$ limit. 
For example, in the single $S^5$ spin point-like string  (BMN) case 
in  type 0  string 
theory setting \ci{kt} there is a non-trivial 
 1-loop string correction  to the energy of the 
 twisted-sector states
 (which is non-analytic in $ \l \ov J^2$ but going to zero when 
 $ {\l \ov J^2}\to 0$)  \ci{gira}. 
 Similar corrections are expected also for extended spinning string
 solutions, complicating direct comparison to perturbative gauge theory
 results.

%%%%%%%%%%%%%%%%%%%%%%%%%%%%%%%%%%%%%%%%%%%%%%
\section*{Acknowledgments}
%%%%%%%%%%%%%%%%%%%%%%%%%%%%%%%%%%%%%%%%%%
We are grateful to G. Arutyunov, N. Beisert, 
S. Frolov,  J. Russo,   M. Staudacher and K. Zarembo  for
collaborations   and many  important discussions.
We  thank   D. Mateos and  A. Mikhailov  
for useful remarks. We acknowledge A. Belitsky, 
 V. Braun,  G. Korchemsky, H. Nastase   and A. Vainshtein 
for useful   explanations,  comments on the literature and discussions.
This work  was supported by the DOE  grant
DE-FG02-91ER40690,  INTAS  99-1590  and the Royal
Society  Wolfson award.

%\pagebreak
%%%%%%%%%%%%%%%%%%%%%%%%%%%%%%%%%%%%%%%%%%%%%%%%%%%%%%%

%\appendix
%%%%%%%%%%%%%%%%%%%%%%%%%%%%%%%%%%%%%%%%%%%%%%%%%%%
%\section{Definitions of elliptic integrals and elliptic functions}
%%%%%%%%%%%%%%%%%%%%%%%%%%%%%%%%%%%%%%%%%%%%%%%%%%%%%%%%%%%%
%\renewcommand{\theequation}{A.\arabic{equation}}
%%%%%%%%%%%%%%%%%%%%%%%%%%%%%%%%%%%%%%%%%%%%%%%%%%%%%%%%%%%
%\section{Second Appendix}
%%%%%%%%%%%%%%%%%%%%%%%%%%%%%%%%%%%%%%%%%%%%%%%%%%%%
%\renewcommand{\theequation}{B.\arabic{equation}}

%%%%%%%%%%%%%%%%%%%%%%%%%%%%%%%%%%%%%%%%%%%%%%%%%%%%%%%%%%%
%\begin{thebibliography}{0}

\end{document}